%% file: ms_axion_search_TH.tex
\documentclass[pteplogo]{ptephy_v2}

\usepackage{bm}
\usepackage{hyperref}
\usepackage{aas_macros}
\definecolor{MyBlue}{rgb}{0.15,0.15,0.70}
\definecolor{Dgreen}{rgb}{0,0.7,0.0}

\hypersetup{
colorlinks=true,
citecolor=MyBlue,
linkcolor=MyBlue,
urlcolor=MyBlue
}

\usepackage{amssymb}
\usepackage{amsmath}
\usepackage{amsfonts}
\usepackage{upgreek}
\usepackage{latexsym}
\usepackage{appendix}
\usepackage{eufrak}
\usepackage{dsfont}

\usepackage[export]{adjustbox}

\newcommand\spart{\;\raise1.0pt\hbox{/}\hskip-6pt\partial}
\newcommand\spartb{\;\overline{\raise1.0pt\hbox{/}\hskip-6pt\partial}}

\newcommand{\bfB}{\boldsymbol{B}}

\newcommand{\bfE}{\boldsymbol{E}}
\newcommand{\bfH}{\boldsymbol{H}}
\newcommand{\bfJ}{\boldsymbol{J}}

\newcommand{\bfr}{\boldsymbol{r}}
\newcommand{\bfX}{\boldsymbol{X}}
\newcommand{\bfY}{\boldsymbol{Y}}
\newcommand{\bfPhi}{\boldsymbol{\Phi}}
\newcommand{\bfPsi}{\boldsymbol{\Psi}}
\newcommand{\gag}{g_{\rm a\gamma}}
\newcommand{\mass}{m_{\rm a}}

\begin{document}
\title{Signature of axion dark matter in low-frequency terrestrial electromagnetic fields: formulation and predictions}

\author[1,2,3]{Atsushi Taruya}
\affil[1]{Center for Gravitational Physics and Quantum Information, Yukawa Institute for Theoretical Physics, Kyoto University, Kyoto 606-8502, Japan \email{ataruya@yukawa.kyoto-u.ac.jp}}

\affil[2]{Kavli Institute for the Physics and Mathematics of the Universe, Todai Institutes for Advanced Study, The University of Tokyo, (Kavli IPMU, WPI), Kashiwa, Chiba 277-8583, Japan}

\affil[3]{Korea Institute for Advanced Study, 85 Hoegiro, Dongdaemun-gu, Seoul 02455, Republic of Korea}

\author[4,5]{Atsushi Nishizawa}

\affil[4]{Physics Program, Graduate School of Advanced Science and Engineering, Hiroshima University, Higashi-Hiroshima, Hiroshima 739-8526, Japan}

\affil[5]{Astrophysical Science Center, Hiroshima University, Higashi-Hiroshima, Hiroshima 739-8526, Japan}

\author[6]{Yoshiaki Himemoto}
\affil[6]{Department of Liberal Arts and Basic Sciences, College of Industrial Technology, Nihon University, Narashino, Chiba 275-8576, Japan}

\begin{abstract}
We develop a theoretical framework for axion dark matter (DM) searches using terrestrial electromagnetic (EM) fields, enabling a global and quantitative characterization of the signal across the Earth. Axions couple to the geomagnetic field and generate a monochromatic EM signal at a frequency set by the axion mass. Incorporating a realistic atmospheric conductivity, we describe the axion-induced EM waves confined in the Earth–ionosphere cavity, avoiding the divergences present in idealized treatments. Our semi-analytical method yields quantitative predictions for the axion-induced magnetic field near the Earth's surface, and we found that (i) at $m_a\gtrsim10^{-14}$ eV, the signal exhibits resonance structures aligned with Schumann resonances, and the magnetic field amplitude is especially enhanced at $\mass\sim3\times10^{-14}$ eV. (ii) The signal amplitude and orientation also vary with geographic location, with Southeast Asia offering the strongest sensitivity. These predictions are insensitive to uncertainties in conductivity models and boundary conditions. Those distinctive features provide a reliable template to distinguish axion-induced signals from natural or anthropogenic EM backgrounds, and the formalism can be extended to other DM candidates such as dark photons.
\end{abstract}
\preprintnumber{YITP-25-144}

\subjectindex{B71, C15, E64, E74}

\maketitle

\section{Introduction}
\label{sec:intro}

The dark matter (DM), invisible matter component, is an essential constituent accounting for $\sim30\%$ of the energy contents of the Universe. Various astronomical observations along with the cosmic structure formation theory suggest that the DM is non baryonic, and have negligible velocity dispersion (e.g., Ref.~\cite{Bertone_Hooper_Silk2005} for a review). While such a hypothetical matter content successfully describes large-scale cosmological observations, including the cosmic microwave background experiments and galaxy surveys, the origin of DM yet remains unclear. It has to be described by the beyond-Standard Model, and there are many possibilities proposed, whose DM candidates span a extremely wide range of mass, from $10^{-22}$\,eV to $10^{66}$\,eV (e.g., Refs.~\cite{Feng_2010,Snowmass_DarkMatter2021}). 

Among various candidates, axions are one of the representative DM having extremely small masses \cite{Preskill:1982cy,Abbott:1982af,Dine:1982ah}. The axion is originally proposed to resolve the strong CP problem in quantum chromodynamics \cite{peccei_quinn, Weinberg:1977ma, Wilczek:1977pj}. Later, it has been pointed out that a plenty of axion-like particles is predicted over a wide range of mass scales in string theory \cite{Svrcek_Witten2006,Arvanitaki_etal2010}. A plausible range of axion DM then reaches as low as $10^{-22}$\,eV, with axion masses around this scale having  attracted particular attention in cosmic structure formation (Refs.~\cite{Marsh_review2016, Niemeyer_review2020, Ferreira_review2020, Hui_review2021} for review, see also Ref.~\cite{Hui_etal2017}), as such DM has been  suggested to alleviate the so-called small-scale crisis (e.g., Refs.~\cite{Bullock_Boylan-Kolchin_review2017,Tulin_Yu_review2018} for review), referred to as the fuzzy dark matter \cite{Hu_Barkana_Gruzinov2000}. 

Although DM is supposed to have only very weak interactions with standard-model particles, axions typically possess a coupling to electromagnetism, which offers a direct route to search for axions in laboratory experiments and from astronomical observations, simultaneously placing a tight constraint on its coupling. A number of such experiments and/or observations are based on axion-photon conversion in strong magnetic fields, including the helioscope and haloscope, which specifically target axions of extraterrestrial origin. The magnetic fields involved in those searches are produced in the laboratory by ferromagnets or solenoids carrying a strong electric current (e.g., Refs.~\cite{Marsh_review2016, Irastorza_Redondo2018, Sikivie_2021review, Adams:2022pbo}). 

In this paper, we consider a novel axion DM search, making use of the terrestrial environment. Unlike dedicated experiments mentioned above, the experimental setup we consider needs only the long-term monitoring data for axion-induced electromagnetic (EM) waves. As pointed out by Ref.~\cite{Arza_etal2022} (see also Refs.~\cite{Sulai:2023zqw, Friel:2024shg,Arza_etal2025}), in the presence of geomagnetic fields as a representative global and static EM field, the coherently oscillating axions produce an effective alternating current, leading to the EM waves with its frequency determined by the mass of axion DM\footnote{Ref.~\cite{Davouldiasl_Huber2006} also considered the coupling of axions to the Earth's magnetic field and pointed out the possibility that solar axions could be  converted into X-ray photons in the magnetosphere. },  whose frequency is given by $f=2.4\,(\mass/10^{-14}{\rm eV})$ Hz. Among a wide range of possible frequencies, the low-frequency bands of $f\lesssim10^2$\,Hz, corresponding to the axion mass of $\mass\lesssim10^{-12}$\,eV, are an optimal window for DM search, less affected by artificial EM waves. Importantly, such low-frequency EM waves, once generated near the Earth's surface, are confined between the Earth's surface and ionosphere. Using the long-term monitoring data of magnetic fields, SuperMAG, Refs.~\cite{Fedderke_etal2021_superMAG,Arza_etal2022,Friel:2024shg} have performed an intensive search for the axion DM or dark photon DM, and they placed a stringent constraint on their coupling parameters in the mass range, $2\times10^{-18}{\rm eV}\lesssim\mass\lesssim4\times10^{-15}{\rm eV}$, corresponding to the frequencies of $5\times10^{-4}\,{\rm Hz} -1\,{\rm Hz}$ (see also Refs.~\cite{Sulai:2023zqw,Arza_etal2025}). 

Here, we are particularly interested in the axion DM search in the frequency range above $1$\,Hz.  The wavelength of such EM waves is comparable to or longer than the Earth circumference. Previous studies~\cite{Arza_etal2022,Friel:2024shg} have focused on the low-frequency regime ($f \lesssim 1$ Hz), 
where a simple calculation based on a spherical vacuum cavity enclosed by perfectly conducting shells provides an adequate description. Using this prediction, Refs.~\cite{Sulai:2023zqw,Arza_etal2025} attempted to extend the analysis beyond $1$\,Hz.
However, at higher frequencies, the resonant behavior of electromagnetic waves becomes significant. In this regime, the simple calculation breaks down, leading to an apparent divergence near the resonant frequencies. Thus, a more refined theoretical treatment that properly incorporates atmospheric conductivity is required to make reliable and quantitative predictions while avoiding singular behavior. 

The aim of this paper is to develop a theoretical framework for calculating axion-induced EM signals, particularly focusing on frequencies higher than those considered in Refs.~\cite{Arza_etal2022,Friel:2024shg}, $f\gtrsim1$\,Hz. 
 This enables us to search for axion signatures and to place a limit on the axion-photon coupling parameter using high-frequency EM dataset, as presenting in Ref.~\cite{NTH_DataAnalysis}. The key findings in Ref.~\cite{NTH_DataAnalysis}, together with the theoretical insights developed in this work, are also highlighted in the companion letter, Ref.~\cite{TNH_Letter}. The formalism presented here is also and extendible, with potential application to other DM candidates. As an illustration, we have applied it to dark photon DM in Ref.~\cite{NNTH_Darkphoton}.  

In this paper, in addition to presenting the formalism and technical details of the computation of axion-induced EM signals, we elucidate the physical properties of the axion-induced EM waves confined between the Earth's surface and ionosphere. With particular focus on their magnetic component, we study their radial structures and geographical dependence, extending previous site-specific predictions (Refs.~\cite{NTH_DataAnalysis,TNH_Letter}) to a global characterization across the Earth. These properties help distinguish the axion signals from other sources that produce similar EM waves and provide a physically grounded basis for interpreting the signal and for designing optimal detection strategies for future searches using low-frequency terrestrial EM observations. We further assess the robustness of the theoretical predictions by examining the impact of the atmospheric conductivity profile, which is a key source of uncertainty, as well as the validity of the assumptions adopted in the calculation.

This paper is organized as follows. In Sec.~\ref{sec:axion_signal}, we begin by reviewing the basic concept of axion DM searches with terrestrial EM waves, and provides a back-of-the-envelope estimate of the expected signal. Section~\ref{sec:formulation} describes the theoretical formalism to compute the axion-induced EM field confined in the Earth-ionosphere cavity, taking the atmospheric conductivity into account. In Sec~\ref{sec:results}, we present results for the axion-induced magnetic field in detail, including its dependence on the axion DM properties as well as  geographical location.  
Section~\ref{sec:discussions} discusses the systematic impacts of the assumptions and treatments in our formalism, and examine the robustness of our calculations with respect to the modeling of atmospheric conductivity. Finally, Sec~\ref{sec:conclusion} provides the conclusions and discusses future prospects for DM searches.

Throughout the paper, unless otherwise explicitly stated, we use natural units with $c=\hbar=1$.

\section{Axion DM signature in the terrestrial EM waves}
\label{sec:axion_signal}

In this section, we begin by reviewing the axion dark matter signature in the terrestrial electromagnetic fields \cite{Arza_etal2022,TNH_Letter} (see also \cite{Fedderke_etal2021} for a general consideration on the terrestrial EM environment in DM searches). 

As described in Sec.~\ref{sec:intro}, one prominent property of the axions is that it is weakly coupled with EM fields through the interaction Lagrangian given by $\mathcal{L}_{\rm int}=(\gag/4)\,a\tilde{F}_{\mu\nu}F_{\mu\nu}$, where $a$ is the axion field and $F_{\mu\nu}$ and $\tilde{F}_{\mu\nu}$ are respectively the EM field strength tensor and its dual defined by $\tilde{F}_{\mu\nu}\equiv\epsilon^{\mu\nu\alpha\beta}F_{\mu\nu}/2$ with $\epsilon^{\mu\nu\alpha\beta}$ being the Levi-Civita symbol. In terms of EM fields $\bfE$ and $\bfB$, this interaction is rewritten in the form as $\mathcal{L}_{\rm int}=-\gag\,a\,\bfE\cdot\bfB$, which leads to modification of Maxwell's equations, resulting in the effective charge and current sources. In particular, if the axion constitutes dark matter, it undergoes coherent oscillation like $a\propto e^{-i\mass t}$, and the corresponding modification appears as an effective alternating current, $\bfJ_{\rm eff}=-\gag\,(\partial_t a)\,\bfB$ (see Sec.~\ref{subsec:mode_equation}). 
This implies that when coupled with a
static magnetic field, the frequency of produced current is determined solely by the axion mass, $f_{\rm a}=\mass/(2\pi)\simeq2.4(\mass/10^{-14}\,{\rm eV})\,$Hz. This can serve as a source of monochromatic EM waves at the same frequency $f_{\rm a}$ via the modified Am\'pere-Maxwell law.  
\begin{figure}[tb]
\centering
 \includegraphics[width=10cm,angle=0]{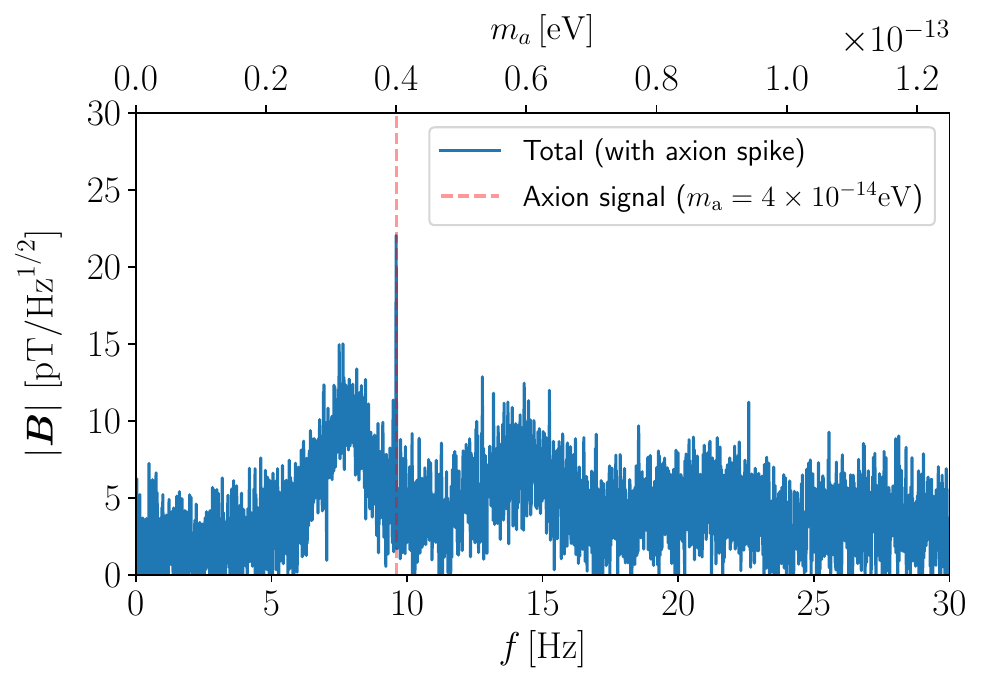}
\caption{Schematic illustration of a magnetic field spectrum showing an axion-induced signal with a mass of $4\times10^{-14}$\,eV,  embedded in  background noise. Unlike the stochastic background, the axion signal is persistent and appears as a sharp spectral spike with an extremely narrow width.
\label{fig:schematic_signal}
}
\end{figure}

Assuming the characteristic scale of $R$ for the axion-induced EM wave, its magnetic field amplitude,  $\bfB_{\rm a}$, is estimated roughly by equating the terms $\nabla\times\bfB$ and $\bfJ_{\rm eff}$, which gives $\bfB_{\rm a}\sim\gag\,\mass\,a_0\,R\,|\bfB_0|$, with $a_0$ and $\bfB_0$ being the amplitude of axion field and static magnetic field, respectively. 

In this paper, we take the geomagnetic field as a typical static magnetic field, with an amplitude of 
$|\bfB_{\rm geo}|\sim20-65\,\mu$\,T and an approximately dipolar configuration. Setting $\bfB_0$ to $\bfB_{\rm geo}$, the induced magnetic field is estimated to give \cite{TNH_Letter,NTH_DataAnalysis}
\begin{align}
    |\bfB_{\rm a}|& \sim 0.3\,{\rm pT} \left(\frac{\gag}{10^{-10}\,{\rm GeV}^{-1}}\right)
    \left(\frac{\rho_{\rm DM}}{0.3\,{\rm GeV}\,{\rm cm}^{-3}}\right)^{1/2}
   \left(\frac{R}{R_{\rm E}}\right)\left(\frac{|\bfB_{\rm geo}|}{50\,\mu{\rm T}}\right),
    \label{eq:rough_estimation_axion_B_field}
\end{align}
where we take the characteristic scale of $R$ to be the radius of the Earth $R_{\rm E}=6371\,$km as the largest possible scale in the terrestrial environment. Further, the axion is assumed to be the DM, with its local density $\rho_{\rm DM}$ related to $(\mass a)^2/2$ averaged over the coherence time, $T_{\rm coh}\sim 2\pi/(\mass v_{\rm DM}^2)$, with $v_{\rm DM}$ being the velocity of DM. 
Since the bandwidth of the produced EM waves is estimated to be $\Delta f /f_{\rm a}\sim v_{\rm DM}^2\sim10^{-6}$, the induced EM fields appear as a sharp frequency spike, persisting over the coherence time (Fig.~\ref{fig:schematic_signal}).

Here, we note that the electric field is also produced at the same frequency $f_{\rm a}$. Given the expression for $\bfB_{\rm a}$ above, its amplitude can be estimated from Faraday's law, by equating $\nabla\bfE$ with $\partial_t \bfB$, leading to $\bfE_{\rm a}\sim (\mass R)\,\bfB_{\rm a}$. The factor $\mass R$ is roughly given by $\mass R\simeq 0.32 (\mass/10^{-14}\,{\rm eV})(R/R_{\rm E})$. Accordingly, for a magnetic field amplitude of $\sim0.3$\,pT, as estimated in Eq.~\eqref{eq:rough_estimation_axion_B_field}, the corresponding induced-electric field can reach a strength of $|\bfE_{\rm a}|\sim0.03\,$mV/m.\,\footnote{In the units adopted here, the magnetic field amplitude of $1$\,pT is converted into the electric field of $0.3\,$mV/m.}

The above back-of-the-envelope calculation suggests that even with the upper bound of $\gag = 10^{-10},{\rm GeV}^{-1}$ set by the ground-based experiment CAST \cite{CAST:2017uph,CAST:2024eil}, the expected amplitude of the induced electromagnetic waves remains extremely small. In particular, the magnetic field component is suppressed by eight orders of magnitude compared to the geomagnetic field. Nevertheless, publicly available datasets of low-frequency magnetic field measurements exist, with sensitivities reaching the $\sim$pT level or below. Magnetic field observations are generally known to offer advantages over electric field measurements, especially in the extremely low-frequency (ELF) range ($3-30$\,Hz), due to greater instrumental stability and reduced susceptibility to environmental noise. Moreover, high-sensitivity magnetometers are commercially available, including the three-axis fluxgate magnetometer (MAG-03MS, Bartington Instruments, UK) and the low-noise induction coil magnetometer (MFS-06e, Metronix GmbH, Germany).

Indeed, Refs.~\cite{Fedderke_etal2021_superMAG, Arza_etal2022, Friel:2024shg} have utilized long-term magnetometer data from the SuperMAG collaboration, which aggregates ground-based measurements across the globe. While these data are primarily used in geoscience research, particularly for studying geomagnetic activity and magnetosphere–ionosphere coupling, they have also been employed to search for signals of axion and dark photon DM. These studies placed constraints on the axion–photon coupling and/or kinetic mixing parameters over a wide frequency range,   
$6\times 10^{-4}\,{\rm Hz}\lesssim f\lesssim 1\,{\rm Hz}$, corresponding to the axion (and dark photon) mass range of $2\times10^{-18}\,{\rm eV}\lesssim \mass\lesssim4\times10^{-15}\,{\rm eV}$. 

In our companion papers \cite{TNH_Letter, NTH_DataAnalysis}, we analyze a geoscience data set maintained by the British Geological Survey (see also Ref.~\cite{NNTH_Darkphoton} for dark photon DM). Specifically, we use the data collected at the Eskdalemuir Observatory between 2012 and 2022. The dataset was acquired with a high sampling rate of $100$\,Hz, enabling the search for axion dark matter in a higher mass range than that accessible with SuperMAG data. As mentioned in Sec.~\ref{sec:intro}, beyond the frequency range considered in  Refs.~\cite{Fedderke_etal2021_superMAG, Arza_etal2022, Friel:2024shg}, the wavelength of induced EM waves becomes comparable to the Earth's  circumference, and resonant behavior is expected to emerge. This phenomenon has been observed in naturally excited EM waves, such as those generated by lighting discharges, and is referred to as the Schumann resonance ~\cite{Schumann1952a, Schumann1952b} (see also  Refs.~\cite{Jackson_1998,Nickolaenko_Hayakawa2002,Simoes_etal2012,Nickolaenko_Hayakawa2013}). Hence, near the resonance frequencies,  Eq.~\eqref{eq:rough_estimation_axion_B_field} should be treated with caution. For reliable and quantitative predictions, a more refined calculation is necessary, taking various complexities into account. Moreover, the non-uniform nature of the geomagnetic field induces a geographical dependence in the axion signal, leading to position-dependent variations in amplitude and directional properties. 
This motivates the systematic framework developed in the following sections, where we compute the signal across the Earth and extend the analysis beyond the site-specific predictions considered in our previous works \cite{TNH_Letter,NTH_DataAnalysis}, providing guidance for future axion searches.

\section{Axion-induced electromagnetic waves in the Earth-ionosphere cavity}
\label{sec:formulation}

In this section, we present a theoretical framework for  quantitatively describing EM waves induced by axion-photon coupling, focusing especially on the ELF band relevant to signals observable near the Earth's surface.  In doing so, a proper account for the conductivity of the Earth atmosphere is important, and it changes significantly the amplitude and spectral features of the propagating EM waves. 

In Sec.~\ref{subsec:conductivity}, we discuss the importance of atmospheric conductivity in computing the EM waves in the ELF bands.
In Sec.~\ref{subsec:setup}, we present a theoretical setup, and write down the basic equations. Then,   
the governing equation for the axion-induced EM wave is derived. Sec.~\ref{subsec:mode_equation} provides a mathematical framework for solving the wave equation, taking properly the geometric configuration into account. In Sec.~\ref{subsec:multi_step}, we describe a semi-analytical method to solve the radial mode equation while the boundary conditions are fulfilled.

\begin{figure}[tb]
\centering
\includegraphics[width=10cm,angle=0]{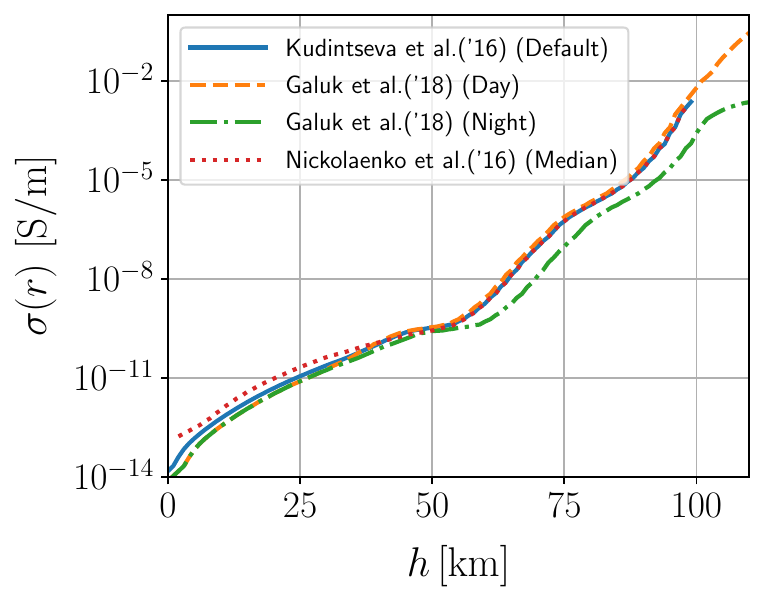}

 \vspace*{-0.3cm}

\caption{Atmospheric conductivity profiles. Several models of the atmospheric conductivity profile are shown, with altitude $h=r-R_{\rm E}$ on the horizontal axis and conductivity $\sigma$ on the vertical axis. The solid line represents the model proposed by Ref.~\cite{Kudintseva_etal2016}, which is consistent with surface electric field observations and upper-atmospheric wave propagation data. The dotted line corresponds to the median profile from Ref.~\cite{Nickolaenko_etal2016}, constructed via numerical optimization to closely match ELF wave propagation and Schumann resonance observations. The two dashed lines, colored orange and green for day and night respectively, are taken from Ref.~\cite{Galuk_etal2018} and represent the conductivity profiles reflecting geographic and temporal variability. Note that the model in Ref.~\cite{Kudintseva_etal2016} spans $0-99$\,km in altitude, while those of Refs.~\cite{Nickolaenko_etal2016} and \cite{Galuk_etal2018} cover $2-98$\,km and $0-110$\,km, respectively. 
\label{fig:conductivity_profiles}}
\end{figure}

\subsection{Atmospheric conductivity}
\label{subsec:conductivity}

At the EM frequencies of our interest, particularly in the ELF band, the atmosphere behaves as a conducting medium characterized by a complex refractive index $n$, which results in lossy EM waves propagation. One crucial aspect of such EM waves is that when generated near the Earth's surface, they become confined between the Earth’s surface and the ionosphere. This is because the Earth's surface and the ionosphere exhibit high conductivity, leading to predominantly imaginary refractive index. As a result, low-frequency EM waves are reflected and absorbed at these boundaries, effectively forming a resonant cavity. Consider a monochromatic EM wave with its frequency given by $f$. The refractive index is given by\footnote{To be precise, the real part of the refractive index in the atmosphere is approximately $1.0003$.  This small deviation from unity has a negligible effect on the axion signal.}\cite{Jackson_1998,Sentman1990,Greifinger_Greifinger1978}: 
\begin{align}
    n^2&=1+i\,\frac{\sigma(r)}{2\pi f}
    \nonumber
    \\
    &\simeq 1+ i\,\cdot 1.8 \,\left(\frac{\sigma(r)}{10^{-10}\,{\rm S/m}}\right)\Bigl(\frac{f}{1\,{\rm Hz}}\Bigr)^{-1}
    \label{eq:refractive_index2}
\end{align}
Here, the quantity $\sigma$ is the conductivity, which we allow to radially vary. The impact of horizontal (i.e., non-spherically symmetric) variations in the conductivity will be later discussed in Sec.~\ref{sec:discussions}. The profile of atmospheric conductivity is typically modeled as a monotonically increasing function of radius $r$, with the center of the Earth taken as the origin of radial coordinate. The conductivity changes from $10^{-14}$ to $10^{-3}$ S/m between the Earth's surface ($r\simeq R_{\rm E}$) and the lower layer of the ionosphere ($r=R_{\rm E}+100$\,km).

Figure~\ref{fig:conductivity_profiles} shows several vertical conductivity profiles taken from Refs.~\cite{Kudintseva_etal2016,Nickolaenko_etal2016,Galuk_etal2018}. These are models constructed in a consistent manner based on observational data, including the propagation characteristics of ELF EM waves and the spectral features of Schumann resonances. Apart from some differences between the models, including day–night variations, the overall behavior of the conductivity is approximately described by a double-exponential function of altitude $h\equiv r-R_{\rm E}$, with a transition in behavior occurring around $h=40-60$\,km (e.g., Refs.~\cite{Greifinger_Greifinger1978, Sentman1990}). Based on the second line of Eq.~\eqref{eq:refractive_index2} and Fig.~\ref{fig:conductivity_profiles}, the imaginary part of the refractive index increases with altitude, indicating that EM wave dissipation, mainly due to absorption, becomes more significant at higher altitudes\footnote{This can be deduced by estimating the skin depth $\delta$. In our case with Eq.~\eqref{eq:refractive_index2}, it is defined by $\delta=\{\omega\,{\rm Im}(n)\}^{-1}$ with $\omega$ being the angular frequency defined by $\omega\equiv 2\pi\,f$, and it gives $\delta=(\sqrt{2}/\omega)\{\sqrt{1+\sigma^2/\omega^2}-1\}^{-1/2}$. }. Reflection also becomes pronounced. Importantly, these effects depend on the EM frequency. 

In what follows, as illustrated in Fig.~\ref{fig:schematic_signal}, we shall focus on the monochromatic EM wave induced by the coherent axion [see Eq.~\eqref{eq:coherent_axion}], with its frequency  determined by the axion mass. In this case, Eq.~\eqref{eq:refractive_index2} is rewritten with
\begin{align}
    n^2&=1+i\,\frac{\sigma(r)}{\mass}
    \nonumber
    \\
    &\simeq 1 + i\cdot0.74\,\left(\frac{\sigma(r)}{10^{-10}\,{\rm S/m}}\right)\Bigl(\frac{\mass}{10^{-14}\,{\rm eV}}\Bigr)^{-1}. 
    \label{eq:refractive_index_axion}
\end{align}

\subsection{Setup}
\label{subsec:setup}

In this subsection, taking the radial-dependent refractive index into account, we present a theoretical setup to quantitatively calculate the axion-induced EM wave.  

We begin by presenting the Lagrangian density for the axion–Maxwell system applicable to the near-surface atmosphere. Since air is effectively non-magnetic, its permeability is taken as that of vacuum. The permittivity then directly relates to the refractive index, and the Lagrangian density is given by: (e.g.,  Refs.~\cite{Millar_etal2017,Sikivie_2021review}):
\begin{align}
    \mathcal{L}=&\frac{1}{2}\bigl(n^2\,\bfE\cdot\bfE-\bfB\cdot\bfB\Bigr)
+\frac{1}{2}\Bigl\{\bigl(\partial_t a\bigr)^2-\bigl(\nabla a\bigr)^2\Bigr\}
-\frac{1}{2}\mass^2a^2-g_{\rm a\gamma}\,a\,\bfE\cdot\bfB.
    \label{eq:Lagrangian_axion_Maxwell}
\end{align}

 Eq.~\eqref{eq:Lagrangian_axion_Maxwell} then leads to the modified Maxwell equations:
\begin{align}
    &\nabla\cdot\bigl(n^2\,\bfE-\,g_{\rm a\gamma}\,a\,\bfB\bigr)=0,
    \label{eq:Maxwell1}
    \\
    &\nabla\times\Bigl(\bfB+g_{\rm a\gamma}\,a\,\bfE\bigr)-\partial_t\Bigl(n^2\,\bfE-g_{\rm a\gamma}\,a\,\bfB\Bigr)=0,
    \label{eq:Maxwell2}
    \\
    &\nabla\cdot\bfB=0,
    \label{eq:Maxwell3}
    \\
    &\nabla\times\bfE + \partial_t\bfB=0.
    \label{eq:Maxwell4}
\end{align}
Since the EM fields, $\bfE$ and $\bfB$, interact with the axion field $a$ through the axion-photon coupling parameter $\gag$, the above equations must be solved together with the axion field equation:
\begin{align}
\partial_t^2\,a -\nabla^2a +\mass^2\,a=-g_{\rm a\gamma}\,\bfE\cdot\bfB.
\label{eq:axion_field_EoM}
\end{align}

The axion-Maxwell system given above is inherently a nonlinear coupled system, but it can be simplified under the assumption that the axion constitutes the DM.
Since the DM is supposed to be non-relativistic, the terms involving the spatial gradient, which is roughly estimated as $|\nabla a|\sim\mass v_{\rm DM}a$ with $v_{\rm DM}\sim10^{-3}$ being the DM velocity, are suppressed compared to those involving the time derivative, $|\dot{a}|\sim \mass a$. Further, the coupling between axion and EM fields is small enough, and the backreaction from EM fields to the axion is ignored\footnote{As we will see later, the axion-induced EM wave is proportional to the coupling parameter $\gag$, and the produced electric and magnetic fields are orthogonal to each other. In this respect, the impact of backreaction, arising from the term $\gag\bfE\cdot\bfB$, is safely neglected. }. Accordingly, while Eqs.~\eqref{eq:Maxwell3} and \eqref{eq:Maxwell4} remain unchanged, the first two  equations, i.e., Eqs.~\eqref{eq:Maxwell1} and \eqref{eq:Maxwell2}, are reduced to the following equations [with a help of Eqs.~\eqref{eq:Maxwell3} and \eqref{eq:Maxwell4}]:
\begin{align}
&    \nabla\cdot\bigl(n^2\,\bfE\bigr)=0,
\label{eq:Maxwell_reduced1}
\\
& \nabla\times \bfB -\partial_t\bigl(n^2\,\bfE\bigr)=-\gag\,(\partial_t a)\,\bfB.
\label{eq:Maxwell_reduced2}
\end{align}
Also, the axion field equation becomes
\begin{align}
    \partial_t^2a+\mass^2a=0. 
    \label{eq:coherent_axion_EoM}
\end{align}

Eqs.~\eqref{eq:Maxwell3}, \eqref{eq:Maxwell4}, \eqref{eq:Maxwell_reduced1}-\eqref{eq:coherent_axion_EoM} serve as our starting point for describing EM waves near the Earth's surface. From Eq.~\eqref{eq:coherent_axion_EoM}, the axion field exhibits a coherent oscillation with the solution given by
\begin{align}
    a=a_0\,e^{-i\,\mass\,t}.
    \label{eq:coherent_axion}
\end{align}
Note that the coherence length of the axion, determined by de Broglie wavelength, is estimated as $L_{\rm coh}=2\pi/(\mass v_{\rm DM})\simeq 0.83\,{\rm AU}\,(10^{-14}\,{\rm eV}/\mass)(10^{-3}/v_{\rm DM})$. Thus, at least for the axion masses relevant to this study, we can safely take the field value $a_0$ to be spatially constant across the Earth's geometry.

As we mentioned in Sec.~\ref{sec:axion_signal}, the axion induces an effective alternating current $\bfJ_{\rm eff}=-\gag (\partial_t a)\bfB$, which appears explicitly in Eq.~\eqref{eq:Maxwell_reduced2}. In particular, when the axion is coupled to the geomagnetic field $\bfB_{\rm geo}$, the frequency of the alternating current is solely determined by the axion mass and is given by $f_{\rm a}=\mass/(2\pi)\simeq2.4\,(\mass/10^{-14}\,{\rm eV})$\,Hz.

Given the setup above, let us consider the wave equation for EM fields. To do so, we take the rotation of Eq.~\eqref{eq:Maxwell4}: 
\begin{align}
& \nabla\times\Bigl\{(\nabla\times\bfE) +\partial_t\bfB\Bigr\}=0. 
\nonumber
\\
&
\Longrightarrow\,
 - \nabla^2\bfE +\nabla(\nabla\cdot\bfE) + \partial_t\bigl(\nabla\times\bfB\bigr) =0.
\end{align}
Assuming the effective current $\bfJ_{\rm eff}=-g_{\rm a\gamma}(\partial_t a)\bfB_{\rm geo}$, we substitute 
Eq.~\eqref{eq:Maxwell_reduced2} into the above to eliminate the magnetic field. We obtain
\begin{align}
    & \nabla^2\bfE - n^2\,\partial^2_t\bfE-\nabla(\nabla\cdot\bfE)= -\gag\,(\partial_t^2\,a)\,\bfB_{\rm geo}.  
\label{eq:wave_eq}
\end{align}
Here, we assumed that the refractive index $n$ is independent of time. Although a similar equation can be derived for the magnetic field, we shall below focus on Eq.\eqref{eq:wave_eq}, as it is more convenient for constructing solutions under the boundary conditions (see below), with $\bfE$ and $\bfB$ related via Eq.~\eqref{eq:Maxwell4}. Note that unlike in vacuum, the divergence-free condition for $\bfE$ does not hold [see Eq.\eqref{eq:Maxwell_reduced1}], which results in an extra term on the left-hand side of Eq.~\eqref{eq:wave_eq}, along with an axion-induced source term on the right. 

Eq.~\eqref{eq:wave_eq} is the basic equation for the axion-induced EM wave. To describe the EM wave trapped near the Earth's surface, the solution must be obtained by imposing the boundary conditions at the Earth's surface ($r=R_{\rm E}$) and the upper atmosphere, with altitude near the lower layer of ionosphere ($h\sim100$\,km). Since the Earth interior is highly conductive, we can treat the Earth surface as a perfect conducting boundary, and impose the condition that the non-radial direction of the electric fields should vanish, that is, $\bfE_{\parallel}=0$ at $r=R_{\rm E}$. On the other hand, as seen in Fig.~\ref{fig:conductivity_profiles}, the conductivity increases significantly at the altitude of lower ionosphere. Although it is still lower than the typical conductivity inside the Earth ($\sim10$ S/m), one expects the low-frequency EM waves to decay rapidly above the ionosphere.  Hence, as a boundary condition, we follow the treatment in the literature (e.g., Refs.~\cite{Greifinger_Greifinger1978, Sentman1990, Nickolaenko_Hayakawa2002}), and allow only upgoing waves at the lower layer of the ionosphere.

\subsection{Radial mode equation}
\label{subsec:mode_equation}

With the setup given above, we now formulate the mathematical framework for solving the wave equation in Eq.~\eqref{eq:wave_eq}, taking into account the boundary conditions and the geometric configuration of the geomagnetic field.

Let us first expand the electric field with the vector spherical harmonics, adopting the conventions in Ref.~\cite{Barrera_etal1985,Carrascal_etal1991,Hill_1954,Weinberg1994}. To fulfill the condition at Eq.~\eqref{eq:Maxwell_reduced1}, 
we express the field $\bfE$, with its frequency determined by $\mass$, as\footnotemark[6]:
\begin{align}
n^2\,\bfE=\frac{1}{\mass}\sum_{\ell,m}\nabla\times\Bigl\{d_{\ell m }(r)\,\bfPhi_{\ell m}(\theta,\phi)\Bigr\}\,e^{-i\mass t}.
\label{eq:E_field_TM-mode}
\end{align}
Here, the quantity $\bfPhi$ is one of the vector spherical harmonics constructed from the scalar spherical harmonics $Y_{\ell m}$, defined by $\bfPhi_{\ell m}\equiv\bfr\times \nabla\,Y_{\ell m}$. The solution of this form is called the transverse-magnetic mode (TM)\footnotemark[7] \cite{Jackson_1998}. 

On the other hand, the geomagnetic field in Eq.~\eqref{eq:wave_eq} is a static field, which can be generally described in terms of the scalar potential $V$, such that $\bfB_{\rm geo}=-\nabla V$. This potential is expanded using the scalar spherical harmonics $Y_{\ell m}$ as follows:
\begin{align}
    V(\bfr)=R_{\rm E}\sum_{\ell,m}C_{\ell m}\,\Bigl(\frac{R_{\rm E}}{r}\Bigr)^{\ell+1}Y_{\ell m}(\theta,\phi).
    \label{eq:V_geo}
\end{align} 
The coefficients $C_{\ell m}$ encodes the structure of the geomagnetic field tilted from the Earth's rotation axis, and a slight deviation from dipole configuration is described by the higher multipole contributions. For a quantitative calculation of the axion-induced EM wave, we adopt the IGRF-13 model of the Earth's magnetic field \cite{IGRF13_2021}, which provides the harmonic coefficients $C_{\ell m}$ up to $\ell=13$ \footnotemark[8]. 

Provided the expansion forms, we write down the electric field and geomagnetic field:
\begin{align}
 \bfE&=\frac{1}{\mass}\sum_{\ell,m }\,\Bigl\{-\frac{\ell(\ell+1)}{r}\,\frac{d_{\ell m}}{n^2}\,\bfY_{\ell m}(\theta,\phi)
-\frac{1}{n^2}\Bigl(\frac{d}{dr}+\frac{1}{r}\Bigr)d_{\ell m}\,\bfPsi_{\ell m}(\theta,\phi)\Bigr\} \,e^{-i\,\mass t},
\label{eq:E_field_TM-mode2}
\\
 \bfB_{\rm geo}&=\sum_{\ell,m }C_{\ell m}\,\Bigl(\frac{R_{\rm E}}{r}\Bigr)^{\ell+2}
\Bigl\{(\ell+1)\,\bfY_{\ell m}(\theta,\phi)-\bfPsi_{\ell m}(\theta,\phi)\Bigr\},
\label{eq:B_geo_field}
\end{align}
where the vector fields $\bfY_{\ell m}$ and $\bfPsi_{\ell m}$ constitute the vector spherical harmonics together with $\bfPhi_{\ell m}$. They are orthogonal to each other, defined respectively as $\bfY_{\ell m}\equiv Y_{\ell m}\,\hat{r}$ and $\bfPsi\equiv r\,\nabla Y_{\ell m}$ \cite{Barrera_etal1985,Carrascal_etal1991,Hill_1954,Weinberg1994}.  
With the background coherent axion given by Eq.~\eqref{eq:coherent_axion}, we substitute Eqs.~\eqref{eq:E_field_TM-mode2} and \eqref{eq:B_geo_field} into the wave equation in Eq.~\eqref{eq:wave_eq}. After some algebra, we obtain the following set of equations: 
\begin{align}
 \Biggl[\frac{d^2}{dr^2}+\Bigl\{\Bigl(n^2\frac{d}{dr}n^{-2}\Bigr)+\frac{2}{r}\Bigr\}\,\frac{d}{dr}+ \frac{1}{r}\Bigl(n^2\frac{d}{dr}n^{-2}\Bigr)+\mass^2n^2-\frac{\ell(\ell+1)}{r^2}  \Biggr]d_{\ell m}=-\gag\,\mass^3 a_0n^2\,\frac{C_{\ell m}}{\ell}\frac{R_{\rm E}^{\ell+2}}{r^{\ell+1}},
\label{eq:mode_eq_dlm}
\end{align}
and
\begin{align}
&\Biggl[\frac{d^3}{dr^3} + \Bigl\{\frac{3}{r}+2\,\Bigl(n^2\frac{d}{dr}n^{-2}\Bigr) \Bigr\}
\frac{d^2}{dr^2} + \Bigl\{\Bigl(n^2\frac{d^2}{dr^2}n^{-2}\Bigr)+\frac{4}{r} \Bigl(n^2\frac{d}{dr}n^{-2}\Bigr) + \mass^2n^2- \frac{\ell(\ell+1)}{r^2}\Bigr\} 
\frac{d}{dr}
\nonumber
\\
&\qquad + \frac{1}{r}\Bigl(n^2\frac{d^2}{dr^2}n^{-2}\Bigr)  + \frac{\ell(\ell+1)}{r^2}\Bigl\{\frac{1}{r}-\Bigl(n^2\frac{d}{dr}n^{-2}\Bigr)\Bigr\}\Biggr] d_{\ell m} = \gag \mass^3 n^2\,a_0\,C_{\ell m}\Bigl(\frac{R_{\rm E}}{r}\Bigr)^{\ell+2}.
\label{eq:mode_eq_dlm2}
\end{align}

\footnotetext[6]{Since the vector spherical harmonics $\bfPhi_{\ell m}$ satisfies $\nabla\cdot(f\,\bfPhi_{\ell m})=0$ for an arbitrarily radial-dependent function $f$, one can construct an alternative expansion that satisfies the divergence-free condition, Eq.~\eqref{eq:Maxwell_reduced1}:
\begin{align}
    n^2\,\bfE=\sum_{\ell,m}f_{\ell m}(r)\,\bfPhi_{\ell m}(\theta,\phi)\,e^{-i\mass t}. 
    \nonumber
\end{align}
This expression leads to the solution independent of Eq.~\eqref{eq:E_field_TM-mode}, referred to as the transverse-electric (TE) mode. However, due to the properties of vector spherical harmonics, the resultant mode equation for $f_{\ell m}$ is source-free, and the TE mode does not couple to the axion field (see also Ref.~\cite{Arza_etal2022}).
}
\footnotetext[7]{This is because the magnetic field solution, derived from Eq.~\eqref{eq:E_field_TM-mode} via the Maxwell-Faraday law [Eq.~\eqref{eq:Maxwell4}], lies in the transverse direction,  i.e., tangent to the Earth' surface (See Eq.~\eqref{eq:B-field_TM-mode}).} 
\footnotetext[8]{Ref.~\cite{IGRF13_2021} provides a table for harmonic coefficients, but basis functions they adopted differ from those in Eq.~\eqref{eq:V_geo}. Following Ref.~\cite{Arza_etal2022}, we converted their coefficients to those in the spherical harmonic basis. }
\setcounter{footnote}{8}

The above two equations are not independent. Let us define
\begin{align}
&P(r)\equiv {\rm LHS\,\,of}\,\,{\rm Eq}.~\eqref{eq:mode_eq_dlm} -  {\rm RHS\,\,of}\,\,{\rm Eq}.~\eqref{eq:mode_eq_dlm},
\nonumber
\\
&Q(r)\equiv {\rm LHS\,\,of}\,\,{\rm Eq}.~\eqref{eq:mode_eq_dlm2} -  {\rm RHS\,\,of}\,\,{\rm Eq}.~\eqref{eq:mode_eq_dlm2}. 
\end{align}
It  can be then shown that the functions $P$ and $Q$ are related by
\begin{align}
\frac{ Q(r)}{n^2 } = \frac{1}{r}\,\frac{d}{dr}\Bigl\{\frac{r\,P(r)}{n^2}\Bigr\}.
\end{align}
That is, if Eq.~\eqref{eq:mode_eq_dlm} is satisfied, Eq.~\eqref{eq:mode_eq_dlm2} automatically follows. Therefore,  Eq.~\eqref{eq:mode_eq_dlm} serves as the master equation to be solved.

In solving Eq.~\eqref{eq:mode_eq_dlm2}, the boundary conditions for the electric field discussed in Sec.~\ref{subsec:setup} can be translated into the conditions on the radial mode function $d_{\ell m}$ as follows. For the boundary condition at the Earth's surface, the vanishing transverse electric field, $\bfE_\parallel=0$, implies, via Eq.~\eqref{eq:E_field_TM-mode2}, that
\begin{align}
   \Bigl( \frac{d}{dr}+\frac{1}{r}\Bigr) d_{\ell m}=0 \quad{\rm at}\,\,r=R_{\rm E}.
    \label{eq:BC_surface}
\end{align}
On the other hand, the upward-propagating wave condition at the upper boundary $r=r_{\rm up}$ implies that the homogeneous part of the solution asymptotically behaves as
\begin{align}
    d_{\ell m}^{\rm hom}\to \frac{e^{i\,k r}}{r} \quad{\rm at}\,\,r= r_{\rm up}.
    \label{eq:BC_ionosphere}
\end{align}
where $k$ is a constant. 

To recap, the radial mode equation for the electric field we solve is given at Eq.~\eqref{eq:mode_eq_dlm}, with the boundary conditions of Eqs.~\eqref{eq:BC_surface} and \eqref{eq:BC_ionosphere}. Introducing the dimensionless radius  $x\equiv \mass r$ and writing the mode function $d_{\ell m}$ as $u_{\ell m}\equiv x\,d_{\ell m}$,
these are written in a rather concise form:
\begin{align}
    &\frac{d}{dx}\Bigl(\frac{1}{n^2}\frac{du_{\ell m}}{dx}\Bigr)+\Bigl\{1-\frac{\ell(\ell+1)}{n^2x^2}\Bigr\}u_{\ell m}
=-\gag a_0\frac{C_{\ell m}}{\ell}\frac{x_0^{\ell+2}}{x^\ell}
    \label{eq:radial_mode_eq_u_ellm}
\end{align}
for the radial mode equation and 
\begin{align}
    \frac{du_{\ell m}}{dx}\Bigr|_{x=x_0}=0,
    \qquad
    u_{\ell m}\stackrel{x\simeq x_{\rm up}}{\longrightarrow} e^{i\,\kappa\,x}
    \label{eq:BC_u_ellm}
\end{align}
for the boundary conditions, where $x_0$ and $x_{\rm up}$ are defined respectively by $x_0\equiv \mass R_{\rm E}$ and $x_{\rm up}\equiv \mass\,r_{\rm up}$.

Finally, once the radial mode function $d_{\ell m}$ for the electric field  is obtained, the corresponding magnetic field can be derived from Eq.~\eqref{eq:Maxwell4}, and 
is expressed in terms of $d_{\ell m}$ or $u_{\ell m}$.  Assuming that the induced magnetic field also takes the form of a monochromatic wave, similar to the electric field, i.e., $\bfB\propto e^{-i\mass t}$, we obtain
\begin{align}
\bfB&=-\frac{i}{\mass}\nabla\times\bfE
\nonumber
\\
&=i\,\sum_{\ell,m}\frac{1}{x}\Bigl[\frac{d}{dx}\Bigl(\frac{1}{n^2}\frac{du_{\ell m}}{dx}\Bigr)-\frac{\ell(\ell+1)}{x^2}\,u_{\ell m}\Bigr]\,\bfPhi_{\ell m}\,e^{-i\mass t}
\nonumber
\\
&=-i\sum_{\ell,m}\frac{1}{x}
\Bigl(u_{\ell m}+\gag a_0\frac{C_{\ell m}}{\ell}\frac{x_0^{\ell+2}}{x^\ell}\Bigr)\,\bfPhi_{\ell m}\,e^{-i\mass t}.
\label{eq:B-field_TM-mode}
\end{align}
Equation \eqref{eq:B-field_TM-mode} shows that the magnetic field lies in the tangential plane to Earth's surface, with no radial component, as expected for a TM mode. 

\subsection{Solving mode equation: semi-analytic treatment}
\label{subsec:multi_step}

\begin{figure}[tb]
    \centering
 \includegraphics[width=9cm,angle=0]{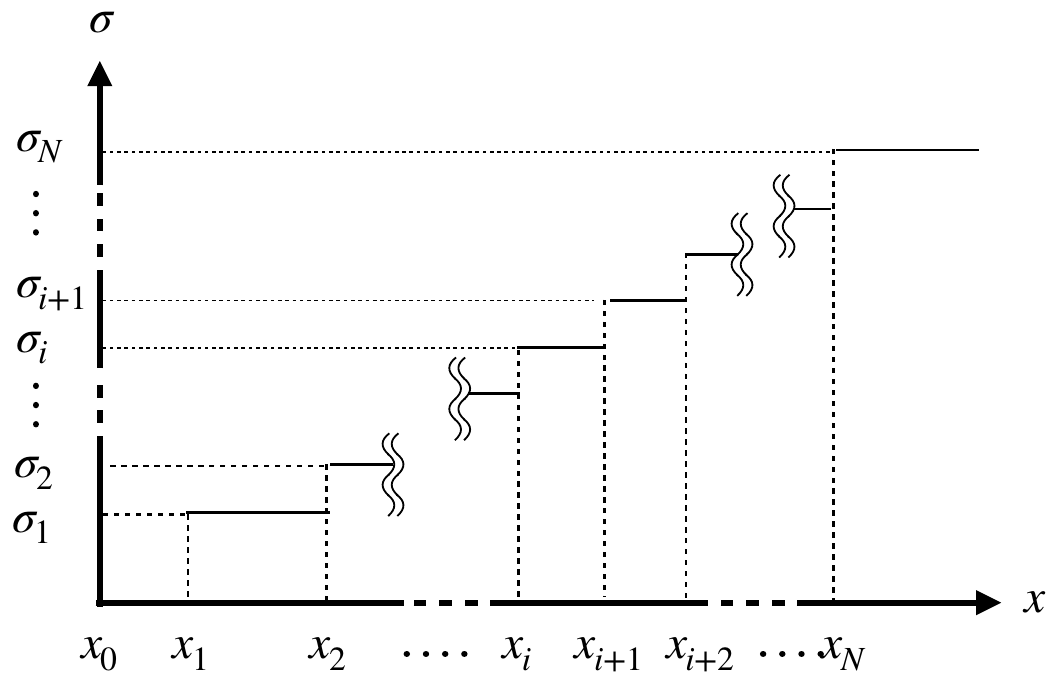}
\caption{Model of step-wise conductivity profile, where the atmospheric conductivity is treated as piecewise constant as a function of dimensionless radius $x$ [see Eq.~\eqref{eq:conductivity_step}]. 
\label{fig:multi-step_conductivity}
}
\end{figure}

In this subsection, we present a semi-analytical treatment for solving the mode equation, Eq.~\eqref{eq:radial_mode_eq_u_ellm}, accounting for radial variations in atmospheric conductivity, $\sigma$, and boundary conditions at Earth's surface and ionosphere. Readers primarily interested in the resulting EM wave solutions may skip the technical details and proceed directly to the next section.

The key idea is to  approximate the radially varying  conductivity by dividing the atmosphere into a series of layers, each with constant conductivity, as illustrated in Fig.~\ref{fig:multi-step_conductivity}:
\begin{align}
\sigma(x)=\left\{
\begin{array}{ll}
     \sigma_k & (x_k\leq x < x_{k+1})\\
      \\
     \sigma_N & (x_N \leq x)
\end{array}
\right.,
\label{eq:conductivity_step}
\end{align}
 where the subscript $k$ runs from $0$ to $N-1$, and we set $\sigma_0$ to 0. Correspondingly, a square of refractive index $n^2$ is given as a sum of step functions: 
\begin{align}
n^2(x)=\left\{
\begin{array}{ll}
     n^2_k & (x_k \leq x < x_{k+1}) \\
     \\
     n^2_N & (x_N \leq x)
\end{array}
\right.
\label{eq:refractive_index_step}
\end{align}
with $n_k^2$ being $1+i\,\sigma_k/\mass$ [see Eq.~\eqref{eq:refractive_index_axion}]. 

This layered representation allows Eq.~\eqref{eq:radial_mode_eq_u_ellm} to be solved analytically within each segment, with solutions matched at the boundaries. This treatment is known as the full-wave solution (Ref.~\cite{wait2013electromagnetic}, see also Refs.~\cite{Bliokh1980,Galuk_etal2019,Malta_Helayel-Neto2022}) and here we extend it to incorporate the axion-induced source term. Specifically, by re-scaling the radius as $x \to \tilde{x} = nx$, the mode equation takes the same form as in vacuum ($n=1$) within each layer of constant refractive index, as follows:
\begin{align}
    \frac{du_{\ell m}}{d\tilde{x}^2}+\Bigl\{1-\frac{\ell(\ell+1)}{\tilde{x}^2}\Bigr\}u_{\ell m}=-\gag a_0\frac{C_{\ell m}}{\ell}\frac{\tilde{x}_0^{\ell+2}}{n^2\,\tilde{x}^{\ell}}. 
\end{align}
One then obtains a general analytic expression for the mode function $d_{\ell m}$ \cite{Arza_etal2022}:
\begin{align}
    d_{\ell m}&=\frac{u_{\ell m}}{x} 
    \nonumber
    \\
    &= p_{\ell m}\,j_\ell (nx) +  q_{\ell m}\, y_\ell (nx) -\gag a_0\frac{C_{\ell m}}{\ell}\,\frac{x_0^{\ell+2}}{x^{\ell+1}}. 
    \label{eq:general_solution_const_n}
\end{align}
with the coefficients $p_{\ell m}$ and $q_{\ell m}$ being  arbitrary constants. Here the functions $j_\ell$ and $y_\ell$ are respectively the spherical Bessel and Neumann functions.

Using the general solution in Eq.~\eqref{eq:general_solution_const_n}, the solution for each layer in 
Eq.~\eqref{eq:conductivity_step} or \eqref{eq:refractive_index_step} is given by 
\begin{align}
    d_{\ell m}^{(k)} = A_{\ell m}^{(k)} \,j_\ell(n_k x) + B_{\ell m}^{(k)} \,y_\ell(n_k x) - \gag a_0 \frac{C_{\ell m}}{\ell}\,\frac{x_0^{\ell+2}}{x^{\ell+1}}
    \label{eq:solution_segment}
\end{align}
for $x_k\leq x< x_{k+1}$ $(k=0,\cdots,N-1)$. In the outermost region $x\geq  x_N$, at which we identify $x_N$ with the upper boundary $x_{\rm up}$, we impose the upward-propagating wave condition at Eq.~\eqref{eq:BC_ionosphere} or the second equation of Eq.~\eqref{eq:BC_u_ellm}. Then the solution leads to 
\begin{align}
    d_{\ell m}^{(N)} = \mathcal{A}_{\ell m} \,h_\ell^{(1)}(n_N x) - \gag a_0 \frac{C_{\ell m}}{\ell}\,\frac{x_0^{\ell+2}}{x^{\ell+1}},
    \label{eq:solution_outermost}
\end{align}
where  $h_\ell^{(1)}$ is the spherical Hankel function of the first kind, with the asymptotic form $h_\ell^{(1)}(z)\sim(-i)^{\ell+1}\,e^{i\,z}/z$ for $z\gg1$.  

A central task in this construction is to determine the coefficients $A_{\ell m}^{(k)}$,  $B_{\ell m}^{(k)}$ $(k=0,1,\dotsc,N-1)$ and $\mathcal{A}_{\ell m}$, which are related to each other through the matching conditions at the interfaces between adjacent layers $x=x_k$ $(k=1,\dotsc,N)$. Denoting the solutions just below and above the interface $x = x_*$ as $d_{\ell m}^-$ and $d_{\ell m}^+$, respectively, the matching conditions are given by\footnote{Eq.~\eqref{eq:matching_condition1} comes from the Maxwell equation $\nabla\cdot(n^2\,\bfE)=0$. Using the Gauss' theorem, the volume integral across the boundary yields $(n_+^2\bfE_+-n_-^2\bfE_-)\cdot\hat{n}=0$, with the unit vector $\hat{n}$ being normal to the boundary surface, pointing to a radial direction in our case, i.e., $\hat{n}=\hat{r}$. Since the radial component of the electric field,  given at Eq.~\eqref{eq:E_field_TM-mode2}, is described by the terms proportional to the vector spherical harmonics $\bfY_{\ell m}$, one obtains Eq.~\eqref{eq:matching_condition1}. On the other hand, Eq.~\eqref{eq:matching_condition2} is derived by integrating Eq.~\eqref{eq:radial_mode_eq_u_ellm} across the boundary layer, $[x_*-\delta,x_*+\delta]$, and taking the limit $\delta\to 0$, under the assumption that the refractive index is non-singular and the mode function is continuous at $x=x_*$. 
}:
\begin{align}
d_{\ell m}^-(x_*) &= d_{\ell m}^+(x_*),
\label{eq:matching_condition1}
\end{align}
and
\begin{align}
\frac{1}{n_-^2}\Bigl\{\frac{d d_{\ell m}^-}{dx} +\frac{d_{\ell m}^-}{x}\Bigr\}\Bigr|_{x=x_*} &=
\frac{1}{n_+^2}\Bigl\{\frac{d d_{\ell m}^+}{dx} +\frac{d_{\ell m}^+}{x}\Bigr\}\Bigr|_{x=x_*}.
\label{eq:matching_condition2}
\end{align}

Applying Eqs.~\eqref{eq:matching_condition1} and \eqref{eq:matching_condition2}  to the solutions \eqref{eq:solution_segment} for each segment, we obtain the recursive relation which relates the coefficients $A_{\ell m}^{(k)}$, $B_{\ell m}^{(k)}$ with those at adjacent segments: 
\begin{align}
            \bfX_{k-1}(x_{k})\left(
    \begin{array}{c}
        A_{\ell m}^{(k-1)}   \\
        \\
        B_{\ell m}^{(k-1)}  
    \end{array}
      \right)
       &= \bfX_{k}(x_{k})\left(
    \begin{array}{c}
        A_{\ell m}^{(k)}   \\
        \\
        B_{\ell m}^{(k)}  
    \end{array}
      \right)
+\gag a_0 C_{\ell m}\,\vec{U}(x_{k})
      \label{eq:matching_condition_ith}
\end{align}
at the matching radius $x_{k}$ $(k=1,\dotsc, N-1)$. Also, for the outermost solution \eqref{eq:solution_outermost}, we have
\begin{align}
      \bfX_{N-1}(x_{N})\left(
    \begin{array}{c}
        A_{\ell m}^{(N-1)}   \\
        \\
        B_{\ell m}^{(N-1)}  
    \end{array}
      \right)&=\mathcal{A}_{\ell m}\,\vec{Y}(x_N)
+\gag a_0 C_{\ell m}\,\vec{U}(x_{N}).
      \label{eq:matching_condition_Nth}
\end{align}
 Here, the matrix $\bfX_j(x_k)$ and the vectors $\vec{U}(x_k)$ and $\vec{Y}(x_N)$ are respectively defined by
\begin{align}
    \bfX_j(x_k)&=\left(
\begin{array}{cc}
   j_\ell(z_{j,k}) & y_\ell(z_{j,k}) 
   \\
    \\
   {\displaystyle   \frac{1}{n_j\,z} \frac{d}{dz}z\, j_\ell(z)\Biggr|_{z_{j,k}} }& {\displaystyle    \frac{1}{n_j\,z} \frac{d}{dz}z\,y_\ell(z)\Biggr|_{z_{j,k}} }
\end{array}
\right),
\label{eq:def_matrix_X}
\\
\vec{U}(x_k)&=\Bigl(\frac{x_0}{x_k}\Bigr)^{\ell+2}
\left(
\begin{array}{c}
0
\\
\\
{\displaystyle -\frac{1}{n_{k-1}^2}+\frac{1}{n_{k}^2} }
\end{array}
\right),
\label{eq:vect_U}
\\
\vec{Y}(x_N)&=\left(
\begin{array}{c}
   h_\ell^{(1)}(z_{N,N}) 
    \\
    \\
   {\displaystyle   \frac{1}{n_N\,z}\frac{d}{dz}z\,h_\ell^{(1)}(z) \Biggr|_{z_{N,N}}}
\end{array}
\right),
\label{eq:vect_Y}
\end{align}
where we introduced a new variable, $z_{j,k}\equiv n_jx_k$. Note that the refractive index $n_0$ is identified with unity.  

Solving Eqs.~\eqref{eq:matching_condition_ith} and \eqref{eq:matching_condition_Nth} recursively, one can express all the coefficients $A_{\ell m}^{(k)}$ and $B_{\ell m}^{(k)}$ in terms of $\mathcal{A}_{\ell m}$ and $C_{\ell m}$. Explicitly, for $k=1,\dotsc,N$, we obtain
\begin{align}
    \left(
    \begin{array}{c}
        A_{\ell m}^{(k-1)}   \\
        \\
        B_{\ell m}^{(k-1)}  
    \end{array}
      \right)&=\bfX_{k-1}^{-1}(x_{k})\Bigl(\vec{M}_{k}\,\mathcal{A}_{\ell m}+\gag a_0C_{\ell m}\,\vec{N}_{k}\Bigr).
      \label{eq:coeff_Alm_Blm_k-1}
\end{align}
The vectors $\vec{M}_k$ and $\vec{N}_k$ are defined as
\begin{align} 
    \vec{M}_k&\equiv \bfH_k^{N-1}\,\vec{Y}(x_N),       
    \label{eq:def_vec_M}
    \\
    \vec{N}_k&\equiv \sum_{m=k}^{N}\bfH_k^{m-1}\vec{U}(x_m),
    \label{eq:def_vec_N}
    \\
\bfH_k^{m-1}&\equiv\left\{
\begin{array}{ll}
\mathbb{I}, & (m=k)
\\
\\
{\displaystyle \prod_{j=k}^{m-1}\bigl[\bfX_j(x_j)\bfX_j^{-1}(x_{j+1})\bigr],} & (m\geq k+1)
\end{array}
\right..
\end{align}

The final step is to eliminate the coefficient $\mathcal{A}_{\ell m}$ from Eq.~\eqref{eq:coeff_Alm_Blm_k-1}, and to express it solely in terms of harmonic coefficients $C_{\ell m}$ of the geomagnetic field, which couples to the coherent axion field. To do so, we apply the boundary condition at the Earth's surface, given at Eq.~\eqref{eq:BC_surface} or first equation in Eq.~\eqref{eq:BC_u_ellm}, which can be rewritten as
\begin{align}
&   \Biggl[ \bfX_0(x_0)\left(
    \begin{array}{c}
        A_{\ell m}^{(0)}   \\
        \\
        B_{\ell m}^{(0)}  
    \end{array}
    \right)\Biggr]_2=-\gag a_0 C_{\ell m},
    \label{eq:boundary_condition_at_ground}
\end{align}
where the subscript $2$ at the left-hand side implies the second component of the vector. 

Substituting Eq.~\eqref{eq:coeff_Alm_Blm_k-1} for $k=1$ into the above, we obtain the explicit expression for $\mathcal{A}_{\ell m}$\footnote{In deriving Eq~\eqref{eq:_curlA_multi-step}, we used the fact from Eqs.~\eqref{eq:def_vec_M} and \eqref{eq:def_vec_N} that $\vec{M}_0=\bfX_{0}(x_{0})\bfX_{0}^{-1}(x_1)\,\vec{M}_{1}$ and $\vec{N}_0=\bfX_{0}(x_{0})\bfX_{0}^{-1}(x_1)\,\vec{N}_{1}$. }:
\begin{align}
    \mathcal{A}_{\ell m}&=-\gag a_0 C_{\ell m}\,\frac{V}{W},
\label{eq:_curlA_multi-step}
\end{align}
with
\begin{align}
    V&= 1+ \Bigl[\vec{N}_0\Bigr]_2, 
\quad
    W=\Bigl[\vec{M}_0\Bigr]_2,
    \label{eq:V_ell_W_ell}
\end{align}
where $V$ and $W$ implicitly depend on the multipole $\ell$ through $\vec{M}_0$ and $\vec{N}_0$, which are constructed from products of the matrices $\bfX_j(x_k)$. Then, plugging Eq.~\eqref{eq:_curlA_multi-step} back into Eq.~\eqref{eq:coeff_Alm_Blm_k-1}, we obtain closed-form expressions for all coefficients in the layer-by-layer solutions for $k=1,\dotsc, N$:
\begin{align}
    \left(
    \begin{array}{c}
        A_{\ell m}^{(k-1)}   \\
        \\
        B_{\ell m}^{(k-1)}  
    \end{array}
      \right) &=\gag \,a_0  C_{\ell m}\,\bfX_{k-1}^{-1}(x_{k})\Bigl(-\frac{V}{W}\,\vec{M}_{k}+\,\vec{N}_{k}\Bigr).
      \label{eq:coeff_Alm_Blm_k-1_final_form}
\end{align}

In summary, when the atmospheric conductivity is discretized into a series of constant layers, the solution to the mode equation [Eq.~\eqref{eq:mode_eq_dlm} or \eqref{eq:radial_mode_eq_u_ellm}] that satisfies the boundary conditions at the Earth's surface and ionosphere is given by Eqs.~\eqref{eq:solution_segment} and \eqref{eq:solution_outermost}, with the coefficients $\mathcal{A}_{\ell m}$, $A_{\ell m}^{(k)}$ and $B_{\ell m}^{(k)}$ respectively determined by Eqs.~\eqref{eq:_curlA_multi-step}  and \eqref{eq:coeff_Alm_Blm_k-1_final_form}.

Finally, in numerically computing the solution of mode function $d_{\ell m}$ as presented above, one frequently needs to evaluate matrix products of the form $\bfX_{k}(x_{k})\bfX_{k}^{-1}(x_{k+1})$, which can become numerically unstable as the imaginary part of the refractive index increases. To mitigate this instability, we adopt the following approximate expression:
\begin{align}
&[\bfX_{k}(x_{k})\bfX_{k}^{-1}(x_{k+1})] \approx \Bigl(\frac{x_{k+1}}{x_{k}}\Bigr)^2
\left(
\begin{array}{cc}
   {\displaystyle 1+\Delta z_k\Bigl(\frac{1}{z_{k,k+1}}-\frac{2}{z_{k,k}}\Bigr) }  &   -n_k\Delta z_k
   \\
   \\
   {\displaystyle \# }  & {\displaystyle 1-\frac{\Delta z_k}{z_{k,k}}}
\end{array}
\right),
\label{eq:X_inv_X_approx}
\end{align}
where the $(2,1)$ component, labeled by $\#$,  is given explicitly by
\begin{align}
\# =\frac{\Delta z_k}{n_k}\Biggl\{1-\frac{\ell(\ell+1)+2}{z_{k,k}^2}+\frac{2}{z_{k,k}\,z_{k,k+1}}\Biggr\}.
\end{align}
with $z_{j,k}$ and $\Delta z_k$ respectively defined as $z_{j,k}\equiv n_{j}x_{k}$ and $\Delta z_k\equiv n_{k}(x_{k+1}-x_{k})$. The above expressions are derived by expanding the spherical Bessel function under the assumption that $z_{k,k+1}$ is close to $z_{k,k}$. Therefore, the approximation in Eq.~\eqref{eq:X_inv_X_approx} becomes increasingly accurate as the number of layers increases.

\section{Results}
\label{sec:results}

In this section, based on the detailed formulation in Sec.~\ref{sec:formulation}, we compute the radial mode function for the axion-induced EM wave and present the resulting properties of the EM signal. In what follows, we focus on the magnetic field component, as it offers observational advantages over the electric field, including the availability of public datasets, which facilitates the search for axion dark matter and enables stringent constraints on the axion-photon coupling.

Substituting the solutions of mode function, Eqs.~\eqref{eq:solution_segment} and \eqref{eq:solution_outermost} into Eq.~\eqref{eq:B-field_TM-mode}, the expression of magnetic field is recast as \cite{TNH_Letter}
\begin{align}
    \bfB=\gag\,a(t)\sum_{\ell, m}b_\ell(x)\,C_{\ell m}\,\bfPhi_{\ell m}(\theta,\phi),
    \label{eq:B_field_vsp_expansion}
\end{align}
where $a$ is the coherent axion field defined in Eq.~\eqref{eq:coherent_axion}. The mode function $b_\ell$ is constructed from the layer-by-layer solutions of $d_{\ell m}$ as follows:
\begin{align}
    b_\ell(x) &=-i\,\left\{
    \begin{array}{ll}
    \widetilde{A}_{\ell}^{(k)}\,j_\ell(n_k x)+\widetilde{B}_{\ell}^{(k)}\,y_\ell(n_k x), & x_k\leq x< x_{k+1}
    \\
    \\
    \widetilde{\mathcal{A}}_{\ell}, \,h_\ell^{(1)}(n_N x), & x\geq x_N
    \end{array}
    \right.
    \label{eq:B-field_axion}
\end{align}
The coefficients $\widetilde{A}_{\ell}^{(k)}$, $\widetilde{B}_{\ell}^{(k)}$ and $\widetilde{\mathcal{A}}_\ell$ correspond to $A_{\ell m}^{(k)}$, $B_{\ell m}^{(k)}$, and $\mathcal{A}_{\ell m}$, but are normalized by the factor $\gag\,a_0\,C_{\ell m}$. With this normalization, the mode function $b_\ell$ becomes independent of both the properties of geomagnetic field and the axion-photon coupling, and is expressed as a function of the dimensionless radius $x=\mass r$, characterized by the multipole index, $\ell$. 

Based on the expression in Eq.~\eqref{eq:B_field_vsp_expansion},  we first present the results for the mode function $b_\ell$ in Sec.~\ref{subsec:mode_function}.  
In Sec.~\ref{subsec:B-field_structure}, we examine the structure of induced magnetic field, focusing on its spectral and geographical characteristics. 

Below, we adopt the atmospheric conductivity profile from Ref.~\cite{Kudintseva_etal2016}, imposing the upper boundary condition at $h=100$ km, above which the conductivity is assumed constant. All results are then obtained with $N=300$ layers, using cubic interpolation for the tabulated conductivity profile.
The impact of our treatment of the conductivity profile as well as the robustness of our predictions are discussed in detail in the next section.

\subsection{Properties of the mode function $b_\ell$}
\label{subsec:mode_function}

\begin{figure}[tb]
  \centering
 \includegraphics[width=9.5cm,angle=0]{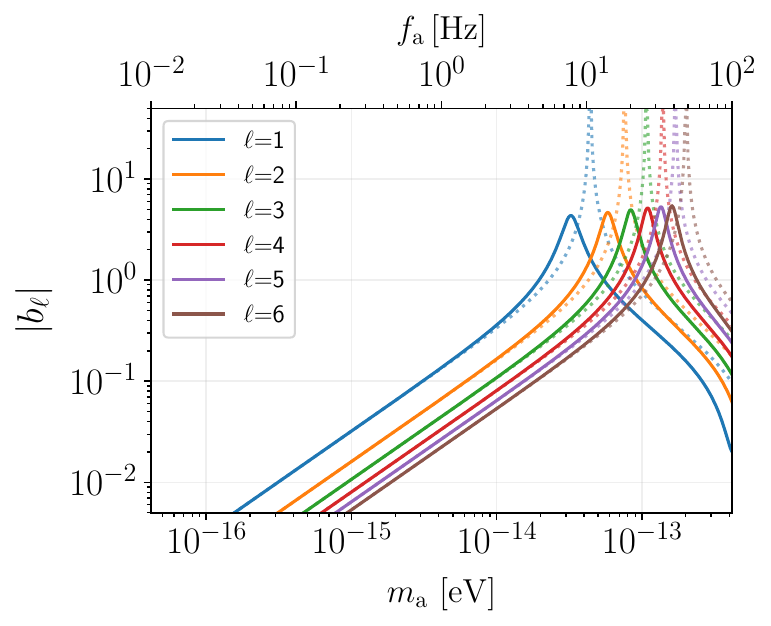}
\caption{Radial mode function for the axion-induced magnetic field $b_\ell$. Fixing $r$ to the Earth radius (i.e., $h=0$\,km), the results for the lowest six multipoles are plotted as a function of the axion mass $\mass$ (lower) and frequency $f_{\rm a}$ (upper). 
\label{fig:mode_functions_B-field}
}
\end{figure}

\subsubsection{Frequency (axion mass) dependence}
\label{subsubsec:spectral_feature}

As mentioned above, the mode function $b_{\ell}$ is characterized by the dimensionless radius $x$ and the multipole $\ell$, and is fully determined by the atmospheric properties and boundary conditions. In the ELF band, the EM waves confined between the Earth's surface and ionosphere are known to exhibit characteristic resonances. It is therefore instructive to first examine how such resonant features appear in the mode function.   

In Fig.~\ref{fig:mode_functions_B-field}, we plot the magnetic mode functions $b_\ell$ for the lowest six multipoles as a function of axion mass $\mass$ (bottom axis) and the corresponding EM wave frequency $f_{\rm a}$ (top axis), fixing the radius $r$ to the Earth's surface $R_{\rm E}$. At low axion masses, $\mass \lesssim 10^{-14}\,\mathrm{eV}$ (or $f_{\rm a} \lesssim 2.4\,\mathrm{Hz}$), the mode functions increase monotonically with $\mass$, and their amplitudes scale approximately as $1/\ell$. This behavior is consistent with the analytical solution in the idealized case that neglects atmospheric conductivity and treats the ionosphere as a perfect conductor, shown as dotted lines. In Appendix~\ref{appendix:EM_wave_perfect_conductor}, the analytical results for this idealized setup are summarized, and they reproduce the low-frequency behavior discussed in Ref.~\cite{Arza_etal2022}.

On the other hand, at higher axion masses, $\mass \gtrsim 10^{-14}\,\mathrm{eV}$, each mode exhibits a resonant peak structure with a finite amplitude around $|b_\ell| \sim 5$. These features correspond to the so-called Schumann resonances \cite{Schumann1952a, Schumann1952b, Nickolaenko_Hayakawa2002, Nickolaenko_Hayakawa2013}. While the idealized calculation predicts divergent resonances at higher frequencies, as given by Eq.~\eqref{eq:Schumann_freq_ideal}, the realistic treatment with finite atmospheric conductivity suppresses the divergence and shifts the resonance to lower frequencies, resulting in well-defined peaks with finite amplitude. The resonance structure, including the width of each peak, indeed resembles those expected from the EM waves excited by lightening discharges. Eliminating the axion-induced effective current, the formalism presented in Sec.~\ref{sec:formulation} also predicts those characteristics, 
i.e., peak resonance frequencies and their width for each multipole $\ell$.  

In Appendix~\ref{appendix:Schumann}, we numerically solve the transcendental equation for the eigenfrequencies, derived from the EM wave solutions in  Sec.~\ref{sec:formulation}. We then present the lowest six Schumann resonance modes in Table~\ref{tab:eigenvalues},  which explain the resonance structures shown in Fig.~\ref{fig:mode_functions_B-field}. The derived eigenfrequencies agree well with the observed properties of Schumann resonances, and reproduce previous theoretical calculations, demonstrating that the present formalism and numerical treatment yield quantitatively accurate and reliable predictions for ELF EM waves.
\begin{figure*}[tb]
\centering
 \includegraphics[width=15cm,angle=0]{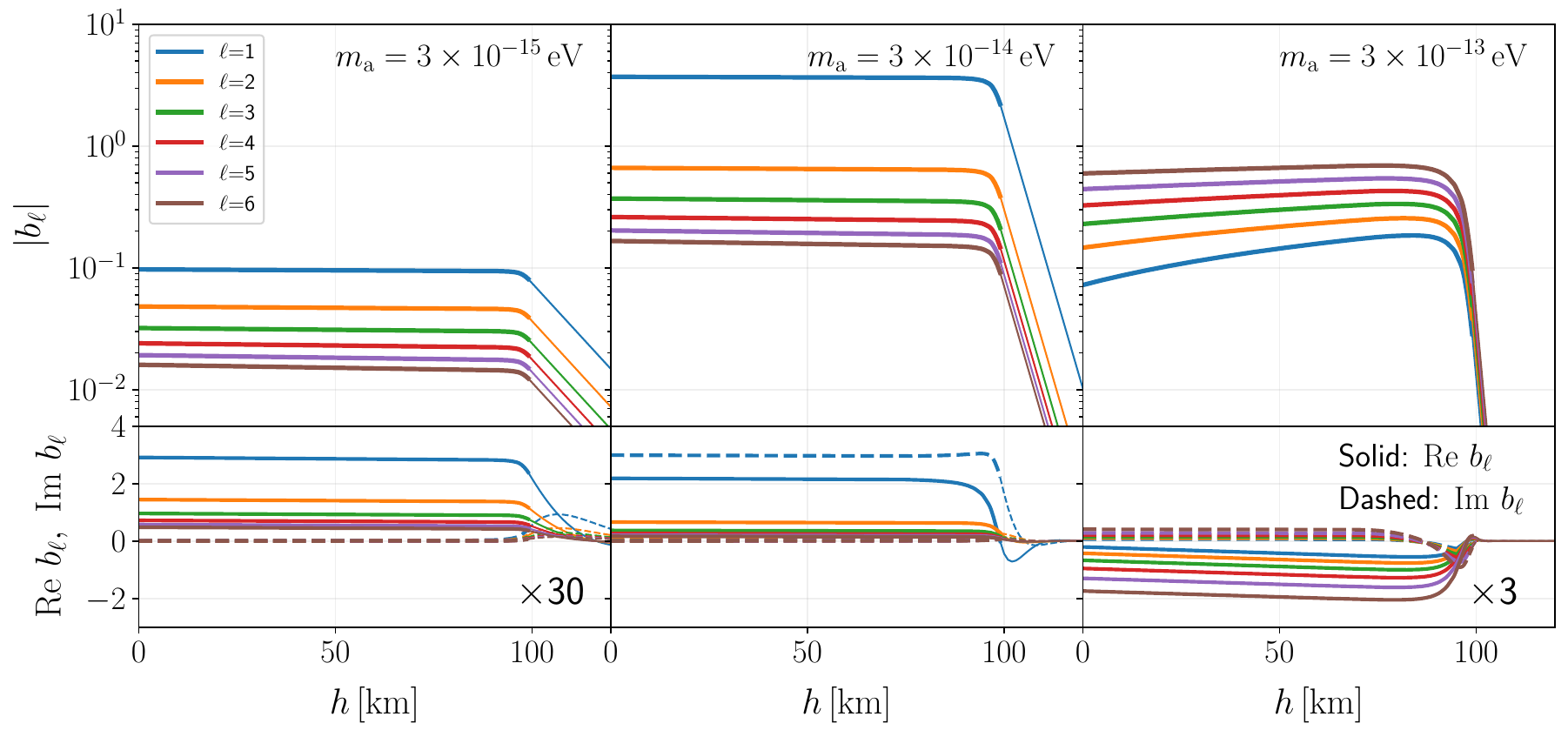}

\caption{Vertical (radial) profile of the magnetic-field mode function, $b_{\ell}$. The results of lowest six multipoles are plotted for specific axion masses of $\mass=3\times10^{-15}$ eV (left), $3\times10^{-14}$ eV (middle) and $3\times10^{-13}$\,eV (right). While the upper main panels show the absolute value, $|b_\ell|$, the lower panels display the real- and imaginary-part of $b_\ell$ on linear scales, indicated by solid and dashed curves, respectively. For ease of comparison, the results in lower-left and lower-right panels are multiplied by factors of $30$ and $3$, respectively. 
\label{fig:radial_profile_mode_func}
}
\end{figure*}
\subsubsection{Vertical profiles}
\label{subsubsec:vertical_profiles}

Next turn to focus on the radial/vertical dependence of the mode function $b_\ell$. Figure~\ref{fig:radial_profile_mode_func} shows the mode functions plotted against altitude $h$ from the Earth's surface, for specific axion masses of $\mass=3\times10^{-15}$\,eV (left), $3\times10^{-14}$\,eV (middle) and $3\times10^{-13}$\,eV (right), corresponding the frequencies of $f_{\rm a}\simeq0.72$, $7.2$ and $72$\,Hz, respectively. The upper panels display the absolute values of the mode functions, while the lower panels separately show their real (solid lines) and imaginary parts (dashed lines), with different colors representing the multipole modes, $\ell=1-6$.  

Overall, the mode functions of the magnetic field remain nearly constant up to $h\simeq100$ km, maintaining amplitudes comparable to those at the Earth's surface, as shown in Fig.~\ref{fig:mode_functions_B-field}. This flat profile appears irrespective of the multipoles and is characteristic of the regime where the axion mass is smaller than the one corresponding to the resonance peak frequency. Note that a similar flat profile is obtained in the idealized setup in Appendix~\ref{appendix:EM_wave_perfect_conductor} [see Eqs.~\eqref{eq:mode_func_B-field_ideal} or \eqref{eq:mode_func_B-field_limit}], but for axion masses $\mass \gtrsim 10^{-14}$\,eV, where resonant behavior becomes significant, the predicted amplitudes begin to deviate from the idealistic case, as shown in Fig.~\ref{fig:mode_functions_B-field}.

Above $h\sim100$ km, the mode functions exhibit rapid decay with height, indicating the reflection boundary.
This characteristic behavior is qualitatively unchanged even when adopting a smoothly varying conductivity profile above $100$ km or modifying the upper boundary condition (see Sec.~\ref{sec:discussions}). This behavior is also consistent with previous studies (e.g., Refs.~\cite{Greifinger_Greifinger1978, Sentman1990}). At these altitudes, where the conductivity exceeds $\sigma\sim 10^{-5}\,$S/m, the refractive index becomes predominantly  imaginary, approximated as $n^2\sim i\,\sigma/\mass$. Accordingly, the skin depth $\delta$ is given by $\delta\simeq\{2/(\sigma\,\mass)\}^{1/2}$, which indicates that the decay becomes sharper for larger axion masses. This trend is clearly illustrated in Fig.~\ref{fig:radial_profile_mode_func}.

\begin{figure*}[tb]
\centering
 \includegraphics[width=7.8cm,angle=0]{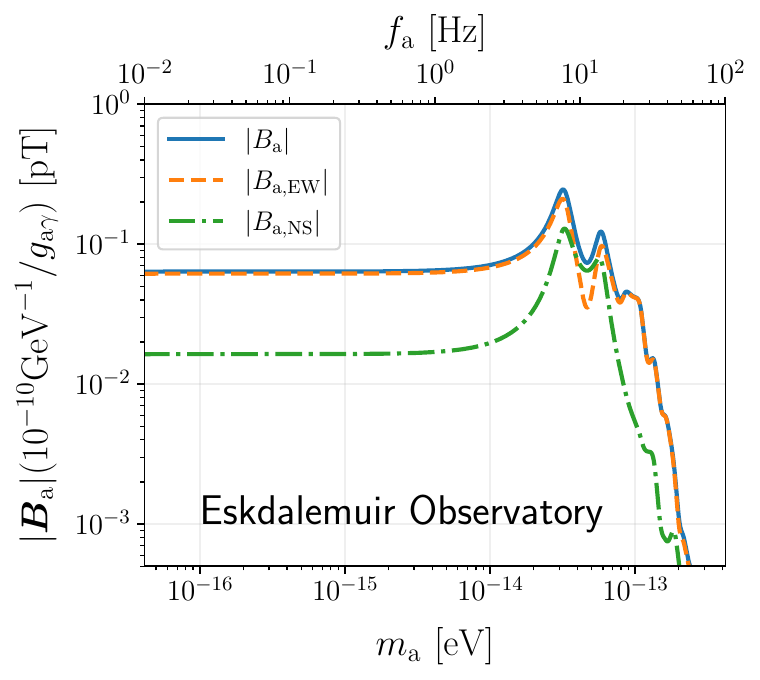}
 \hspace*{-0.5cm}
 \includegraphics[width=7.8cm,angle=0]{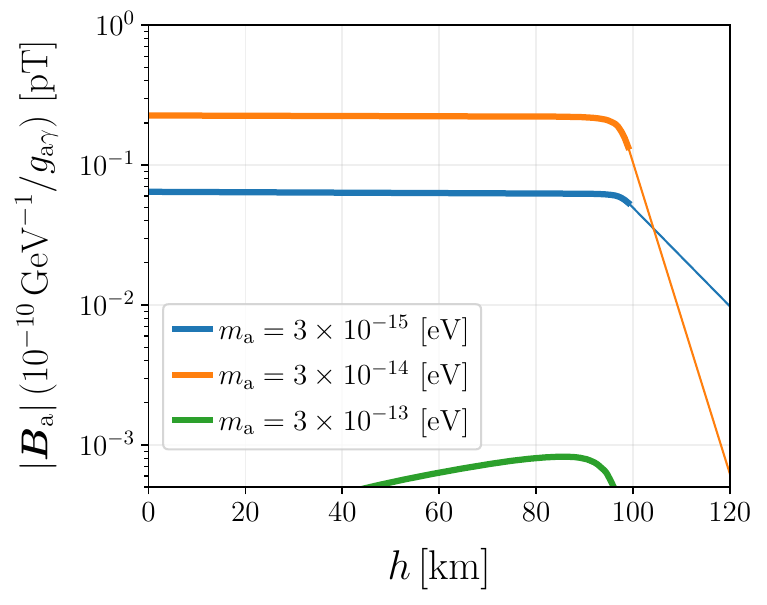}
\caption{Predictions of axion-induced magnetic field at Eskdalemuir observatory, ($55.31^\circ$ N, $3.21^\circ$ W). Left panel shows the frequency (axion mass) dependence of the magnetic field amplitude. The solid line shows the total magnetic field amplitude, while the orange and green dashed lines represent its east-west and north-south components, respectively. Right panel shows the altitude dependence of the total amplitude $|\bfB_{\rm a}|$ for specific axion masses of $\mass=3\times10^{-15}$, $3\times10^{-14}$ and $3\times10^{-13}$ eV, depicted respectively as blue, orange and green. Note that the induced magnetic field possesses only transverse components and has no vertical (radial) component.
\label{fig:B-field_at_Eskdalemuir}
}
\end{figure*}

\subsection{Magnetic field structure}
\label{subsec:B-field_structure}

Having studied the behaviors of mode functions $b_\ell$, we now convolve it with the geomagnetic field and vector spherical harmonics to compute the axion-induced magnetic field $\bfB_{\rm a}$. Assuming the coherent axion as the DM, Eq.~\eqref{eq:B_field_vsp_expansion} can be recast as 
\begin{align}
    \bfB_{\rm a}&=1.06\,{\rm pT}\,\Bigl(\frac{\gag}{10^{-10}\,{\rm GeV}^{-1}}\Big)\Bigl(\frac{\rho_{\rm DM}}{0.3\,{\rm GeV}\,{\rm cm}^{-3}}\Bigr)^{1/2}
    \Bigl(\frac{\mass}{10^{-14}\,{\rm eV}}\Bigr)^{-1}
    \nonumber
\\
    &\times e^{-i \mass t}\sum_{\ell, m}b_\ell(x)\,\Bigl(\frac{C_{\ell m}}{50\,{\rm \mu T}}\Bigr)\,\bfPhi_{\ell m}(\theta,\phi),
    \label{eq:B_field_vsp_expansion2}
\end{align}
where the harmonic coefficients of the magnetic scalar potential $C_{\ell m}$ is given by 
the IGRF-13 model \cite{IGRF13_2021}, with available multipole information up to $\ell=13$. 

As indicated in Eq.~\eqref{eq:B_field_vsp_expansion2} and  Sec.~\ref{sec:axion_signal}, the axion-induced magnetic field is a monochromatic signal with a sharp peak at the frequency $f_{\rm a}=\mass/(2\pi)$. The field lies entirely in the tangential plane to Earth’s surface, with no radial component.

In what follows, adopting the nominal value of the local DM density of $\rho_{\rm DM}=0.3\,{\rm GeV}\,{\rm cm}^{-3}$ (e.g., see Refs.~\cite{Weber_deBoer2010,Sivertsson_etal2018,Sofue2020}), we focus on the amplitude of this peak and examine its dependence on the axion mass and geographic location. In Sec.~\ref{subsubsec:Eskdalemuir_observatory}, we begin by examining the magnetic field amplitude at the Eskdalemuir Observatory, a representative location previously analyzed in our earlier works \cite{TNH_Letter, NTH_DataAnalysis}. We then examine the geographic dependence of the axion-induced magnetic field, including its behavior at representative extreme locations in Sec.~\ref{subsubsec:geography}.

\subsubsection{Local magnetic field at Eskdalemuir Observatory}
\label{subsubsec:Eskdalemuir_observatory}

Choosing the location at Eskdalemuir Observatory in the UK, ($55.31^\circ$N,\,$3.21^\circ$W), 
left panel of Fig.~\ref{fig:B-field_at_Eskdalemuir} shows the magnetic field amplitude at the Earth's surface given as a function of axion mass, $\mass$. In addition to the total amplitude $|\bfB_{\rm a}|$ (solid), we also plot the field strength along the east-west and north-south directions, which we respectively denote by $B_{\rm a, EW}$ (dashed) and $B_{\rm a, NS}$ (dot-dashed). 

As expected from Fig.~\ref{fig:mode_functions_B-field}, the induced magnetic field exhibits a peak at $\mass\sim3\times10^{-14}$\,eV, corresponding to the first Schumann resonance of the frequency $f_{\rm a} \sim 7.8$\,Hz. However, at $\mass\gtrsim 10^{-13}$\,eV, its amplitude is sharply suppressed, in marked contrast to the behavior in the mode function. This is solely due to the dipole nature of the geomagnetic fields, where the higher multipoles are sufficiently small. We have checked that the behavior at $\mass\lesssim10^{-13}$\,eV is hardly changed even if we truncate the contributions at $\ell=5$, consistent with Ref.~\cite{Arza_etal2022}. Although the suppressed behavior above $\mass=10^{-13}$\,eV becomes mildly sensitive to the higher multipoles, we confirmed that it is well-converged with the contributions up to $\ell=13$ over the plotted mass range in the left panel of Fig.~\ref{fig:mode_functions_B-field}.   
That is, the axion-induced magnetic field mediated by the geomagnetic field generically has a sharp cutoff around $\mass\gtrsim10^{-13}$\,eV regardless of the geographic location, and the behavior is mainly determined by the mode function at lower multipoles $\ell=1-3$. On the other hand, the axion-induced magnetic field is dominantly aligned along the east–west axis, with its north–south component suppressed. As we will see later, this directly comes from the geomagnetic field configuration, and reflects its local structure at the Eskdalemuir Observatory.

Turning to focus on the low-mass scale or low-frequency regime, we observe a plateau in the magnetic field amplitude. For the axion-photon coupling of $\gag=10^{-10}$\,GeV$^{-1}$ shown here, the field has an amplitude of approximately $|\bfB_{\rm a}|\sim0.07$\,pT. This behavior matches the low-mass or low-frequency limit of the idealized setup in Appendix~\ref{appendix:EM_wave_perfect_conductor}, 
again consistent with Ref.~\cite{Arza_etal2022}. As shown in Fig.~\ref{fig:mode_functions_B-field}, the mode function $b_\ell$ becomes linearly proportional to the axion mass at the Earth surface [see also Eq.~\eqref{eq:mode_func_B-field_limit}]. Substituting this dependence into Eq.~\eqref{eq:B_field_vsp_expansion2}, the magnetic field is found to be independent of the axion mass, although its amplitude still depends on not only the axion DM properties (i.e., local DM density and axion-photon coupling) but also the geomagnetic field, implying that the amplitude of plateau level can vary depending on geographical location, which we will discuss in more detail in Sec.~\ref{subsubsec:geography}.

Finally, the right panel of Fig.~\ref{fig:B-field_at_Eskdalemuir} shows the vertical profile of the magnetic field for selected axion masses. Since the mode function remains nearly constant up to the lower ionosphere, the alignment of the magnetic field also remains the same as at the Earth's surface. We therefore plot only the total amplitude, $|\bfB_{\rm a}|$. The magnetic field profiles exhibit behavior similar to that of the mode function (see Fig.~\ref{fig:radial_profile_mode_func}). It stays constant up to $h \simeq 100$\,km, above which the amplitude begins to decrease, with the characteristic scale of the fall-off depending on the axion mass.

\begin{figure}[tb]
\centering
\hspace*{-0.2cm}
\includegraphics[height=5.2cm,angle=0]{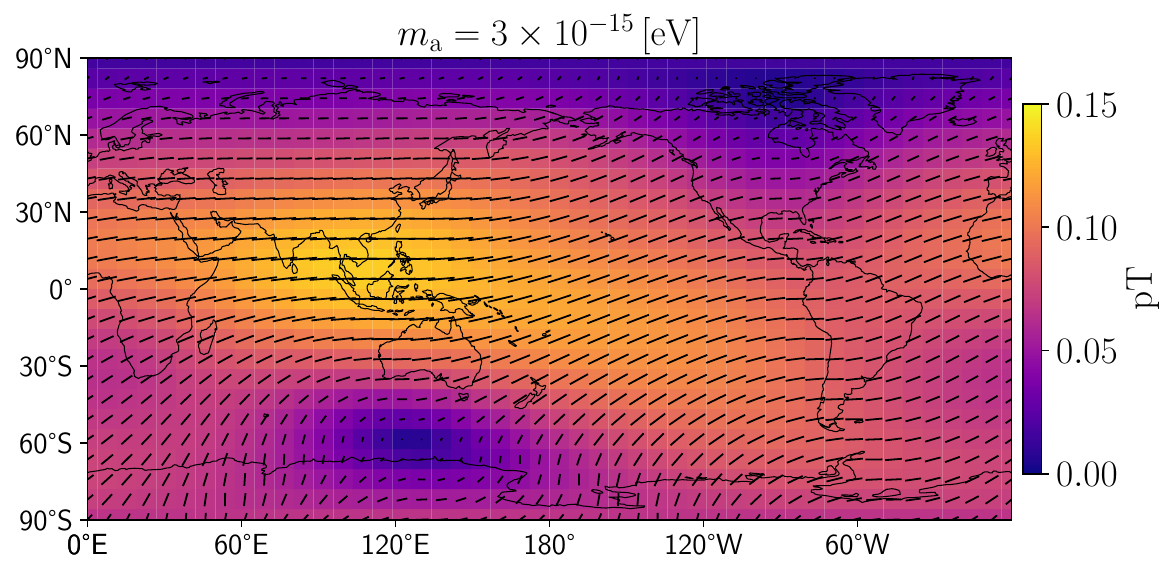}
\hspace*{-0.2cm}
\includegraphics[height=5.2cm,angle=0]{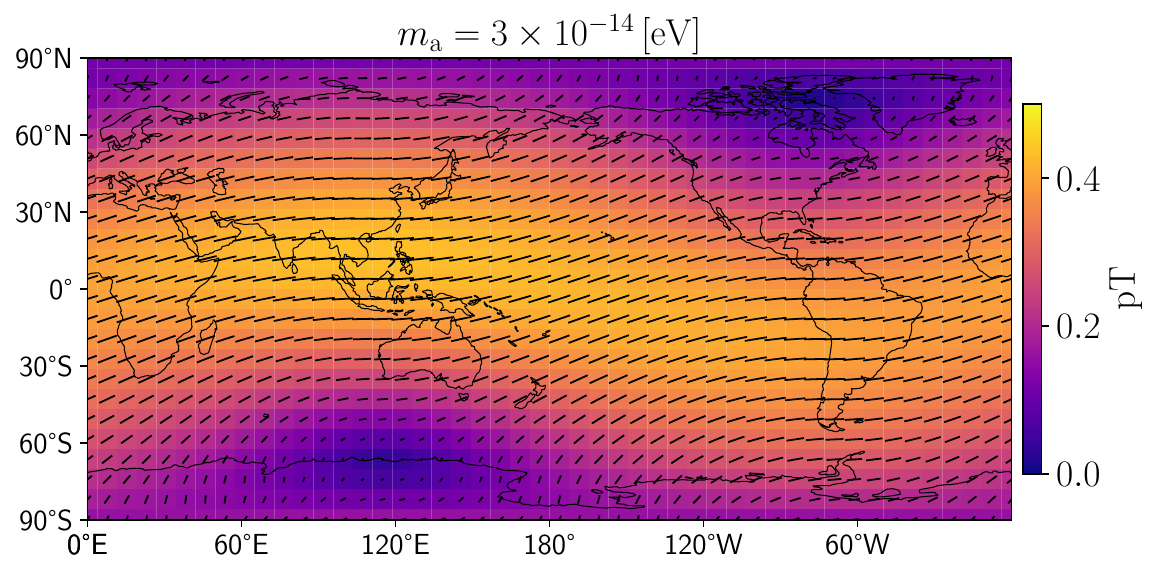}
\hspace*{-0.2cm}
\includegraphics[height=5.2cm,angle=0]{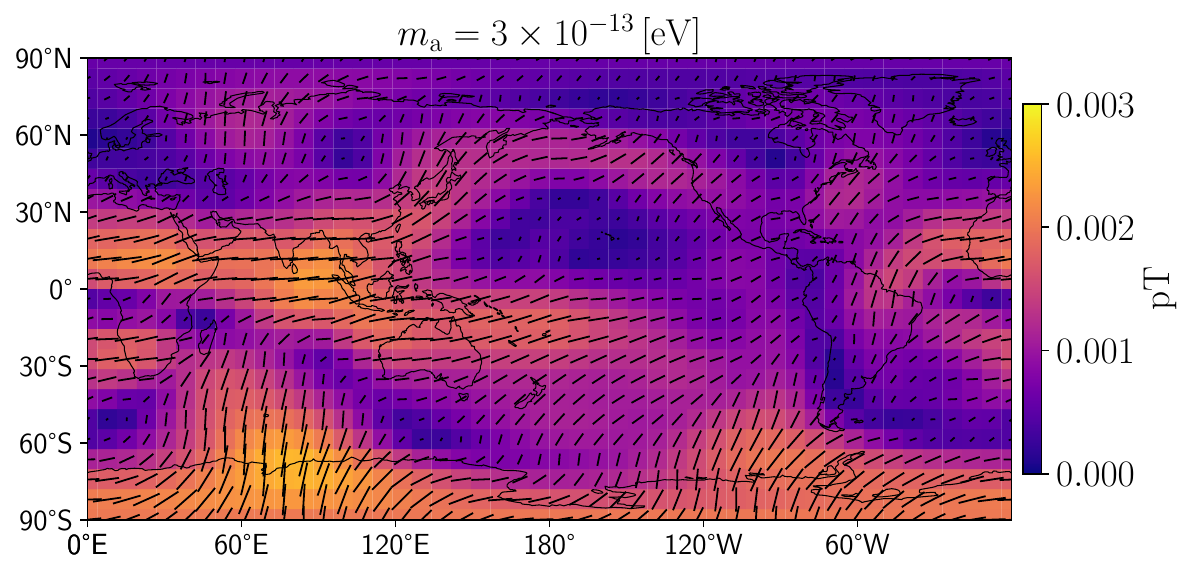}

\caption{World map of the axion-induced magnetic field for specific axion masses,  $\mass=\times3\times10^{-15}$\,eV (top), $3\times10^{-14}$\,eV (middle) and $3\times10^{-13}$\,eV (bottom). The color represents the total amplitude of magnetic field $|\bfB_{\rm a}|$, and black bar indicate the magnetic field alignment estimated from the field strength along the east-west and north-south directions. These are computed at each pixel of $7.5^\circ\times7.5^\circ$. 
\label{fig:worldmap_B-field}
}
\end{figure}

\subsubsection{Geographical dependence and extreme locations}
\label{subsubsec:geography}

Let us next consider the spatial structure of the axion-induced magnetic field. Fixing the radius at the Earth's surface, Fig.~\ref{fig:worldmap_B-field} shows a global map in the Plate Carr\'ee projection, illustrating the geographical dependence of the field amplitude and orientation. For each pixelized cell of $7.5^\circ \times 7.5^\circ$, the field amplitude is represented by a color scale, and the orientation is indicated by a black bar. Since the amplitude varies significantly with the axion mass, the color scale is adjusted in each panel.

For axion masses corresponding to the first Schumann resonance and lower, as shown in the top and middle panels of Fig.~\ref{fig:worldmap_B-field}, the magnetic field amplitude is enhanced near the equator and is generally aligned along the east–west direction. This behavior can be understood from the Amp\`ere-Maxwell law at \eqref{eq:Maxwell_reduced2}, which suggests that the axion-induced magnetic field tends to align perpendicular to the background geomagnetic field. As a result, for axion masses below $\mass = 10^{-13}$\,eV, where the dipole component of the geomagnetic field dominates, the induced magnetic field is weakest near the magnetic poles and strongest in Southeast Asia. From this map, we see that the Eskdalemuir Observatory, examined in Sec.~\ref{subsubsec:Eskdalemuir_observatory}, lies in a region with an intermediate field amplitude between these two extremes. 

On the other hand, for axion masses above $\mass = 10^{-13}$\,eV, higher multipole contributions to the geomagnetic field become non-negligible. As a result, the spatial pattern of the axion-induced magnetic field, shown in the bottom panel, becomes featureless and incoherent compared to the low-mass cases, with its amplitude significantly suppressed.

\begin{figure*}[tb]
\centering
\hspace*{-0.5cm}
 \includegraphics[width=16cm,angle=0]{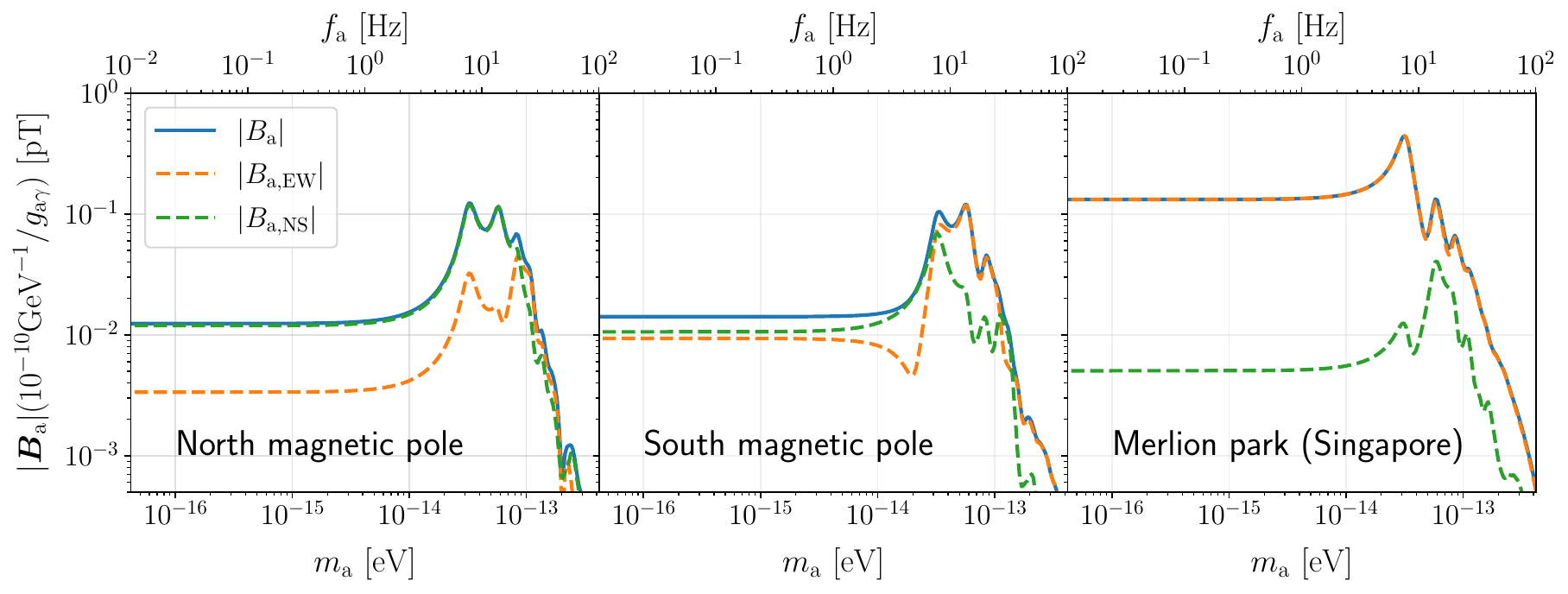}
\caption{Amplitude of the axion-induced magnetic field at the three extreme sites, plotted against the axion mass (lower) and frequency (upper). The left and middle panels respectively show the results at north and south magnetic poles, while the right panel plot the axion-induced magnetic field at Merlion park in Singapore, as a representative location of Southeast Asia. Plotted results are at Earth's surface. Meaning of line types are the same as those in the left panel of Fig.~\ref{fig:B-field_at_Eskdalemuir}. 
\label{fig:B-field_at_representative}
}
\end{figure*}

Having understood the overall trend of the axion-induced signal, we now examine its behavior at specific characteristic locations, and compare them with the result at the Eskdalemuir Observatory. 
In Fig.~\ref{fig:B-field_at_representative}, we plot the axion mass dependence of the axion-induced magnetic field at the north magnetic pole (left), the south magnetic pole (middle), and Merlion Park in Singapore (right), the latter of which serves as a representative location in Southeast Asia. The geographic coordinates are, respectively, ($85.02^\circ$N, $132.83^\circ$W), ($64.43^\circ$S, $137.33^\circ$E), and ($10.79^\circ$N, $106.70^\circ$E). As expected, the magnetic field at the first two sites has a suppressed amplitude. In particular, at the low-mass plateau, it reaches $|\bfB_{\rm a}|\sim0.01$\,pT for an axion-photon coupling of $\gag=10^{-10}$\,GeV$^{-1}$. Since the dipole contribution becomes small, higher-order multipole moments of the geomagnetic field, such as the quadrupole and beyond, start to play a role. Consequently, above $10^{-14}$\,eV, instead of a prominent resonant peak, multiple ridge-like features appear. In addition, the magnetic field alignment changes across the first resonant peak, reflecting a transition in the dominant multipole contribution. By contrast, the magnetic field at the Southeast Asian site (Merlion Park in Singapore) is dominated by the dipole component and shows a clear resonant peak structure. The field is coherently aligned in the east–west direction across the plotted axion mass range. Compared to the magnetic poles, the amplitude on the low-mass plateau is nearly an order of magnitude larger, and the resonant peaks are significantly higher. These features make Southeast Asia, including Merlion Park, a promising region for axion DM searches. Moreover, combining the field alignment information helps discriminate axion signals from systematics enabling us more robust detection.

\section{Discussions}
\label{sec:discussions}

In the previous section, we computed the axion signal that is expected to be embedded in the terrestrial magnetic fields,  and investigated its properties based on the formulation in Sec.~\ref{sec:formulation}. In this section, we assess the impact of assumptions or simplifications in the formulation and discuss the robustness of our results. 

Among the assumptions involved in our setup, the treatment of atmospheric conductivity plays a particularly important role, as it can affect the quantitative predictions of axion signal especially at the mass above $\mass\gtrsim10^{-14}$\,eV. Specifically, the atmospheric conductivity is assumed to be spherically symmetric, and to vary only with radius. We adopt a specific conductivity profile from Ref.~\cite{Kudintseva_etal2016}, using their numerical table for $\sigma$ up to the altitude of $h=100$km, at which we impose an upward-propagating boundary condition. Furthermore, we assume a constant conductivity above this point. 
\begin{figure*}[tp]
\centering
\includegraphics[width=7.8cm,angle=0]{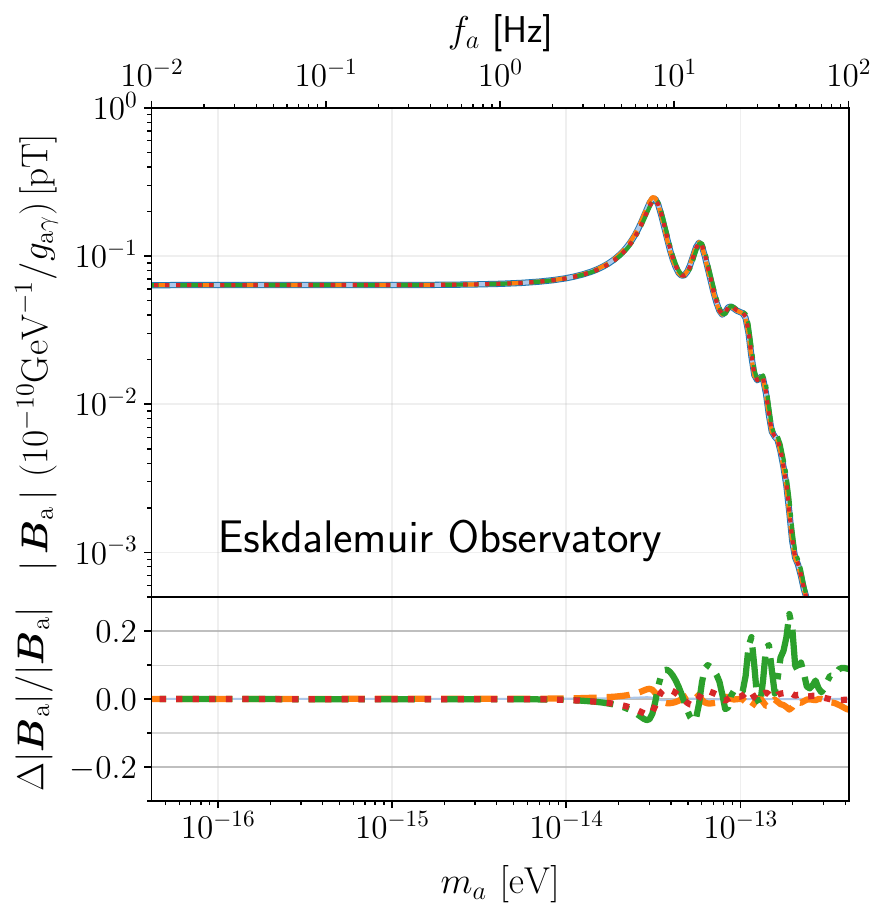}
\hspace*{-0.5cm}
\includegraphics[width=7.8cm,angle=0]{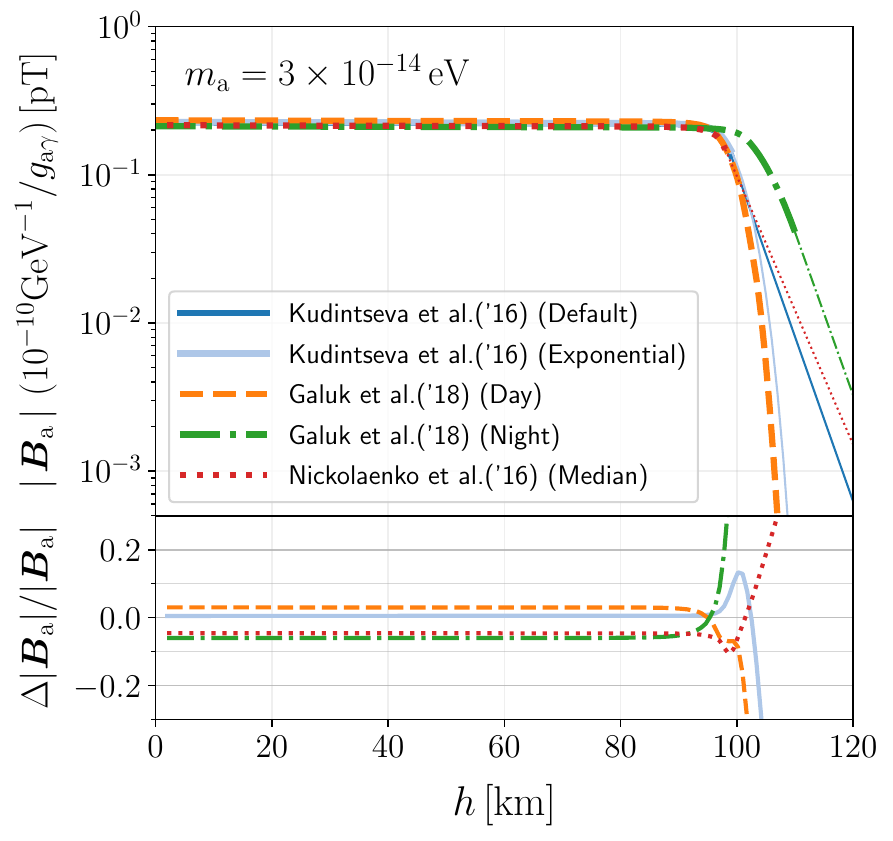}

\caption{
Dependence of the atmospheric conductivity profile and lower-ionosphere boundary condition on the axion-induced magnetic field amplitude, $|\bfB_{\rm a}|$, at the Eskdalemuir Observatory. 
The left panel shows results at the Earth's surface plotted against axion mass (lower) and frequency (upper), and the right panel shows vertical profiles as a function of  altitude $h$. In both panels, lower subpanels give fractional differences from a reference model (blue solid), which adopts the conductivity profile of Ref.\cite{Kudintseva_etal2016} with constant conductivity above the upper boundary. Dashed, dash-dotted, and dotted curves use the day-side and night-side profiles of Ref.\cite{Galuk_etal2018} and the median profile of Ref.~\cite{Nickolaenko_etal2016}, respectively, sharing the same upper-boundary condition as the reference model. 
The faint blue curves use the same conductivity profile as the reference model, with the constant conductivity above the lower-ionosphere boundary replaced by an exponential profile. The modifications to the boundary condition in this treatment are described in Appendix~\ref{appendix:exponential_profile}.
Note that in all cases, the upper-boundary condition is applied at the maximum altitude of conductivity profile in each model (see Fig.~\ref{fig:conductivity_profiles}).
\label{fig:Eskdalemuir_model_difference}
}
\end{figure*}

At altitudes $h\lesssim 90$\,km, the radial variation in the atmospheric conductivity dominates over the transverse variation (e.g., \cite{Holzworth_etal1985,Takeda_Araki1985,Richmond_Thayer2000}), making the spherical symmetry a reasonable approximation\footnote{To be precise, observational studies show that near-surface atmospheric conductivity is lower over the ocean and a few times higher over land \cite{Cobb_Wells1970,Kamra_Deshpande1995}. However, this contrast rapidly decreases with altitude. Because the dominant effect mainly arises from the strong vertical variation of conductivity \cite{Harrison_Carslaw2003}, such near-surface inhomogeneities are expected to have only a negligible effect on the axion-induced magnetic fields.}. This explains why most previous studies of terrestrial low-frequency EM waves considered only the radial dependence \cite{Simoes_etal2012}. Nevertheless, experimental data at higher altitudes are limited, and there is still non-negligible uncertainty in the radial conductivity profile. Furthermore, the ionosphere exhibits daily variations due to solar irradiation, causing asphericity between day and night sides, as shown in Fig.~\ref{fig:conductivity_profiles}. 

On the other hand, above the altitude $h\sim90$\,km, the conductivity becomes further increasing, developing also anisotropies due to the influence of the geomagnetic fields. In this respect, our boundary condition, the constant conductivity above the upward-propagating boundary, is simplistic. Changing the upper conductivity can significantly impact the axion-induced magnetic fields.

In general, non-uniformities in the atmospheric conductivity and ionosphere boundary can induce not only the $m$-dependent frequency shifts also a coupling between the TE and TM modes, potentially leading to non-zero vertical magnetic fields \cite{Bliokh1980,Nickolaenko_Hayakawa2002,Nickolaenko_Hayakawa2013}, in contrast to the case studied in Sec.~\ref{sec:formulation} and \ref{sec:results}. There have been a number of studies examining their impacts on the Schumann resonances in both observational and theoretical contexts, the latter of which typically relying on full 3D simulations (e.g., \cite{Toledo-Redondo_2016,Salinas_etal2024,Niknam_Simpson2021}). Currently, existing observational studies have not provided unambiguous evidence for either effect (but see Refs.~\cite{Nickolaenko_Sentman2007,Navarro_etal2008} for possible $m$-dependence of the Schumann resonance and Ref.~\cite{Silber_etal2015} for reports of vertical magnetic fields). While such effects may be present under certain conditions, they are generally not expected to play a dominant role at the level relevant for the present work.

Although our formalism allows only radial variation in the atmospheric conductivity and a full assessment of non-sphericity effects on the axion-induced signal is  beyond the scope of this work, we can still partially gauge their impact by comparing different conductivity profiles and boundary conditions.

Figure~\ref{fig:Eskdalemuir_model_difference} shows the axion-induced magnetic fields at the location of the Eskdalemuir Observatory, adopting different atmospheric conductivity models and boundary conditions. Left panel examines the axion mass dependence at the Earth's surface, while right panel plots the radial dependence for the specific axion mass of $\mass=3\times10^{-14}$\,eV. In both cases, the top panels present the total amplitude $|\bfB_{\rm a}|$, while the bottom panels plot the fractional difference relative to the reference result adopting the conductivity profile of Ref.~\cite{Kudintseva_etal2016}, depicted as the blue solid curve in the upper panels. Here, apart from the reference result, we examine four additional cases: the results adopting the conductivity profiles of Ref.~\cite{Nickolaenko_etal2016} (red dotted) and day-side (orange dashed) and night-side (green dot-dashed) profiles of Ref~\cite{Galuk_etal2018}, and an exponential extrapolation above the upper boundary (faint blue). In all cases, except for the exponential extrapolation, a constant conductivity is assumed above the upper boundary. The treatment of the exponential extrapolation and the modifications in the calculation are described in detail in Appendix~\ref{appendix:exponential_profile}.

The left panel of Fig.~\ref{fig:Eskdalemuir_model_difference} demonstrates that variations in the conductivity profile or boundary condition begin to influence the predicted amplitude of $|\bfB_{\rm a}|$ above the first Schumann resonance, corresponding to an axion mass of $\mass\simeq3\times10^{-14}$\,eV. However, at the Earth's surface, the impact remains small, reaching at most $\sim20$\%, where the magnetic field is already suppressed. Appendix \ref{appendix:Schumann} further confirms that the resonance structure of the Schumann resonances, i.e., the peak frequencies and $Q$ values, exhibits only minor model dependence\footnote{The results summarized in Table~\ref{tab:eigenvalues} assume constant conductivity above the upper boundary, but we confirmed that they remain almost unchanged even when an exponential profile is adopted.}. In contrast, the right panel of Fig.~\ref{fig:Eskdalemuir_model_difference} shows that these variations can significantly alter the radial structure of the magnetic field near the upper boundary ($h\gtrsim90$\,km), highlighting the need for careful treatment in the EM calculations, including the anisotropies in atmospheric conductivity \cite{Richmond_Thayer2000,Simoes_etal2012}. Nevertheless, for axion signals near the Earth's surface, our formalism provides robust predictions that can be used with relatively low uncertainty to estimate the low-frequency magnetic field.

\section{Conclusion}
\label{sec:conclusion}

The axions or axion-like particles are a representative candidate of ultra-light dark matter (DM), and its salient feature is their coupling to electromagnetism (EM), which offer a direct route to search for DM via its EM signal. In particular, as pointed out by Refs.~\cite{Arza_etal2022}, near the Earth’s surface, axions couple with the geomagnetic fields, and generate a monochromatic and persistent EM signal as illustrated in Fig.~\ref{fig:schematic_signal}, whose characteristic frequency is determined solely by the axion mass  (see also Ref.~\cite{Fedderke_etal2021} for dark photon DM).

In this paper, we have developed a theoretical framework to describe axion-induced EM waves confined in the Earth–ionosphere cavity, focusing on the low-frequency regime relevant for axion masses of $m_a \lesssim 3\times10^{-13}$ eV (corresponding to $f \lesssim 100$ Hz). Our formulation is valid across this entire low-frequency range, including frequencies below 1 Hz. Extending early studies by Refs.~\cite{Arza_etal2022,Sulai:2023zqw,Friel:2024shg} (see also Ref.~\cite{Arza_etal2025}), we have properly taken into account the atmospheric conductivity, which leads to a dissipative behavior of the axion-induced EM waves. This treatment becomes particularly important at frequencies above 1 Hz, where resonant features emerge. In contrast, a naive calculation neglecting the atmospheric conductivity, as presented in Appendix~\ref{appendix:EM_wave_perfect_conductor}, exhibits divergent behavior at the resonant frequencies or for certain axion masses.

Incorporating a quantitative model for the radial-dependent atmospheric conductivity, we have presented a semi-analytical method to solve the EM wave equation while fulfilling the boundary conditions at the Earth's surface ($h=0$\,km) and lower ionosphere ($h\sim100$\,km). This enables us to obtain quantitative predictions for the axion-induced EM signal with finite amplitude, thereby overcoming the divergences encountered in the idealized treatments. Focusing particularly on the amplitude of produced magnetic fields, our analysis has revealed several robust features of the axion-induced signals: 

\begin{itemize}
\item The axion-induced magnetic field is expressed in a harmonic expansion, as given in Eq.~\eqref{eq:B_field_vsp_expansion} [or \eqref{eq:B_field_vsp_expansion2}]. It is generally characterized by a transverse mode with a vanishing vertical (radial) component.
\item For the low-mass limit $\mass\ll10^{-14}$\,eV, the spectral features (i.e., axion mass dependence) of the magnetic-field mode function $b_\ell$ recover those obtained from an idealized setup with perfect conducting boundaries neglecting the atmospheric conductivity (Fig.~\ref{fig:mode_functions_B-field}). Hence, the predicted amplitude of the magnetic field becomes consistent with Refs.~\cite{Arza_etal2022,Sulai:2023zqw,Friel:2024shg} (see also Ref.~\cite{Arza_etal2025}). 
\item Above $\mass\simeq10^{-14}$\,eV, the magnetic-field mode functions exhibit a resonant behavior in its axion mass dependence, and their peak frequencies are predicted to be the same as those of the Schumann resonances in natural EM waves. However, due to the dipole nature of the geomagnetic field, the axion-induced magnetic fields are found to be suppressed at $\mass\gtrsim10^{-13}$\,eV (Fig.~\ref{fig:B-field_at_Eskdalemuir}). The highest amplitude is obtained around the mass of $\mass\sim3\times10^{-14}$\,eV, accompanying minor resonant peaks above that mass.
\item The amplitude and orientation of  the axion-induced magnetic field vary not only with axion DM properties (axion mass $\mass$,  axion-photon coupling $\gag$, and local DM density $\rho_{\rm DM}$), but also with geographical location (Fig.~\ref{fig:worldmap_B-field}). For the masses of $\mass\lesssim10^{-13}$\,eV, the axion-induced magnetic field tends to align perpendicular to geomagnetic field. In particular, the region around Southeast Asia is expected to have the largest signal, exhibiting also an enhanced resonant amplitude at $\mass\sim3\times10^{-14}$\,eV (Fig.~\ref{fig:B-field_at_representative}). 
\end{itemize}

The features summarized above provide distinctive signatures of axion DM and extend our previous analyses, namely those presented in Refs.~\cite{NTH_DataAnalysis,TNH_Letter}, to a global and geographically dependent characterization of the signal. These features help discriminate an axion signal from other natural or anthropogenic sources of low-frequency EM fields. We have verified that the predicted axion signal at the Earth's surface remains almost unchanged irrespective of the atmospheric conductivity profile or the choice of upper boundary condition in the ionosphere. Hence, the formalism presented in this paper offers reliable and quantitative predictions, and our results can serve as a template for searching for axion DM signals in terrestrial EM fields. In this spirit, Ref.~\cite{NTH_DataAnalysis} has analyzed public geoscience data from the Eskdalemuir Observatory and placed stringent constraints on $\gag$, in the mass range $10^{-15}{\rm eV}\lesssim\mass\lesssim10^{-13}{\rm eV}$, identifying also potential signal candidates (see also Ref.~\cite{TNH_Letter}).

Throughout this paper, we have focused on axion DM and computed the associated EM signal, but the present formalism can also be applied to other types of DM coupled to EM fields. A notable example is dark photon DM \cite{Nelson_Scholtz2011, Fabbrichesi_etal2020} (see also Ref.~\cite{Arza_millichargedDM2026} for millicharged DM), which excites EM waves even without the geomagnetic fields and is thus distinguishable from axion-induced signals. Refs.~\cite{Fedderke_etal2021_superMAG, Friel:2024shg} have searched in the mass range of dark photon DM below $4\times 10^{-15}$\,eV, and placed constraints on the kinetic mixing parameter (see also Refs.~\cite{Sulai:2023zqw, Arza_etal2025} for  attempts to extend to heavier mass range). In a similar manner, as reported in Ref.~\cite{NNTH_Darkphoton}, we have extended the present formalism to dark photon DM and performed a search for masses beyond previous studies, obtaining stronger constraints on the kinetic mixing parameter in certain mass regions.

Finally, we note that the actual DM signal cannot remain strictly stationary with a constant amplitude on time scales exceeding the coherence time. This arises from the superposition of field modes with random phases, which causes the amplitude of the axion-induced magnetic field to fluctuate in time in a stochastic manner. Although our present formalism does not explicitly account for such stochastic effects \cite{Centers_etal2019, Lisanti_etal2021}, our predictions can still be reliably applied to data analysis, provided that observational data are stacked over time scales sufficiently long compared to the coherence time (see Ref.~\cite{NTH_DataAnalysis}). This smooths out short-term fluctuations and thus reduces the influence of stochastic effects in the analysis (see Ref.~\cite{Nakatsuka:2022gaf} for a quantitative assessment).

\bigskip 
\noindent{\bf Acknowledgements} 
We thank Kimihiro Nomura for discussion and his useful comments. 
We are also grateful to the KAGRA Physical Environment Monitoring group for their valuable information and insights on low-frequency magnetic field measurements. This work was supported in part by JSPS KAKENHI Grant Numbers 
JP23K20844, JP23K25868 and JP26H02044 (AT), JP23K03408, JP23H00110, and JP23H04893 (AN), JP21K03580 and JP25K07288 (YH). 

\appendix
\section{Axion-induced EM wave in an idealized setup}
\label{appendix:EM_wave_perfect_conductor}

In this appendix, we present an analytical solution for the axion-induced EM wave in an idealized setup, where the ionosphere is treated as a perfect conductor and atmospheric conductivity is neglected. In the low-frequency limit, Ref.~\cite{Arza_etal2022} derived an analytical expression for the axion-induced signal, which has been used in axion DM searches well below the resonance frequencies \cite{Arza_etal2022, Sulai:2023zqw, Friel:2024shg,Arza_etal2025}. Here, for comparison with our more realistic calculations that include finite conductivity, we extend the analysis without taking the low-frequency approximation.

\subsection{Radial mode function}

Similar to the setup described in Sec.~\ref{sec:formulation}, the governing equation for the EM waves is Eq.~\eqref{eq:wave_eq}, which is expanded using the vector spherical harmonics as in Eq.~\eqref{eq:E_field_TM-mode}. This reduces the system to the radial mode equation, given by Eq.~\eqref{eq:mode_eq_dlm} or \eqref{eq:radial_mode_eq_u_ellm}. Neglecting atmospheric conductivity, the refractive index becomes unity, and the mode equation can be solved analytically in the same manner as Eq.~\eqref{eq:general_solution_const_n}. Within the Earth-ionosphere cavity, the solution takes the form:
\begin{align}
    d_{\ell m}^{\rm ideal}=A_{\ell m}^{\rm ideal}\,j_\ell(x) + B_{\ell m}^{\rm ideal}\,y_\ell(x)-\gag\,a_0\frac{C_{\ell m}}{\ell}\,\frac{x_0^{\ell+2}}{x^{\ell+1}}.
    \label{eq:mode_func_idealized}
\end{align}
The coefficients $A_{\ell m}^{\rm idel}$ and $B_{\ell m}^{\rm idel}$ are determined by the boundary conditions at the Earth's surface and ionosphere, both of which are now regarded as a perfect conducting boundary. Assuming that the upper boundary is located at $r=R_{\rm E}+L$, the boundary conditions becomes [see Eq.~\eqref{eq:BC_surface}]
\begin{align}
  \Bigl(\frac{d}{dr}+\frac{1}{r}\Bigr)g_{\ell m}\Biggr|_{r=R_{\rm E},\,R_{\rm E}+L} =0, 
\label{eq:boundary_conditions_idealized}
\end{align}
which leads to 
\begin{align}
 & \left(
\begin{array}{cc}
K(x_0) & L(x_0)\\
\\
K(x_1) & L(x_1)\\
\end{array}
\right)\,\left(
\begin{array}{c}
 A_{\ell m}^{\rm ideal} \\
 \\
 B_{\ell m}^{\rm ideal}
\end{array}
\right) 
 = -\gag\,a_0\,C_{\ell m}\left(
\begin{array}{c}
 1 \\
 \\
 \bigl(x_0/x_1\bigr)^{\ell+2}
\end{array}
\right),
\label{eq:boundary_conditions1.5}
\end{align}
where we define $K(x)\equiv j_\ell'(x)+j_\ell(x)/x$ and $L(x)\equiv y_\ell'(x)+y_\ell(x)/x$. The dimensionless radius $x_0$ and $x_1$ are respectively given by $x_0\equiv\mass R_{\rm E}$ and $x_1\equiv\mass (R_{\rm E}+L)$. 
This is solved to give 
\begin{align}
& \left(\begin{array}{c}
  A_{\ell m}^{\rm ideal}\\
  \\
  B_{\ell m}^{\rm ideal}
 \end{array}
 \right)= -\frac{\gag \,a_0\,C_{\ell m}}{K(x_0)L(x_1)-K(x_1)L(x_0)} 
\left(
\begin{array}{cc}
 L(x_1)& -L(x_0)\\
 \\
 -K(x_1)& K(x_0)
\end{array}
\right)\left(
\begin{array}{c}
 1\\
 \\
 \bigl(x_0/x_1\bigr)^{\ell+2}
\end{array}\right).
\label{eq:coefficients_ideal}
\end{align}

\subsection{Magnetic field}

Substituting the expression in Eq.~\eqref{eq:mode_func_idealized} into Eq.~\eqref{eq:B-field_TM-mode}, the induced magnetic field takes the same harmonic expansion form as in Eq.~\eqref{eq:B_field_vsp_expansion} or \eqref{eq:B_field_vsp_expansion2}, but with the mode function $b_\ell$ now given by 
\begin{align}
    b_\ell(x) = \widetilde{A}_\ell^{\rm ideal}\,j_\ell(x) + \widetilde{B}_\ell^{\rm ideal}\,y_\ell(x),\quad x_0\leq x\leq x_1,
    \label{eq:mode_func_B-field_ideal}
\end{align}
where the coefficients $\widetilde{A}_\ell^{\rm idel}$ and $\widetilde{B}_\ell^{\rm idel}$ are defined in Eqs.~\eqref{eq:coefficients_ideal} and normalized by the factor $\gag \,a_0\,C_{\ell m}$. That is, the coefficients are independent of axion-coupling and geomagnetic field properties, characterized solely by dimensionless radii $x_0$ and $x_1$. 

Assuming the upper boundary at $r=R_{\rm E}+L$ represents the lower ionosphere (D or E layer), we can take the ratio $L/R_{\rm E}$ to be small. In this limit ($L/R_{\rm E}\ll1$), the coefficients is simplified and become independent of $L$:  
\begin{align}
    \left(
\begin{array}{c}
 \widetilde{A}_{\ell}^{\rm ideal}
 \\
 \\
 \widetilde{B}_{\ell}^{\rm ideal}
\end{array}
\right) &\simeq - \frac{x_0^4}{x_0^2-\ell(\ell+1)} 
\left(
\begin{array}{c}
 {\displaystyle \frac{\ell+1}{x_0}\,y_{\ell-1}(x_0) - y_{\ell}(x_0) }
 \\
 \\
{\displaystyle - \frac{\ell+1}{x_0}\,j_{\ell-1}(x_0) + j_{\ell}(x_0) }
\end{array}
\right).
\label{eq:A_ellm_B_ellm_L_R}
\end{align}
This expression indicates that the magnetic field diverges when $x_0=\sqrt{\ell(\ell+1)}$, corresponding to the Schumann resonance condition \cite{Schumann1952a,Schumann1952b}. Expressed in terms of the frequency $f_{\rm a}=\mass /(2\pi)$, the resonance occurs at  
\begin{align}
    f_{\rm a}^{\rm Res}&=\frac{1}{2\pi\,R_{\rm E}}\sqrt{\ell(\ell+1)}
    \nonumber
    \\
    &\simeq 7.5 \sqrt{\ell(\ell+1)}\,\,{\rm Hz}.
    \label{eq:Schumann_freq_ideal}
\end{align}
As shown in Fig.~\ref{fig:mode_functions_B-field}, these resonance frequencies are slightly higher than the peak frequencies obtained from the more realistic model incorporating finite atmospheric conductivity and the upward-propagating wave condition at the ionosphere.

Since the dimensionless radius $x$ is evaluated as $x\simeq 0.032 \,(\mass/10^{-15}\,{\rm eV})(r/R_{\rm E})$,  a further simplification is possible in the small-mass (low-frequency)  regime.  In this limit ($x\ll1$), the spherical Bessel functions admit the following asymptotic forms:
\begin{align}
 j_\ell(x) \simeq \frac{x^\ell}{(2\ell+1)!!},
\qquad
 y_\ell(x) \simeq  -(2\ell-1)!!\,\,x^{-\ell-1},
\end{align}
from which the coefficients are approximated as
\begin{align}
        \left(
\begin{array}{c}
 \widetilde{A}_{\ell}^{\rm ideal}
 \\
 \\
 \widetilde{B}_{\ell}^{\rm ideal}
\end{array}
\right) &\simeq \left(
\begin{array}{c}
{\displaystyle \frac{(\ell-2)(2\ell-3)!!}{\ell(\ell+1)}\,x_0^{-\ell+3} }
\\
\\
{\displaystyle -\frac{1}{\ell(2\ell-1)!!}x_0^{\ell+2} }
\end{array}
\right)
\end{align}
Substituting this into Eq.~\eqref{eq:mode_func_B-field_ideal}, we find, to leading order, that $b_\ell$ is dominated by the Neumann contribution, and we obtain
\begin{align}
    b_\ell(x)\simeq\,\widetilde{B}_\ell^{\rm ideal}\,y_\ell(x)\simeq \frac{x_0}{\ell}\Bigl(\frac{x_0}{x}\Bigr)^{\ell+1}, 
    \label{eq:mode_func_B-field_limit}
\end{align}
which reproduces the result in Ref.~\cite{Arza_etal2022}. 
Eq.~\eqref{eq:mode_func_B-field_limit} shows that the amplitude of the mode function $b_\ell$ at the Earth's surface (i.e., $x=x_0$) increases linearly with $x_0$, and thus with the axion mass $\mass$. Its strength, however, decreases as multipole order $\ell$ increases, as illustrated in Fig.~\ref{fig:mode_functions_B-field}.

\section{Schumann resonance modes}
\label{appendix:Schumann}

In this Appendix, based on the formalism in Sec.~\ref{sec:formulation}, we provide a quantitative evaluation of the Schumann resonance structure, namely, the resonance frequencies and peak widths, which appear not only in axion-induced EM waves but also in natural EM environment excited by the lightning discharges. 

The Schumann resonance arises from propagating EM waves confined within the Earth-ionosphere cavity, and can be described by the wave equation without the axion-photon coupling. Specifically, these resonances correspond to the solutions of the radial mode equation in Eq.~\eqref{eq:radial_mode_eq_u_ellm} with the coupling $\gag$ set to zero. In the absence of a source term, the axion mass, which originally enters through the dimensionless radius $x$, becomes physically irrelevant, and should be replaced with the EM frequency through $\mass\to 2\pi\,f$. With $\gag=0$, the mode equation reduces a Strum-Liouville form, which permits only discrete frequencies for each multipole mode. We denote these as $f_\ell^{\rm Sch}$. These discrete values are determined by solving the mode equation subject to the boundary condition in Eq.~\eqref{eq:BC_u_ellm}. 

Following the method in Sec.~\ref{sec:formulation}, we obtain  the source-free solutions from   Eq.~\eqref{eq:solution_segment} and \eqref{eq:solution_outermost}, with coefficients determined by Eqs.~\eqref{eq:_curlA_multi-step} and \eqref{eq:coeff_Alm_Blm_k-1_final_form}, by setting $\gag$ to zero. Since these linearly depend on $\gag$, the resulting solution becomes trivial (i.e., $d_{\ell m}=0$) unless the quantity $W$, defined in Eq.~\eqref{eq:V_ell_W_ell}, vanishes. In other words, a non-trivial EM wave solution exists only if the following condition is satisfied:    
\begin{align}
    \bigl[\vec{M}_0\bigr]_2\Bigr|_{\mass\to2\pi \,f_\ell^{\rm Sch}}=0,
    \label{eq:Schumann_resonance_condition}
\end{align}
where the subscript $2$ indicates the second component of the vector, and $\vec{M}_0$ is a multipole-dependent vector defined in Eq.~\eqref{eq:def_vec_M}.

Eq.~\eqref{eq:Schumann_resonance_condition} represents a transcendental equation for the resonance frequency $f_\ell^{\rm Sch}$, whose discrete values must be obtained numerically for each multipole mode. Due to the finite conductivity of the atmosphere, the solution for  $f_\ell^{\rm Sch}$ are generally complex. Since the derivative of Eq.~\eqref{eq:Schumann_resonance_condition} is difficult to evaluate analytically, standard root-finding algorithms such as Newton's method are not suitable for solving this transcendental equation. 

Here, we employ the Muller method \cite{Numerical_Recipes2002} to find the resonance frequency $f_\ell^{\rm Sch}$ that satisfies Eq.~\eqref{eq:Schumann_resonance_condition}.  Table~\ref{tab:eigenvalues} shows the results for the lowest six resonance frequencies, including both their real and imaginary parts. The imaginary part is expressed in terms of the quality factor $Q_\ell$, which characterizes the sharpness of the resonance peak, and is defined as  $Q_\ell\equiv {\rm Re}\,f_\ell^{\rm Sch}/(2\,{\rm Im}\,f_\ell^{\rm Sch})$. The inverse of the quality factor is related to the width of the resonance peak, and for the multipole mode $\ell$, the full width at half maximum $\Delta f_\ell$ can be approximated as $\Delta f_\ell^{\rm Sch}/f_\ell^{\rm Sch}\approx 1/Q_\ell$.

Table~\ref{tab:eigenvalues} presents the Schumann resonance characteristics for different conductivity profiles shown in Fig.~\ref{fig:conductivity_profiles}. Despite differences in the conductivity models, the resultant peak frequencies (${\rm Re}\,f_\ell^{\rm Sch}$) and quality factors ($Q_\ell$) are in close agreement with each other. They are also consistent with actual observations (e.g., Table~1 of Ref.~\cite{Balser_Wagner1960}, Table~2.2 of Ref.~\cite{Bliokh1980}, Table~4 of Ref.~\cite{Ondraskova_etal2011}, see also Refs.~\cite{ Satori_Zieger1996, Nickolaenko_Hayakawa2013, Tatsis_etal2020}) 
and previous theoretical studies adopting other conductivity profiles (e.g., Table 4 of Ref.~\cite{Mushtak_Williams2002}, Table~1 of Ref.~\cite{Williams_Mushtak_Nickolaenko2006}). The characteristic behavior summarized in Table ~\ref{tab:eigenvalues}, namely that the peak frequencies $f_\ell^{\rm Sch}$ scale roughly linearly with $\ell$, while the quality factors remain almost constant regardless of $\ell$, is well explained by the simple two-exponential model proposed in  Refs.~\cite{Greifinger_Greifinger1978,Sentman1990}.
This agreement confirms that our theoretical formulation provides quantitatively reliable calculations for ELF EM waves and is applicable to predicting axion-induced EM signals. Furthermore, the weak dependence on the conductivity profile suggests that such predictions are robust against uncertainties in atmospheric conductivity, as discussed in more detail in Sec.~\ref{sec:discussions}.

\begin{table*}[tb]
 \caption{Peak frequencies and quality factor of the Schumann resonances for different conductivity profiles, obtained by solving Eq.~\eqref{eq:Schumann_resonance_condition} for $f_\ell^{\rm Sch}$. The results of the lowest six modes are shown. The peak frequencies correspond to the real part of $f_\ell^{\rm Sch}$, while the quality factors, given in the parentheses, are computed as $Q_\ell\equiv {\rm Re}\,f_\ell^{\rm Sch}/(2\,{\rm Im}\,f_\ell^{\rm Sch})$. All results are obtained by interpolating the tabulated conductivity profiles using $500$ segments. 
\label{tab:eigenvalues}}
\vspace*{0.3cm}
\begin{center}
\begin{tabular}{lcccc}
\hline
  $\ell$ &  Ref.~\cite{Kudintseva_etal2016} (Default) & Ref.~\cite{Nickolaenko_etal2016} (Median) & Ref.~\cite{Galuk_etal2018} (Day) & Ref.~\cite{Galuk_etal2018} (Night) 
\\
\hline
$1$  & 7.72\,Hz\, (3.91) & 7.80\,Hz\, (3.83) & 7.69\,Hz\, (3.99) & 7.87\,Hz\, (3.83) 
\\
$2$  & 13.9\,Hz\, (4.86) & 14.0\,Hz\, (4.88) & 13.9\,Hz\, (4.86) & 14.2\,Hz\, (4.87) 
\\
$3$  & 20.1\,Hz\, (5.44) & 20.1\,Hz\, (5.53) & 20.0\,Hz\, (5.45) & 20.4\,Hz\, (5.58) 
\\
$4$  & 26.2\,Hz\, (5.85) & 26.3\,Hz\, (5.96) & 26.1\,Hz\, (5.86) & 26.7\,Hz\, (6.07) 
\\
$5$  & 32.4\,Hz\, (6.15) & 32.4\,Hz\, (6.25) & 32.3\,Hz\, (6.14) & 32.9\,Hz\, (6.42) 
\\
$6$  & 38.6\,Hz\, (6.36) & 38.6\,Hz\, (6.46) & 38.5\,Hz\, (6.34) & 39.2\,Hz\, (6.68) 
\\

\hline
\end{tabular}
\end{center}
\end{table*}

\section{Upward-wave boundary condition in exponential conductivity}
\label{appendix:exponential_profile}

In Sec.~\ref{sec:formulation}, the solution of axion-induced EM wave was obtained by imposing the boundary conditions at the Earth's surface ($r=R_{\rm E}$) and lower ionosphere ($r=R_{\rm E}+h$ with $h\simeq100$\,km). In particular, at the altitude of lower ionosphere,  we require the homogeneous part of the solution to correspond to an upward-propagating wave, assuming constant conductivity above that altitude.

In this Appendix, as another boundary condition at the lower ionosphere, we consider an upward-propagating-wave condition assuming an exponential conductivity profile above the boundary. The resulting solution for the axion-induced magnetic field is discussed in Sec.~\ref{sec:discussions}, with the specific case at Eskdalemuir Observatory in Fig.~\ref{fig:Eskdalemuir_model_difference} (see faint blue curves). 

Following the layered representation of the radial mode function $d_{\ell m}$ as illustrated in Fig.~\ref{fig:multi-step_conductivity}, we denote the radius at which we impose the boundary condition by $r_N=x_N/\mass$. Above $r_N$, instead of the constant conductivity in Eq.~\eqref{eq:conductivity_step}, we assume that the conductivity behaves as
\begin{align}
\sigma(r)=\sigma_N\,e^{(r-r_N)/\zeta},\quad r>r_N
\label{eq:exponential_profile}
\end{align}

In imposing the boundary condition, having an analytic form for the mode function is helpful, facilitating the straightforward construction of the upward-propagating solution. At $r=r_N$, the conductivity is sufficiently large, and the square of refractive index $n^2$ is approximately described by $n^2\simeq i\,(\sigma_N/\mass)\gg1$. Furthermore, the characteristic scale $\zeta$ of the exponential profile in Eq.~\eqref{eq:exponential_profile} is much smaller than the Earth radius, $\zeta\ll R_{\rm E}$ (typically $\zeta\sim3-10$\,km, see e.g., Refs.~\cite{Greifinger_Greifinger1978, Sentman1990}).
Using these properties, the radial mode equation given in Eq.~\eqref{eq:mode_eq_dlm} or \eqref{eq:radial_mode_eq_u_ellm} can be  simplified. Introducing the new (dimensionless) variable $y$,
\begin{align}
y\equiv 2\,(\mass \zeta)\,\Bigl(\frac{\sigma_N}{\mass}\Bigr)^{1/2}\,e^{(r-r_N)/(2\zeta)}
\label{eq:variable_y}
\end{align}
and neglecting subdominant terms, we obtain~\cite{Greifinger_Greifinger1978, Sentman1990}
\begin{align}
\Bigl(\frac{d^2}{dy^2}-\frac{1}{y}\frac{d}{dy}+i\,\Bigr)\,d_{\ell m}=-i\,\gag\,a_0\,\frac{C_{\ell m}}{\ell}\frac{x_0^{\ell+2}}{x^{\ell+1}}.
\label{eq:mode_equation_exp_profile}
\end{align}
The right-hand side of Eq.~\eqref{eq:mode_equation_exp_profile} is still expressed in terms of $x$, but since $y$ varies much more rapidly than $x$, it can be approximated as constant. Then, the solution is analytically expressed, with its homogeneous part representing the upward-propagating wave:
\begin{align}
d_{\ell m}=\mathcal{A}_{\ell m}\,y\,H_1^{(1)}\Bigl(e^{i(\pi/4)}\,y\Bigr)-\gag\,a_0\,\frac{C_{\ell m}}{\ell}\frac{x_0^{\ell+2}}{x^{\ell+1}},
\label{eq:upgoing_solution_exp_profile}
\end{align}
where the function $H_1^{(1)}$ is the Hankel function of the first kind. Compared with the constant-conductivity case in Eq.~\eqref{eq:solution_outermost}, the argument of the Hankel function here involves $y$, leading to a faster radial decay of the mode function.

In Eq.~\eqref{eq:upgoing_solution_exp_profile}, the coefficient $\mathcal{A}_{\ell m}$ must be determined consistently by the matching condition with the solution below $r_N$ together with the lower boundary condition at the Earth's surface. This can be done by following the procedure in Sec.~\ref{subsec:multi_step}, with only a minor modification. Specifically, the vector $\vec{Y}(x_N)$, defined at Eq.~\eqref{eq:vect_Y} is replaced by the following expression: 
\begin{align}
\vec{Y}(x_N)&=\left(
\begin{array}{c}
   y_N \,H_1^{(1)}(e^{i(\pi/4)}y_N) 
    \\
    \\
   {\displaystyle   \frac{1}{n_N^2}\Bigl(\frac{d}{dx}+\frac{1}{x}\Bigr)\Bigl\{y\,H_1^{(1)}\Bigl(e^{i(\pi/4)}\,y\Bigr)\Bigr\}\Biggr|_{x=x_N} }
\end{array}
\right),
\label{eq:vect_Y_exp}
\end{align}
with $y_N\equiv2(\mass\zeta)\,(\sigma_N/\mass)^{1/2}$. Note that the second component can be recast as 
\begin{align}
    \frac{y_N^2}{n_N^2}
    \Bigl\{
    \frac{H_1^{(1)}\Bigl(e^{i(\pi/4)}y_N\Bigr)}{x_Ny_N}+
    \frac{e^{i(\pi/4)}H_0^{(1)}\Bigl(e^{i(\pi/4)}y_N\Bigr)}{2\mass\zeta}\,\,\Bigr\}.
\end{align}

\bibliographystyle{apsrev4-1}
\input{references.bbl}

\end{document}

%% file: references.bbl
%

%% file: ms_axion_search_TH.bbl
\begin{thebibliography}{83}%
\makeatletter
\providecommand \@ifxundefined [1]{%
 \@ifx{#1\undefined}
}%
\providecommand \@ifnum [1]{%
 \ifnum #1\expandafter \@firstoftwo
 \else \expandafter \@secondoftwo
 \fi
}%
\providecommand \@ifx [1]{%
 \ifx #1\expandafter \@firstoftwo
 \else \expandafter \@secondoftwo
 \fi
}%
\providecommand \natexlab [1]{#1}%
\providecommand \enquote  [1]{``#1''}%
\providecommand \bibnamefont  [1]{#1}%
\providecommand \bibfnamefont [1]{#1}%
\providecommand \citenamefont [1]{#1}%
\providecommand \href@noop [0]{\@secondoftwo}%
\providecommand \href [0]{\begingroup \@sanitize@url \@href}%
\providecommand \@href[1]{\@@startlink{#1}\@@href}%
\providecommand \@@href[1]{\endgroup#1\@@endlink}%
\providecommand \@sanitize@url [0]{\catcode `\\12\catcode `\$12\catcode
  `\&12\catcode `\#12\catcode `\^12\catcode `\_12\catcode `\%12\relax}%
\providecommand \@@startlink[1]{}%
\providecommand \@@endlink[0]{}%
\providecommand \url  [0]{\begingroup\@sanitize@url \@url }%
\providecommand \@url [1]{\endgroup\@href {#1}{\urlprefix }}%
\providecommand \urlprefix  [0]{URL }%
\providecommand \Eprint [0]{\href }%
\providecommand \doibase [0]{http://dx.doi.org/}%
\providecommand \selectlanguage [0]{\@gobble}%
\providecommand \bibinfo  [0]{\@secondoftwo}%
\providecommand \bibfield  [0]{\@secondoftwo}%
\providecommand \translation [1]{[#1]}%
\providecommand \BibitemOpen [0]{}%
\providecommand \bibitemStop [0]{}%
\providecommand \bibitemNoStop [0]{.\EOS\space}%
\providecommand \EOS [0]{\spacefactor3000\relax}%
\providecommand \BibitemShut  [1]{\csname bibitem#1\endcsname}%
\let\auto@bib@innerbib\@empty
\bibitem [{\citenamefont {{Bertone}}\ \emph {et~al.}(2005)\citenamefont
  {{Bertone}}, \citenamefont {{Hooper}},\ and\ \citenamefont
  {{Silk}}}]{Bertone_Hooper_Silk2005}%
  \BibitemOpen
  \bibfield  {author} {\bibinfo {author} {\bibfnamefont {G.}~\bibnamefont
  {{Bertone}}}, \bibinfo {author} {\bibfnamefont {D.}~\bibnamefont {{Hooper}}},
  \ and\ \bibinfo {author} {\bibfnamefont {J.}~\bibnamefont {{Silk}}},\ }\href
  {\doibase 10.1016/j.physrep.2004.08.031} {\bibfield  {journal} {\bibinfo
  {journal} {\physrep}\ }\textbf {\bibinfo {volume} {405}},\ \bibinfo {pages}
  {279} (\bibinfo {year} {2005})},\ \Eprint
  {http://arxiv.org/abs/hep-ph/0404175} {arXiv:hep-ph/0404175 [hep-ph]}
  \BibitemShut {NoStop}%
\bibitem [{\citenamefont {{Feng}}(2010)}]{Feng_2010}%
  \BibitemOpen
  \bibfield  {author} {\bibinfo {author} {\bibfnamefont {J.~L.}\ \bibnamefont
  {{Feng}}},\ }\href {\doibase 10.1146/annurev-astro-082708-101659} {\bibfield
  {journal} {\bibinfo  {journal} {\araa}\ }\textbf {\bibinfo {volume} {48}},\
  \bibinfo {pages} {495} (\bibinfo {year} {2010})},\ \Eprint
  {http://arxiv.org/abs/1003.0904} {arXiv:1003.0904 [astro-ph.CO]} \BibitemShut
  {NoStop}%
\bibitem [{\citenamefont {{Drlica-Wagner}}\ \emph {et~al.}(2022)\citenamefont
  {{Drlica-Wagner}}, \citenamefont {{Prescod-Weinstein}}, \citenamefont {{Yu}},
  \citenamefont {{Albert}}, \citenamefont {{Amin}}, \citenamefont {{Banerjee}},
  \citenamefont {{Baryakhtar}}, \citenamefont {{Bechtol}}, \citenamefont
  {{Bird}}, \citenamefont {{Birrer}}, \citenamefont {{Bringmann}},
  \citenamefont {{Caputo}}, \citenamefont {{Chakrabarti}}, \citenamefont
  {{Chen}}, \citenamefont {{Croon}}, \citenamefont {{Cyr-Racine}},
  \citenamefont {{Dawson}}, \citenamefont {{Dvorkin}}, \citenamefont
  {{Gluscevic}}, \citenamefont {{Gilman}}, \citenamefont {{Grin}},
  \citenamefont {{Hlo{\v{z}}ek}}, \citenamefont {{Leane}}, \citenamefont
  {{Li}}, \citenamefont {{Mao}}, \citenamefont {{Meyers}}, \citenamefont
  {{Mishra-Sharma}}, \citenamefont {{Mu{\~n}oz}}, \citenamefont {{Munshi}},
  \citenamefont {{Nadler}}, \citenamefont {{Parikh}}, \citenamefont {{Perez}},
  \citenamefont {{Peter}}, \citenamefont {{Profumo}}, \citenamefont {{Schutz}},
  \citenamefont {{Sehgal}}, \citenamefont {{Simon}}, \citenamefont {{Sinha}},
  \citenamefont {{Valluri}},\ and\ \citenamefont
  {{Wechsler}}}]{Snowmass_DarkMatter2021}%
  \BibitemOpen
  \bibfield  {author} {\bibinfo {author} {\bibfnamefont {A.}~\bibnamefont
  {{Drlica-Wagner}}}, \bibinfo {author} {\bibfnamefont {C.}~\bibnamefont
  {{Prescod-Weinstein}}}, \bibinfo {author} {\bibfnamefont {H.-B.}\
  \bibnamefont {{Yu}}}, \bibinfo {author} {\bibfnamefont {A.}~\bibnamefont
  {{Albert}}}, \bibinfo {author} {\bibfnamefont {M.}~\bibnamefont {{Amin}}},
  \bibinfo {author} {\bibfnamefont {A.}~\bibnamefont {{Banerjee}}}, \bibinfo
  {author} {\bibfnamefont {M.}~\bibnamefont {{Baryakhtar}}}, \bibinfo {author}
  {\bibfnamefont {K.}~\bibnamefont {{Bechtol}}}, \bibinfo {author}
  {\bibfnamefont {S.}~\bibnamefont {{Bird}}}, \bibinfo {author} {\bibfnamefont
  {S.}~\bibnamefont {{Birrer}}}, \bibinfo {author} {\bibfnamefont
  {T.}~\bibnamefont {{Bringmann}}}, \bibinfo {author} {\bibfnamefont
  {R.}~\bibnamefont {{Caputo}}}, \bibinfo {author} {\bibfnamefont
  {S.}~\bibnamefont {{Chakrabarti}}}, \bibinfo {author} {\bibfnamefont {T.~Y.}\
  \bibnamefont {{Chen}}}, \bibinfo {author} {\bibfnamefont {D.}~\bibnamefont
  {{Croon}}}, \bibinfo {author} {\bibfnamefont {F.-Y.}\ \bibnamefont
  {{Cyr-Racine}}}, \bibinfo {author} {\bibfnamefont {W.~A.}\ \bibnamefont
  {{Dawson}}}, \bibinfo {author} {\bibfnamefont {C.}~\bibnamefont {{Dvorkin}}},
  \bibinfo {author} {\bibfnamefont {V.}~\bibnamefont {{Gluscevic}}}, \bibinfo
  {author} {\bibfnamefont {D.}~\bibnamefont {{Gilman}}}, \bibinfo {author}
  {\bibfnamefont {D.}~\bibnamefont {{Grin}}}, \bibinfo {author} {\bibfnamefont
  {R.}~\bibnamefont {{Hlo{\v{z}}ek}}}, \bibinfo {author} {\bibfnamefont
  {R.~K.}\ \bibnamefont {{Leane}}}, \bibinfo {author} {\bibfnamefont {T.~S.}\
  \bibnamefont {{Li}}}, \bibinfo {author} {\bibfnamefont {Y.-Y.}\ \bibnamefont
  {{Mao}}}, \bibinfo {author} {\bibfnamefont {J.}~\bibnamefont {{Meyers}}},
  \bibinfo {author} {\bibfnamefont {S.}~\bibnamefont {{Mishra-Sharma}}},
  \bibinfo {author} {\bibfnamefont {J.~B.}\ \bibnamefont {{Mu{\~n}oz}}},
  \bibinfo {author} {\bibfnamefont {F.}~\bibnamefont {{Munshi}}}, \bibinfo
  {author} {\bibfnamefont {E.~O.}\ \bibnamefont {{Nadler}}}, \bibinfo {author}
  {\bibfnamefont {A.}~\bibnamefont {{Parikh}}}, \bibinfo {author}
  {\bibfnamefont {K.}~\bibnamefont {{Perez}}}, \bibinfo {author} {\bibfnamefont
  {A.~H.~G.}\ \bibnamefont {{Peter}}}, \bibinfo {author} {\bibfnamefont
  {S.}~\bibnamefont {{Profumo}}}, \bibinfo {author} {\bibfnamefont
  {K.}~\bibnamefont {{Schutz}}}, \bibinfo {author} {\bibfnamefont
  {N.}~\bibnamefont {{Sehgal}}}, \bibinfo {author} {\bibfnamefont {J.~D.}\
  \bibnamefont {{Simon}}}, \bibinfo {author} {\bibfnamefont {K.}~\bibnamefont
  {{Sinha}}}, \bibinfo {author} {\bibfnamefont {M.}~\bibnamefont {{Valluri}}},
  \ and\ \bibinfo {author} {\bibfnamefont {R.~H.}\ \bibnamefont {{Wechsler}}},\
  }\href@noop {} {\bibfield  {journal} {\bibinfo  {journal} {arXiv e-prints}\ }
  (\bibinfo {year} {2022})},\ \Eprint {http://arxiv.org/abs/2209.08215}
  {arXiv:2209.08215 [hep-ph]} \BibitemShut {NoStop}%
\bibitem [{\citenamefont {Preskill}\ \emph {et~al.}(1983)\citenamefont
  {Preskill}, \citenamefont {Wise},\ and\ \citenamefont
  {Wilczek}}]{Preskill:1982cy}%
  \BibitemOpen
  \bibfield  {author} {\bibinfo {author} {\bibfnamefont {J.}~\bibnamefont
  {Preskill}}, \bibinfo {author} {\bibfnamefont {M.~B.}\ \bibnamefont {Wise}},
  \ and\ \bibinfo {author} {\bibfnamefont {F.}~\bibnamefont {Wilczek}},\ }\href
  {\doibase 10.1016/0370-2693(83)90637-8} {\bibfield  {journal} {\bibinfo
  {journal} {Phys. Lett. B}\ }\textbf {\bibinfo {volume} {120}},\ \bibinfo
  {pages} {127} (\bibinfo {year} {1983})}\BibitemShut {NoStop}%
\bibitem [{\citenamefont {Abbott}\ and\ \citenamefont
  {Sikivie}(1983)}]{Abbott:1982af}%
  \BibitemOpen
  \bibfield  {author} {\bibinfo {author} {\bibfnamefont {L.~F.}\ \bibnamefont
  {Abbott}}\ and\ \bibinfo {author} {\bibfnamefont {P.}~\bibnamefont
  {Sikivie}},\ }\href {\doibase 10.1016/0370-2693(83)90638-X} {\bibfield
  {journal} {\bibinfo  {journal} {Phys. Lett. B}\ }\textbf {\bibinfo {volume}
  {120}},\ \bibinfo {pages} {133} (\bibinfo {year} {1983})}\BibitemShut
  {NoStop}%
\bibitem [{\citenamefont {Dine}\ and\ \citenamefont
  {Fischler}(1983)}]{Dine:1982ah}%
  \BibitemOpen
  \bibfield  {author} {\bibinfo {author} {\bibfnamefont {M.}~\bibnamefont
  {Dine}}\ and\ \bibinfo {author} {\bibfnamefont {W.}~\bibnamefont
  {Fischler}},\ }\href {\doibase 10.1016/0370-2693(83)90639-1} {\bibfield
  {journal} {\bibinfo  {journal} {Phys. Lett. B}\ }\textbf {\bibinfo {volume}
  {120}},\ \bibinfo {pages} {137} (\bibinfo {year} {1983})}\BibitemShut
  {NoStop}%
\bibitem [{\citenamefont {Peccei}\ and\ \citenamefont
  {Quinn}(1977)}]{peccei_quinn}%
  \BibitemOpen
  \bibfield  {author} {\bibinfo {author} {\bibfnamefont {R.~D.}\ \bibnamefont
  {Peccei}}\ and\ \bibinfo {author} {\bibfnamefont {H.~R.}\ \bibnamefont
  {Quinn}},\ }\href {\doibase 10.1103/PhysRevLett.38.1440} {\bibfield
  {journal} {\bibinfo  {journal} {Phys. Rev. Lett.}\ }\textbf {\bibinfo
  {volume} {38}},\ \bibinfo {pages} {1440} (\bibinfo {year}
  {1977})}\BibitemShut {NoStop}%
\bibitem [{\citenamefont {Weinberg}(1978)}]{Weinberg:1977ma}%
  \BibitemOpen
  \bibfield  {author} {\bibinfo {author} {\bibfnamefont {S.}~\bibnamefont
  {Weinberg}},\ }\href {\doibase 10.1103/PhysRevLett.40.223} {\bibfield
  {journal} {\bibinfo  {journal} {Phys. Rev. Lett.}\ }\textbf {\bibinfo
  {volume} {40}},\ \bibinfo {pages} {223} (\bibinfo {year} {1978})}\BibitemShut
  {NoStop}%
\bibitem [{\citenamefont {Wilczek}(1978)}]{Wilczek:1977pj}%
  \BibitemOpen
  \bibfield  {author} {\bibinfo {author} {\bibfnamefont {F.}~\bibnamefont
  {Wilczek}},\ }\href {\doibase 10.1103/PhysRevLett.40.279} {\bibfield
  {journal} {\bibinfo  {journal} {Phys. Rev. Lett.}\ }\textbf {\bibinfo
  {volume} {40}},\ \bibinfo {pages} {279} (\bibinfo {year} {1978})}\BibitemShut
  {NoStop}%
\bibitem [{\citenamefont {{Svrcek}}\ and\ \citenamefont
  {{Witten}}(2006)}]{Svrcek_Witten2006}%
  \BibitemOpen
  \bibfield  {author} {\bibinfo {author} {\bibfnamefont {P.}~\bibnamefont
  {{Svrcek}}}\ and\ \bibinfo {author} {\bibfnamefont {E.}~\bibnamefont
  {{Witten}}},\ }\href {\doibase 10.1088/1126-6708/2006/06/051} {\bibfield
  {journal} {\bibinfo  {journal} {Journal of High Energy Physics}\ }\textbf
  {\bibinfo {volume} {2006}},\ \bibinfo {eid} {051} (\bibinfo {year} {2006})},\
  \Eprint {http://arxiv.org/abs/hep-th/0605206} {arXiv:hep-th/0605206 [hep-th]}
  \BibitemShut {NoStop}%
\bibitem [{\citenamefont {{Arvanitaki}}\ \emph {et~al.}(2010)\citenamefont
  {{Arvanitaki}}, \citenamefont {{Dimopoulos}}, \citenamefont {{Dubovsky}},
  \citenamefont {{Kaloper}},\ and\ \citenamefont
  {{March-Russell}}}]{Arvanitaki_etal2010}%
  \BibitemOpen
  \bibfield  {author} {\bibinfo {author} {\bibfnamefont {A.}~\bibnamefont
  {{Arvanitaki}}}, \bibinfo {author} {\bibfnamefont {S.}~\bibnamefont
  {{Dimopoulos}}}, \bibinfo {author} {\bibfnamefont {S.}~\bibnamefont
  {{Dubovsky}}}, \bibinfo {author} {\bibfnamefont {N.}~\bibnamefont
  {{Kaloper}}}, \ and\ \bibinfo {author} {\bibfnamefont {J.}~\bibnamefont
  {{March-Russell}}},\ }\href {\doibase 10.1103/PhysRevD.81.123530} {\bibfield
  {journal} {\bibinfo  {journal} {\prd}\ }\textbf {\bibinfo {volume} {81}},\
  \bibinfo {eid} {123530} (\bibinfo {year} {2010})},\ \Eprint
  {http://arxiv.org/abs/0905.4720} {arXiv:0905.4720 [hep-th]} \BibitemShut
  {NoStop}%
\bibitem [{\citenamefont {{Marsh}}(2016)}]{Marsh_review2016}%
  \BibitemOpen
  \bibfield  {author} {\bibinfo {author} {\bibfnamefont {D.~J.~E.}\
  \bibnamefont {{Marsh}}},\ }\href {\doibase 10.1016/j.physrep.2016.06.005}
  {\bibfield  {journal} {\bibinfo  {journal} {\physrep}\ }\textbf {\bibinfo
  {volume} {643}},\ \bibinfo {pages} {1} (\bibinfo {year} {2016})},\ \Eprint
  {http://arxiv.org/abs/1510.07633} {arXiv:1510.07633 [astro-ph.CO]}
  \BibitemShut {NoStop}%
\bibitem [{\citenamefont {{Niemeyer}}(2020)}]{Niemeyer_review2020}%
  \BibitemOpen
  \bibfield  {author} {\bibinfo {author} {\bibfnamefont {J.~C.}\ \bibnamefont
  {{Niemeyer}}},\ }\href {\doibase 10.1016/j.ppnp.2020.103787} {\bibfield
  {journal} {\bibinfo  {journal} {Progress in Particle and Nuclear Physics}\
  }\textbf {\bibinfo {volume} {113}},\ \bibinfo {eid} {103787} (\bibinfo {year}
  {2020})},\ \Eprint {http://arxiv.org/abs/1912.07064} {arXiv:1912.07064
  [astro-ph.CO]} \BibitemShut {NoStop}%
\bibitem [{\citenamefont {{Ferreira}}(2021)}]{Ferreira_review2020}%
  \BibitemOpen
  \bibfield  {author} {\bibinfo {author} {\bibfnamefont {E.~G.~M.}\
  \bibnamefont {{Ferreira}}},\ }\href {\doibase 10.1007/s00159-021-00135-6}
  {\bibfield  {journal} {\bibinfo  {journal} {\aapr}\ }\textbf {\bibinfo
  {volume} {29}},\ \bibinfo {eid} {7} (\bibinfo {year} {2021})},\ \Eprint
  {http://arxiv.org/abs/2005.03254} {arXiv:2005.03254 [astro-ph.CO]}
  \BibitemShut {NoStop}%
\bibitem [{\citenamefont {{Hui}}(2021)}]{Hui_review2021}%
  \BibitemOpen
  \bibfield  {author} {\bibinfo {author} {\bibfnamefont {L.}~\bibnamefont
  {{Hui}}},\ }\href {\doibase 10.1146/annurev-astro-120920-010024} {\bibfield
  {journal} {\bibinfo  {journal} {\araa}\ }\textbf {\bibinfo {volume} {59}}
  (\bibinfo {year} {2021}),\ 10.1146/annurev-astro-120920-010024},\ \Eprint
  {http://arxiv.org/abs/2101.11735} {arXiv:2101.11735 [astro-ph.CO]}
  \BibitemShut {NoStop}%
\bibitem [{\citenamefont {{Hui}}\ \emph {et~al.}(2017)\citenamefont {{Hui}},
  \citenamefont {{Ostriker}}, \citenamefont {{Tremaine}},\ and\ \citenamefont
  {{Witten}}}]{Hui_etal2017}%
  \BibitemOpen
  \bibfield  {author} {\bibinfo {author} {\bibfnamefont {L.}~\bibnamefont
  {{Hui}}}, \bibinfo {author} {\bibfnamefont {J.~P.}\ \bibnamefont
  {{Ostriker}}}, \bibinfo {author} {\bibfnamefont {S.}~\bibnamefont
  {{Tremaine}}}, \ and\ \bibinfo {author} {\bibfnamefont {E.}~\bibnamefont
  {{Witten}}},\ }\href {\doibase 10.1103/PhysRevD.95.043541} {\bibfield
  {journal} {\bibinfo  {journal} {\prd}\ }\textbf {\bibinfo {volume} {95}},\
  \bibinfo {eid} {043541} (\bibinfo {year} {2017})},\ \Eprint
  {http://arxiv.org/abs/1610.08297} {arXiv:1610.08297 [astro-ph.CO]}
  \BibitemShut {NoStop}%
\bibitem [{\citenamefont {{Bullock}}\ and\ \citenamefont
  {{Boylan-Kolchin}}(2017)}]{Bullock_Boylan-Kolchin_review2017}%
  \BibitemOpen
  \bibfield  {author} {\bibinfo {author} {\bibfnamefont {J.~S.}\ \bibnamefont
  {{Bullock}}}\ and\ \bibinfo {author} {\bibfnamefont {M.}~\bibnamefont
  {{Boylan-Kolchin}}},\ }\href {\doibase 10.1146/annurev-astro-091916-055313}
  {\bibfield  {journal} {\bibinfo  {journal} {\araa}\ }\textbf {\bibinfo
  {volume} {55}},\ \bibinfo {pages} {343} (\bibinfo {year} {2017})},\ \Eprint
  {http://arxiv.org/abs/1707.04256} {arXiv:1707.04256 [astro-ph.CO]}
  \BibitemShut {NoStop}%
\bibitem [{\citenamefont {{Tulin}}\ and\ \citenamefont
  {{Yu}}(2018)}]{Tulin_Yu_review2018}%
  \BibitemOpen
  \bibfield  {author} {\bibinfo {author} {\bibfnamefont {S.}~\bibnamefont
  {{Tulin}}}\ and\ \bibinfo {author} {\bibfnamefont {H.-B.}\ \bibnamefont
  {{Yu}}},\ }\href {\doibase 10.1016/j.physrep.2017.11.004} {\bibfield
  {journal} {\bibinfo  {journal} {\physrep}\ }\textbf {\bibinfo {volume}
  {730}},\ \bibinfo {pages} {1} (\bibinfo {year} {2018})},\ \Eprint
  {http://arxiv.org/abs/1705.02358} {arXiv:1705.02358 [hep-ph]} \BibitemShut
  {NoStop}%
\bibitem [{\citenamefont {{Hu}}\ \emph {et~al.}(2000)\citenamefont {{Hu}},
  \citenamefont {{Barkana}},\ and\ \citenamefont
  {{Gruzinov}}}]{Hu_Barkana_Gruzinov2000}%
  \BibitemOpen
  \bibfield  {author} {\bibinfo {author} {\bibfnamefont {W.}~\bibnamefont
  {{Hu}}}, \bibinfo {author} {\bibfnamefont {R.}~\bibnamefont {{Barkana}}}, \
  and\ \bibinfo {author} {\bibfnamefont {A.}~\bibnamefont {{Gruzinov}}},\
  }\href {\doibase 10.1103/PhysRevLett.85.1158} {\bibfield  {journal} {\bibinfo
   {journal} {\prl}\ }\textbf {\bibinfo {volume} {85}},\ \bibinfo {pages}
  {1158} (\bibinfo {year} {2000})},\ \Eprint
  {http://arxiv.org/abs/astro-ph/0003365} {arXiv:astro-ph/0003365 [astro-ph]}
  \BibitemShut {NoStop}%
\bibitem [{\citenamefont {{Irastorza}}\ and\ \citenamefont
  {{Redondo}}(2018)}]{Irastorza_Redondo2018}%
  \BibitemOpen
  \bibfield  {author} {\bibinfo {author} {\bibfnamefont {I.~G.}\ \bibnamefont
  {{Irastorza}}}\ and\ \bibinfo {author} {\bibfnamefont {J.}~\bibnamefont
  {{Redondo}}},\ }\href {\doibase 10.1016/j.ppnp.2018.05.003} {\bibfield
  {journal} {\bibinfo  {journal} {Progress in Particle and Nuclear Physics}\
  }\textbf {\bibinfo {volume} {102}},\ \bibinfo {pages} {89} (\bibinfo {year}
  {2018})},\ \Eprint {http://arxiv.org/abs/1801.08127} {arXiv:1801.08127
  [hep-ph]} \BibitemShut {NoStop}%
\bibitem [{\citenamefont {{Sikivie}}(2021)}]{Sikivie_2021review}%
  \BibitemOpen
  \bibfield  {author} {\bibinfo {author} {\bibfnamefont {P.}~\bibnamefont
  {{Sikivie}}},\ }\href {\doibase 10.1103/RevModPhys.93.015004} {\bibfield
  {journal} {\bibinfo  {journal} {Reviews of Modern Physics}\ }\textbf
  {\bibinfo {volume} {93}},\ \bibinfo {eid} {015004} (\bibinfo {year}
  {2021})},\ \Eprint {http://arxiv.org/abs/2003.02206} {arXiv:2003.02206
  [hep-ph]} \BibitemShut {NoStop}%
\bibitem [{\citenamefont {Adams}\ \emph {et~al.}(2022)\citenamefont {Adams}
  \emph {et~al.}}]{Adams:2022pbo}%
  \BibitemOpen
  \bibfield  {author} {\bibinfo {author} {\bibfnamefont {C.~B.}\ \bibnamefont
  {Adams}} \emph {et~al.},\ }in\ \href@noop {} {\emph {\bibinfo {booktitle}
  {{Snowmass 2021}}}}\ (\bibinfo {year} {2022})\ \Eprint
  {http://arxiv.org/abs/2203.14923} {arXiv:2203.14923 [hep-ex]} \BibitemShut
  {NoStop}%
\bibitem [{\citenamefont {{Arza}}\ \emph {et~al.}(2022)\citenamefont {{Arza}},
  \citenamefont {{Fedderke}}, \citenamefont {{Graham}}, \citenamefont {{Jackson
  Kimball}},\ and\ \citenamefont {{Kalia}}}]{Arza_etal2022}%
  \BibitemOpen
  \bibfield  {author} {\bibinfo {author} {\bibfnamefont {A.}~\bibnamefont
  {{Arza}}}, \bibinfo {author} {\bibfnamefont {M.~A.}\ \bibnamefont
  {{Fedderke}}}, \bibinfo {author} {\bibfnamefont {P.~W.}\ \bibnamefont
  {{Graham}}}, \bibinfo {author} {\bibfnamefont {D.~F.}\ \bibnamefont {{Jackson
  Kimball}}}, \ and\ \bibinfo {author} {\bibfnamefont {S.}~\bibnamefont
  {{Kalia}}},\ }\href {\doibase 10.1103/PhysRevD.105.095007} {\bibfield
  {journal} {\bibinfo  {journal} {\prd}\ }\textbf {\bibinfo {volume} {105}},\
  \bibinfo {eid} {095007} (\bibinfo {year} {2022})},\ \Eprint
  {http://arxiv.org/abs/2112.09620} {arXiv:2112.09620 [hep-ph]} \BibitemShut
  {NoStop}%
\bibitem [{\citenamefont {Sulai}\ \emph {et~al.}(2023)\citenamefont {Sulai}
  \emph {et~al.}}]{Sulai:2023zqw}%
  \BibitemOpen
  \bibfield  {author} {\bibinfo {author} {\bibfnamefont {I.~A.}\ \bibnamefont
  {Sulai}} \emph {et~al.},\ }\href {\doibase 10.1103/PhysRevD.108.096026}
  {\bibfield  {journal} {\bibinfo  {journal} {Phys. Rev. D}\ }\textbf {\bibinfo
  {volume} {108}},\ \bibinfo {pages} {096026} (\bibinfo {year} {2023})},\
  \Eprint {http://arxiv.org/abs/2306.11575} {arXiv:2306.11575 [hep-ph]}
  \BibitemShut {NoStop}%
\bibitem [{\citenamefont {{Friel}}\ \emph {et~al.}(2024)\citenamefont
  {{Friel}}, \citenamefont {{Gjerloev}}, \citenamefont {{Kalia}},\ and\
  \citenamefont {{Zamora}}}]{Friel:2024shg}%
  \BibitemOpen
  \bibfield  {author} {\bibinfo {author} {\bibfnamefont {M.}~\bibnamefont
  {{Friel}}}, \bibinfo {author} {\bibfnamefont {J.~W.}\ \bibnamefont
  {{Gjerloev}}}, \bibinfo {author} {\bibfnamefont {S.}~\bibnamefont {{Kalia}}},
  \ and\ \bibinfo {author} {\bibfnamefont {A.}~\bibnamefont {{Zamora}}},\
  }\href {\doibase 10.1103/PhysRevD.110.115036} {\bibfield  {journal} {\bibinfo
   {journal} {\prd}\ }\textbf {\bibinfo {volume} {110}},\ \bibinfo {eid}
  {115036} (\bibinfo {year} {2024})},\ \Eprint
  {http://arxiv.org/abs/2408.16045} {arXiv:2408.16045 [hep-ph]} \BibitemShut
  {NoStop}%
\bibitem [{\citenamefont {{Arza}}\ \emph {et~al.}(2025)\citenamefont {{Arza}},
  \citenamefont {{Gong}}, \citenamefont {{Guo}}, \citenamefont {{Huang}},
  \citenamefont {{Shu}}, \citenamefont {{Tian}}, \citenamefont {{Wang}},
  \citenamefont {{Wei}}, \citenamefont {{Wu}}, \citenamefont {{Xia}},
  \citenamefont {{Yang}}, \citenamefont {{Yuan}}, \citenamefont {{Zhang}},
  \citenamefont {{Zhang}},\ and\ \citenamefont {{Zhu}}}]{Arza_etal2025}%
  \BibitemOpen
  \bibfield  {author} {\bibinfo {author} {\bibfnamefont {A.}~\bibnamefont
  {{Arza}}}, \bibinfo {author} {\bibfnamefont {Y.}~\bibnamefont {{Gong}}},
  \bibinfo {author} {\bibfnamefont {J.}~\bibnamefont {{Guo}}}, \bibinfo
  {author} {\bibfnamefont {X.}~\bibnamefont {{Huang}}}, \bibinfo {author}
  {\bibfnamefont {J.}~\bibnamefont {{Shu}}}, \bibinfo {author} {\bibfnamefont
  {H.}~\bibnamefont {{Tian}}}, \bibinfo {author} {\bibfnamefont
  {W.}~\bibnamefont {{Wang}}}, \bibinfo {author} {\bibfnamefont
  {K.}~\bibnamefont {{Wei}}}, \bibinfo {author} {\bibfnamefont
  {L.}~\bibnamefont {{Wu}}}, \bibinfo {author} {\bibfnamefont {M.}~\bibnamefont
  {{Xia}}}, \bibinfo {author} {\bibfnamefont {J.~M.}\ \bibnamefont {{Yang}}},
  \bibinfo {author} {\bibfnamefont {Q.}~\bibnamefont {{Yuan}}}, \bibinfo
  {author} {\bibfnamefont {Y.}~\bibnamefont {{Zhang}}}, \bibinfo {author}
  {\bibfnamefont {Y.}~\bibnamefont {{Zhang}}}, \ and\ \bibinfo {author}
  {\bibfnamefont {B.}~\bibnamefont {{Zhu}}},\ }\href {\doibase
  10.48550/arXiv.2511.16553} {\bibfield  {journal} {\bibinfo  {journal} {arXiv
  e-prints}\ ,\ \bibinfo {eid} {arXiv:2511.16553}} (\bibinfo {year} {2025})},\
  \Eprint {http://arxiv.org/abs/2511.16553} {arXiv:2511.16553 [hep-ph]}
  \BibitemShut {NoStop}%
\bibitem [{\citenamefont {{Davoudiasl}}\ and\ \citenamefont
  {{Huber}}(2006)}]{Davouldiasl_Huber2006}%
  \BibitemOpen
  \bibfield  {author} {\bibinfo {author} {\bibfnamefont {H.}~\bibnamefont
  {{Davoudiasl}}}\ and\ \bibinfo {author} {\bibfnamefont {P.}~\bibnamefont
  {{Huber}}},\ }\href {\doibase 10.1103/PhysRevLett.97.141302} {\bibfield
  {journal} {\bibinfo  {journal} {\prl}\ }\textbf {\bibinfo {volume} {97}},\
  \bibinfo {eid} {141302} (\bibinfo {year} {2006})},\ \Eprint
  {http://arxiv.org/abs/hep-ph/0509293} {arXiv:hep-ph/0509293 [hep-ph]}
  \BibitemShut {NoStop}%
\bibitem [{\citenamefont {{Fedderke}}\ \emph
  {et~al.}(2021{\natexlab{a}})\citenamefont {{Fedderke}}, \citenamefont
  {{Graham}}, \citenamefont {{Jackson Kimball}},\ and\ \citenamefont
  {{Kalia}}}]{Fedderke_etal2021_superMAG}%
  \BibitemOpen
  \bibfield  {author} {\bibinfo {author} {\bibfnamefont {M.~A.}\ \bibnamefont
  {{Fedderke}}}, \bibinfo {author} {\bibfnamefont {P.~W.}\ \bibnamefont
  {{Graham}}}, \bibinfo {author} {\bibfnamefont {D.~F.}\ \bibnamefont {{Jackson
  Kimball}}}, \ and\ \bibinfo {author} {\bibfnamefont {S.}~\bibnamefont
  {{Kalia}}},\ }\href {\doibase 10.1103/PhysRevD.104.095032} {\bibfield
  {journal} {\bibinfo  {journal} {\prd}\ }\textbf {\bibinfo {volume} {104}},\
  \bibinfo {eid} {095032} (\bibinfo {year} {2021}{\natexlab{a}})},\ \Eprint
  {http://arxiv.org/abs/2108.08852} {arXiv:2108.08852 [hep-ph]} \BibitemShut
  {NoStop}%
\bibitem [{\citenamefont {{Nishizawa}}\ \emph {et~al.}(2026)\citenamefont
  {{Nishizawa}}, \citenamefont {{Taruya}},\ and\ \citenamefont
  {{Himemoto}}}]{NTH_DataAnalysis}%
  \BibitemOpen
  \bibfield  {author} {\bibinfo {author} {\bibfnamefont {A.}~\bibnamefont
  {{Nishizawa}}}, \bibinfo {author} {\bibfnamefont {A.}~\bibnamefont
  {{Taruya}}}, \ and\ \bibinfo {author} {\bibfnamefont {Y.}~\bibnamefont
  {{Himemoto}}},\ }\href {\doibase 10.1093/ptep/ptag108} {\bibfield  {journal}
  {\bibinfo  {journal} {Progress of Theoretical and Experimental Physics}\
  }\textbf {\bibinfo {volume} {2026}},\ \bibinfo {eid} {073E02} (\bibinfo
  {year} {2026})},\ \Eprint {http://arxiv.org/abs/2504.07559} {arXiv:2504.07559
  [hep-ph]} \BibitemShut {NoStop}%
\bibitem [{\citenamefont {{Taruya}}\ \emph {et~al.}(2025)\citenamefont
  {{Taruya}}, \citenamefont {{Nishizawa}},\ and\ \citenamefont
  {{Himemoto}}}]{TNH_Letter}%
  \BibitemOpen
  \bibfield  {author} {\bibinfo {author} {\bibfnamefont {A.}~\bibnamefont
  {{Taruya}}}, \bibinfo {author} {\bibfnamefont {A.}~\bibnamefont
  {{Nishizawa}}}, \ and\ \bibinfo {author} {\bibfnamefont {Y.}~\bibnamefont
  {{Himemoto}}},\ }\href {\doibase 10.1093/ptep/ptaf136} {\bibfield  {journal}
  {\bibinfo  {journal} {Progress of Theoretical and Experimental Physics}\
  }\textbf {\bibinfo {volume} {2025}},\ \bibinfo {pages} {111E01} (\bibinfo
  {year} {2025})},\ \Eprint {http://arxiv.org/abs/2504.06653} {arXiv:2504.06653
  [hep-ph]} \BibitemShut {NoStop}%
\bibitem [{\citenamefont {{Nomura}}\ \emph {et~al.}(2026)\citenamefont
  {{Nomura}}, \citenamefont {{Nishizawa}}, \citenamefont {{Taruya}},\ and\
  \citenamefont {{Himemoto}}}]{NNTH_Darkphoton}%
  \BibitemOpen
  \bibfield  {author} {\bibinfo {author} {\bibfnamefont {K.}~\bibnamefont
  {{Nomura}}}, \bibinfo {author} {\bibfnamefont {A.}~\bibnamefont
  {{Nishizawa}}}, \bibinfo {author} {\bibfnamefont {A.}~\bibnamefont
  {{Taruya}}}, \ and\ \bibinfo {author} {\bibfnamefont {Y.}~\bibnamefont
  {{Himemoto}}},\ }\href {\doibase 10.1103/kw4j-8v12} {\bibfield  {journal}
  {\bibinfo  {journal} {\prd}\ }\textbf {\bibinfo {volume} {113}},\ \bibinfo
  {eid} {L101303} (\bibinfo {year} {2026})},\ \Eprint
  {http://arxiv.org/abs/2509.15783} {arXiv:2509.15783 [hep-ph]} \BibitemShut
  {NoStop}%
\bibitem [{\citenamefont {{Fedderke}}\ \emph
  {et~al.}(2021{\natexlab{b}})\citenamefont {{Fedderke}}, \citenamefont
  {{Graham}}, \citenamefont {{Jackson Kimball}},\ and\ \citenamefont
  {{Kalia}}}]{Fedderke_etal2021}%
  \BibitemOpen
  \bibfield  {author} {\bibinfo {author} {\bibfnamefont {M.~A.}\ \bibnamefont
  {{Fedderke}}}, \bibinfo {author} {\bibfnamefont {P.~W.}\ \bibnamefont
  {{Graham}}}, \bibinfo {author} {\bibfnamefont {D.~F.}\ \bibnamefont {{Jackson
  Kimball}}}, \ and\ \bibinfo {author} {\bibfnamefont {S.}~\bibnamefont
  {{Kalia}}},\ }\href {\doibase 10.1103/PhysRevD.104.075023} {\bibfield
  {journal} {\bibinfo  {journal} {\prd}\ }\textbf {\bibinfo {volume} {104}},\
  \bibinfo {eid} {075023} (\bibinfo {year} {2021}{\natexlab{b}})},\ \Eprint
  {http://arxiv.org/abs/2106.00022} {arXiv:2106.00022 [hep-ph]} \BibitemShut
  {NoStop}%
\bibitem [{\citenamefont {Anastassopoulos}\ \emph {et~al.}(2017)\citenamefont
  {Anastassopoulos} \emph {et~al.}}]{CAST:2017uph}%
  \BibitemOpen
  \bibfield  {author} {\bibinfo {author} {\bibfnamefont {V.}~\bibnamefont
  {Anastassopoulos}} \emph {et~al.} (\bibinfo {collaboration} {CAST}),\ }\href
  {\doibase 10.1038/nphys4109} {\bibfield  {journal} {\bibinfo  {journal}
  {Nature Phys.}\ }\textbf {\bibinfo {volume} {13}},\ \bibinfo {pages} {584}
  (\bibinfo {year} {2017})},\ \Eprint {http://arxiv.org/abs/1705.02290}
  {arXiv:1705.02290 [hep-ex]} \BibitemShut {NoStop}%
\bibitem [{\citenamefont {{Altenm{\"u}ller}}\ \emph {et~al.}(2024)\citenamefont
  {{Altenm{\"u}ller}}, \citenamefont {{Anastassopoulos}}, \citenamefont
  {{Arguedas-Cuendis}}, \citenamefont {{Aune}}, \citenamefont {{Baier}},
  \citenamefont {{Barth}}, \citenamefont {{Br{\"a}uninger}}, \citenamefont
  {{Cantatore}}, \citenamefont {{Caspers}}, \citenamefont {{Castel}},
  \citenamefont {{{\c{C}}etin}}, \citenamefont {{Christensen}}, \citenamefont
  {{Cogollos}}, \citenamefont {{Dafni}}, \citenamefont {{Davenport}},
  \citenamefont {{Decker}}, \citenamefont {{Desch}}, \citenamefont
  {{D{\'\i}ez-Ib{\'a}{\~n}ez}}, \citenamefont {{D{\"o}brich}}, \citenamefont
  {{Ferrer-Ribas}}, \citenamefont {{Fischer}}, \citenamefont {{Funk}},
  \citenamefont {{Gal{\'a}n}}, \citenamefont {{Garc{\'\i}a}}, \citenamefont
  {{Gardikiotis}}, \citenamefont {{Giomataris}}, \citenamefont {{Golm}},
  \citenamefont {{Hailey}}, \citenamefont {{Hasinoff}}, \citenamefont
  {{Hoffmann}}, \citenamefont {{Irastorza}}, \citenamefont {{Jacoby}},
  \citenamefont {{Jakobsen}}, \citenamefont {{Jakov{\v{c}}i{\'c}}},
  \citenamefont {{Kaminski}}, \citenamefont {{Karuza}}, \citenamefont
  {{Kostoglou}}, \citenamefont {{Krieger}}, \citenamefont {{Laki{\'c}}},
  \citenamefont {{Laurent}}, \citenamefont {{Luz{\'o}n}}, \citenamefont
  {{Malbrunot}}, \citenamefont {{Margalejo}}, \citenamefont {{Maroudas}},
  \citenamefont {{Miceli}}, \citenamefont {{Mirallas}}, \citenamefont
  {{Navarro}}, \citenamefont {{Obis}}, \citenamefont {{{\"O}zbey}},
  \citenamefont {{{\"O}zbozduman}}, \citenamefont {{Papaevangelou}},
  \citenamefont {{P{\'e}rez}}, \citenamefont {{Pivovaroff}}, \citenamefont
  {{Rosu}}, \citenamefont {{Ruiz-Ch{\'o}liz}}, \citenamefont {{Ruz}},
  \citenamefont {{Schmidt}}, \citenamefont {{Schumann}}, \citenamefont
  {{Semertzidis}}, \citenamefont {{Solanki}}, \citenamefont {{Stewart}},
  \citenamefont {{Vafeiadis}}, \citenamefont {{Vogel}}, \citenamefont
  {{Zioutas}},\ and\ \citenamefont {{CAST Collaboration}}}]{CAST:2024eil}%
  \BibitemOpen
  \bibfield  {author} {\bibinfo {author} {\bibfnamefont {K.}~\bibnamefont
  {{Altenm{\"u}ller}}}, \bibinfo {author} {\bibfnamefont {V.}~\bibnamefont
  {{Anastassopoulos}}}, \bibinfo {author} {\bibfnamefont {S.}~\bibnamefont
  {{Arguedas-Cuendis}}}, \bibinfo {author} {\bibfnamefont {S.}~\bibnamefont
  {{Aune}}}, \bibinfo {author} {\bibfnamefont {J.}~\bibnamefont {{Baier}}},
  \bibinfo {author} {\bibfnamefont {K.}~\bibnamefont {{Barth}}}, \bibinfo
  {author} {\bibfnamefont {H.}~\bibnamefont {{Br{\"a}uninger}}}, \bibinfo
  {author} {\bibfnamefont {G.}~\bibnamefont {{Cantatore}}}, \bibinfo {author}
  {\bibfnamefont {F.}~\bibnamefont {{Caspers}}}, \bibinfo {author}
  {\bibfnamefont {J.~F.}\ \bibnamefont {{Castel}}}, \bibinfo {author}
  {\bibfnamefont {S.~A.}\ \bibnamefont {{{\c{C}}etin}}}, \bibinfo {author}
  {\bibfnamefont {F.}~\bibnamefont {{Christensen}}}, \bibinfo {author}
  {\bibfnamefont {C.}~\bibnamefont {{Cogollos}}}, \bibinfo {author}
  {\bibfnamefont {T.}~\bibnamefont {{Dafni}}}, \bibinfo {author} {\bibfnamefont
  {M.}~\bibnamefont {{Davenport}}}, \bibinfo {author} {\bibfnamefont {T.~A.}\
  \bibnamefont {{Decker}}}, \bibinfo {author} {\bibfnamefont {K.}~\bibnamefont
  {{Desch}}}, \bibinfo {author} {\bibfnamefont {D.}~\bibnamefont
  {{D{\'\i}ez-Ib{\'a}{\~n}ez}}}, \bibinfo {author} {\bibfnamefont
  {B.}~\bibnamefont {{D{\"o}brich}}}, \bibinfo {author} {\bibfnamefont
  {E.}~\bibnamefont {{Ferrer-Ribas}}}, \bibinfo {author} {\bibfnamefont
  {H.}~\bibnamefont {{Fischer}}}, \bibinfo {author} {\bibfnamefont
  {W.}~\bibnamefont {{Funk}}}, \bibinfo {author} {\bibfnamefont
  {J.}~\bibnamefont {{Gal{\'a}n}}}, \bibinfo {author} {\bibfnamefont {J.~A.}\
  \bibnamefont {{Garc{\'\i}a}}}, \bibinfo {author} {\bibfnamefont
  {A.}~\bibnamefont {{Gardikiotis}}}, \bibinfo {author} {\bibfnamefont
  {I.}~\bibnamefont {{Giomataris}}}, \bibinfo {author} {\bibfnamefont
  {J.}~\bibnamefont {{Golm}}}, \bibinfo {author} {\bibfnamefont {C.~H.}\
  \bibnamefont {{Hailey}}}, \bibinfo {author} {\bibfnamefont {M.~D.}\
  \bibnamefont {{Hasinoff}}}, \bibinfo {author} {\bibfnamefont {D.~H.~H.}\
  \bibnamefont {{Hoffmann}}}, \bibinfo {author} {\bibfnamefont {I.~G.}\
  \bibnamefont {{Irastorza}}}, \bibinfo {author} {\bibfnamefont
  {J.}~\bibnamefont {{Jacoby}}}, \bibinfo {author} {\bibfnamefont {A.~C.}\
  \bibnamefont {{Jakobsen}}}, \bibinfo {author} {\bibfnamefont
  {K.}~\bibnamefont {{Jakov{\v{c}}i{\'c}}}}, \bibinfo {author} {\bibfnamefont
  {J.}~\bibnamefont {{Kaminski}}}, \bibinfo {author} {\bibfnamefont
  {M.}~\bibnamefont {{Karuza}}}, \bibinfo {author} {\bibfnamefont
  {S.}~\bibnamefont {{Kostoglou}}}, \bibinfo {author} {\bibfnamefont
  {C.}~\bibnamefont {{Krieger}}}, \bibinfo {author} {\bibfnamefont
  {B.}~\bibnamefont {{Laki{\'c}}}}, \bibinfo {author} {\bibfnamefont {J.~M.}\
  \bibnamefont {{Laurent}}}, \bibinfo {author} {\bibfnamefont {G.}~\bibnamefont
  {{Luz{\'o}n}}}, \bibinfo {author} {\bibfnamefont {C.}~\bibnamefont
  {{Malbrunot}}}, \bibinfo {author} {\bibfnamefont {C.}~\bibnamefont
  {{Margalejo}}}, \bibinfo {author} {\bibfnamefont {M.}~\bibnamefont
  {{Maroudas}}}, \bibinfo {author} {\bibfnamefont {L.}~\bibnamefont
  {{Miceli}}}, \bibinfo {author} {\bibfnamefont {H.}~\bibnamefont
  {{Mirallas}}}, \bibinfo {author} {\bibfnamefont {P.}~\bibnamefont
  {{Navarro}}}, \bibinfo {author} {\bibfnamefont {L.}~\bibnamefont {{Obis}}},
  \bibinfo {author} {\bibfnamefont {A.}~\bibnamefont {{{\"O}zbey}}}, \bibinfo
  {author} {\bibfnamefont {K.}~\bibnamefont {{{\"O}zbozduman}}}, \bibinfo
  {author} {\bibfnamefont {T.}~\bibnamefont {{Papaevangelou}}}, \bibinfo
  {author} {\bibfnamefont {O.}~\bibnamefont {{P{\'e}rez}}}, \bibinfo {author}
  {\bibfnamefont {M.~J.}\ \bibnamefont {{Pivovaroff}}}, \bibinfo {author}
  {\bibfnamefont {M.}~\bibnamefont {{Rosu}}}, \bibinfo {author} {\bibfnamefont
  {E.}~\bibnamefont {{Ruiz-Ch{\'o}liz}}}, \bibinfo {author} {\bibfnamefont
  {J.}~\bibnamefont {{Ruz}}}, \bibinfo {author} {\bibfnamefont
  {S.}~\bibnamefont {{Schmidt}}}, \bibinfo {author} {\bibfnamefont
  {M.}~\bibnamefont {{Schumann}}}, \bibinfo {author} {\bibfnamefont {Y.~K.}\
  \bibnamefont {{Semertzidis}}}, \bibinfo {author} {\bibfnamefont {S.~K.}\
  \bibnamefont {{Solanki}}}, \bibinfo {author} {\bibfnamefont {L.}~\bibnamefont
  {{Stewart}}}, \bibinfo {author} {\bibfnamefont {T.}~\bibnamefont
  {{Vafeiadis}}}, \bibinfo {author} {\bibfnamefont {J.~K.}\ \bibnamefont
  {{Vogel}}}, \bibinfo {author} {\bibfnamefont {K.}~\bibnamefont {{Zioutas}}},
  \ and\ \bibinfo {author} {\bibnamefont {{CAST Collaboration}}},\ }\href
  {\doibase 10.1103/PhysRevLett.133.221005} {\bibfield  {journal} {\bibinfo
  {journal} {\prl}\ }\textbf {\bibinfo {volume} {133}},\ \bibinfo {eid}
  {221005} (\bibinfo {year} {2024})},\ \Eprint
  {http://arxiv.org/abs/2406.16840} {arXiv:2406.16840 [hep-ex]} \BibitemShut
  {NoStop}%
\bibitem [{\citenamefont {{Schumann}}(1952{\natexlab{a}})}]{Schumann1952a}%
  \BibitemOpen
  \bibfield  {author} {\bibinfo {author} {\bibfnamefont {W.~O.}\ \bibnamefont
  {{Schumann}}},\ }\href {\doibase 10.1515/zna-1952-0202} {\bibfield  {journal}
  {\bibinfo  {journal} {Zeitschrift Naturforschung Teil A}\ }\textbf {\bibinfo
  {volume} {7}},\ \bibinfo {pages} {149} (\bibinfo {year}
  {1952}{\natexlab{a}})}\BibitemShut {NoStop}%
\bibitem [{\citenamefont {{Schumann}}(1952{\natexlab{b}})}]{Schumann1952b}%
  \BibitemOpen
  \bibfield  {author} {\bibinfo {author} {\bibfnamefont {W.~O.}\ \bibnamefont
  {{Schumann}}},\ }\href {\doibase 10.1515/zna-1952-3-404} {\bibfield
  {journal} {\bibinfo  {journal} {Zeitschrift Naturforschung Teil A}\ }\textbf
  {\bibinfo {volume} {7}},\ \bibinfo {pages} {250} (\bibinfo {year}
  {1952}{\natexlab{b}})}\BibitemShut {NoStop}%
\bibitem [{\citenamefont {{Jackson}}(1998)}]{Jackson_1998}%
  \BibitemOpen
  \bibfield  {author} {\bibinfo {author} {\bibfnamefont {J.~D.}\ \bibnamefont
  {{Jackson}}},\ }\href@noop {} {\emph {\bibinfo {title} {{Classical
  Electrodynamics, 3rd Edition}}}}\ (\bibinfo  {publisher} {John Wiley \&
  Sons},\ \bibinfo {year} {1998})\BibitemShut {NoStop}%
\bibitem [{\citenamefont {{Nickolaenko}}\ and\ \citenamefont
  {{Hayakawa}}(2002)}]{Nickolaenko_Hayakawa2002}%
  \BibitemOpen
  \bibfield  {author} {\bibinfo {author} {\bibfnamefont {A.~P.}\ \bibnamefont
  {{Nickolaenko}}}\ and\ \bibinfo {author} {\bibfnamefont {M.}~\bibnamefont
  {{Hayakawa}}},\ }\href@noop {} {\emph {\bibinfo {title} {{Resonances in the
  Earth-Ionosphere Cavity}}}}\ (\bibinfo  {publisher} {Springer Dordrecht},\
  \bibinfo {year} {2002})\BibitemShut {NoStop}%
\bibitem [{\citenamefont {{Sim{\~o}es}}\ \emph {et~al.}(2012)\citenamefont
  {{Sim{\~o}es}}, \citenamefont {{Pfaff}}, \citenamefont {{Berthelier}},\ and\
  \citenamefont {{Klenzing}}}]{Simoes_etal2012}%
  \BibitemOpen
  \bibfield  {author} {\bibinfo {author} {\bibfnamefont {F.}~\bibnamefont
  {{Sim{\~o}es}}}, \bibinfo {author} {\bibfnamefont {R.}~\bibnamefont
  {{Pfaff}}}, \bibinfo {author} {\bibfnamefont {J.-J.}\ \bibnamefont
  {{Berthelier}}}, \ and\ \bibinfo {author} {\bibfnamefont {J.}~\bibnamefont
  {{Klenzing}}},\ }\href {\doibase 10.1007/s11214-011-9854-0} {\bibfield
  {journal} {\bibinfo  {journal} {\ssr}\ }\textbf {\bibinfo {volume} {168}},\
  \bibinfo {pages} {551} (\bibinfo {year} {2012})}\BibitemShut {NoStop}%
\bibitem [{\citenamefont {{Nickolaenko}}\ and\ \citenamefont
  {{Hayakawa}}(2013)}]{Nickolaenko_Hayakawa2013}%
  \BibitemOpen
  \bibfield  {author} {\bibinfo {author} {\bibfnamefont {A.~P.}\ \bibnamefont
  {{Nickolaenko}}}\ and\ \bibinfo {author} {\bibfnamefont {M.}~\bibnamefont
  {{Hayakawa}}},\ }\href {\doibase 10.1007/978-4-431-54358-9} {\emph {\bibinfo
  {title} {{Schumann Resonance for Tyros}}}}\ (\bibinfo  {publisher} {Springer
  Tokyo},\ \bibinfo {year} {2013})\BibitemShut {NoStop}%
\bibitem [{\citenamefont {{Kudintseva}}\ \emph {et~al.}(2016)\citenamefont
  {{Kudintseva}}, \citenamefont {{Nickolaenko}}, \citenamefont {{Rycroft}},\
  and\ \citenamefont {{Odzimek}}}]{Kudintseva_etal2016}%
  \BibitemOpen
  \bibfield  {author} {\bibinfo {author} {\bibfnamefont {I.~G.}\ \bibnamefont
  {{Kudintseva}}}, \bibinfo {author} {\bibfnamefont {A.~P.}\ \bibnamefont
  {{Nickolaenko}}}, \bibinfo {author} {\bibfnamefont {M.~J.}\ \bibnamefont
  {{Rycroft}}}, \ and\ \bibinfo {author} {\bibfnamefont {A.}~\bibnamefont
  {{Odzimek}}},\ }\href {\doibase 10.1109/MSMW.2016.7538030} {\bibfield
  {journal} {\bibinfo  {journal} {Ann. Geophys.}\ }\textbf {\bibinfo {volume}
  {59}},\ \bibinfo {pages} {5} (\bibinfo {year} {2016})}\BibitemShut {NoStop}%
\bibitem [{\citenamefont {Nickolaenko}\ \emph {et~al.}(2016)\citenamefont
  {Nickolaenko}, \citenamefont {Galuk},\ and\ \citenamefont
  {Hayakawa}}]{Nickolaenko_etal2016}%
  \BibitemOpen
  \bibfield  {author} {\bibinfo {author} {\bibfnamefont {A.~P.}\ \bibnamefont
  {Nickolaenko}}, \bibinfo {author} {\bibfnamefont {Y.~P.}\ \bibnamefont
  {Galuk}}, \ and\ \bibinfo {author} {\bibfnamefont {M.}~\bibnamefont
  {Hayakawa}},\ }\href {\doibase 10.1186/s40064-016-1742-3} {\bibfield
  {journal} {\bibinfo  {journal} {Springer Plus}\ }\textbf {\bibinfo {volume}
  {5}} (\bibinfo {year} {2016}),\ 10.1186/s40064-016-1742-3}\BibitemShut
  {NoStop}%
\bibitem [{\citenamefont {{Galuk}}\ \emph {et~al.}(2018)\citenamefont
  {{Galuk}}, \citenamefont {{Nickolaenko}},\ and\ \citenamefont
  {{Hayakawa}}}]{Galuk_etal2018}%
  \BibitemOpen
  \bibfield  {author} {\bibinfo {author} {\bibfnamefont {Y.~P.}\ \bibnamefont
  {{Galuk}}}, \bibinfo {author} {\bibfnamefont {A.~P.}\ \bibnamefont
  {{Nickolaenko}}}, \ and\ \bibinfo {author} {\bibfnamefont {M.}~\bibnamefont
  {{Hayakawa}}},\ }\href {\doibase 10.1016/j.jastp.2018.01.001} {\bibfield
  {journal} {\bibinfo  {journal} {Journal of Atmospheric and Solar-Terrestrial
  Physics}\ }\textbf {\bibinfo {volume} {169}},\ \bibinfo {pages} {23}
  (\bibinfo {year} {2018})}\BibitemShut {NoStop}%
\bibitem [{\citenamefont {{Sentman}}(1990)}]{Sentman1990}%
  \BibitemOpen
  \bibfield  {author} {\bibinfo {author} {\bibfnamefont {D.~D.}\ \bibnamefont
  {{Sentman}}},\ }\href {\doibase 10.1016/0021-9169(90)90113-2} {\bibfield
  {journal} {\bibinfo  {journal} {Journal of Atmospheric and Terrestrial
  Physics}\ }\textbf {\bibinfo {volume} {52}},\ \bibinfo {pages} {35} (\bibinfo
  {year} {1990})}\BibitemShut {NoStop}%
\bibitem [{\citenamefont {{Greifinger}}\ and\ \citenamefont
  {{Greifinger}}(1978)}]{Greifinger_Greifinger1978}%
  \BibitemOpen
  \bibfield  {author} {\bibinfo {author} {\bibfnamefont {C.}~\bibnamefont
  {{Greifinger}}}\ and\ \bibinfo {author} {\bibfnamefont {P.}~\bibnamefont
  {{Greifinger}}},\ }\href {\doibase 10.1029/RS013i005p00831} {\bibfield
  {journal} {\bibinfo  {journal} {Radio Science}\ }\textbf {\bibinfo {volume}
  {13}},\ \bibinfo {pages} {831} (\bibinfo {year} {1978})}\BibitemShut
  {NoStop}%
\bibitem [{\citenamefont {{Millar}}\ \emph {et~al.}(2017)\citenamefont
  {{Millar}}, \citenamefont {{Raffelt}}, \citenamefont {{Redondo}},\ and\
  \citenamefont {{Steffen}}}]{Millar_etal2017}%
  \BibitemOpen
  \bibfield  {author} {\bibinfo {author} {\bibfnamefont {A.~J.}\ \bibnamefont
  {{Millar}}}, \bibinfo {author} {\bibfnamefont {G.~G.}\ \bibnamefont
  {{Raffelt}}}, \bibinfo {author} {\bibfnamefont {J.}~\bibnamefont
  {{Redondo}}}, \ and\ \bibinfo {author} {\bibfnamefont {F.~D.}\ \bibnamefont
  {{Steffen}}},\ }\href {\doibase 10.1088/1475-7516/2017/01/061} {\bibfield
  {journal} {\bibinfo  {journal} {\jcap}\ }\textbf {\bibinfo {volume} {2017}},\
  \bibinfo {eid} {061} (\bibinfo {year} {2017})},\ \Eprint
  {http://arxiv.org/abs/1612.07057} {arXiv:1612.07057 [hep-ph]} \BibitemShut
  {NoStop}%
\bibitem [{\citenamefont {{Barrera}}\ \emph {et~al.}(1985)\citenamefont
  {{Barrera}}, \citenamefont {{Estevez}},\ and\ \citenamefont
  {{Giraldo}}}]{Barrera_etal1985}%
  \BibitemOpen
  \bibfield  {author} {\bibinfo {author} {\bibfnamefont {R.~G.}\ \bibnamefont
  {{Barrera}}}, \bibinfo {author} {\bibfnamefont {G.~A.}\ \bibnamefont
  {{Estevez}}}, \ and\ \bibinfo {author} {\bibfnamefont {J.}~\bibnamefont
  {{Giraldo}}},\ }\href {\doibase 10.1088/0143-0807/6/4/014} {\bibfield
  {journal} {\bibinfo  {journal} {European Journal of Physics}\ }\textbf
  {\bibinfo {volume} {6}},\ \bibinfo {pages} {287} (\bibinfo {year}
  {1985})}\BibitemShut {NoStop}%
\bibitem [{\citenamefont {{Carrascal}}\ \emph {et~al.}(1991)\citenamefont
  {{Carrascal}}, \citenamefont {{Estevez}}, \citenamefont {{Lee}},\ and\
  \citenamefont {{Lorenzo}}}]{Carrascal_etal1991}%
  \BibitemOpen
  \bibfield  {author} {\bibinfo {author} {\bibfnamefont {B.}~\bibnamefont
  {{Carrascal}}}, \bibinfo {author} {\bibfnamefont {G.~A.}\ \bibnamefont
  {{Estevez}}}, \bibinfo {author} {\bibfnamefont {P.}~\bibnamefont {{Lee}}}, \
  and\ \bibinfo {author} {\bibfnamefont {V.}~\bibnamefont {{Lorenzo}}},\ }\href
  {\doibase 10.1088/0143-0807/12/4/007} {\bibfield  {journal} {\bibinfo
  {journal} {European Journal of Physics}\ }\textbf {\bibinfo {volume} {12}},\
  \bibinfo {pages} {184} (\bibinfo {year} {1991})}\BibitemShut {NoStop}%
\bibitem [{\citenamefont {{Hill}}(1954)}]{Hill_1954}%
  \BibitemOpen
  \bibfield  {author} {\bibinfo {author} {\bibfnamefont {E.~L.}\ \bibnamefont
  {{Hill}}},\ }\href {\doibase 10.1119/1.1933682} {\bibfield  {journal}
  {\bibinfo  {journal} {American Journal of Physics}\ }\textbf {\bibinfo
  {volume} {22}},\ \bibinfo {pages} {211} (\bibinfo {year} {1954})}\BibitemShut
  {NoStop}%
\bibitem [{\citenamefont {{Weinberg}}(1994)}]{Weinberg1994}%
  \BibitemOpen
  \bibfield  {author} {\bibinfo {author} {\bibfnamefont {E.~J.}\ \bibnamefont
  {{Weinberg}}},\ }\href {\doibase 10.1103/PhysRevD.49.1086} {\bibfield
  {journal} {\bibinfo  {journal} {\prd}\ }\textbf {\bibinfo {volume} {49}},\
  \bibinfo {pages} {1086} (\bibinfo {year} {1994})},\ \Eprint
  {http://arxiv.org/abs/hep-th/9308054} {arXiv:hep-th/9308054 [hep-th]}
  \BibitemShut {NoStop}%
\bibitem [{\citenamefont {{Alken}}\ \emph {et~al.}(2021)\citenamefont
  {{Alken}}, \citenamefont {{Th{\'e}bault}}, \citenamefont {{Beggan}},
  \citenamefont {{Amit}}, \citenamefont {{Aubert}}, \citenamefont
  {{Baerenzung}}, \citenamefont {{Bondar}}, \citenamefont {{Brown}},
  \citenamefont {{Califf}}, \citenamefont {{Chambodut}}, \citenamefont
  {{Chulliat}}, \citenamefont {{Cox}}, \citenamefont {{Finlay}}, \citenamefont
  {{Fournier}}, \citenamefont {{Gillet}}, \citenamefont {{Grayver}},
  \citenamefont {{Hammer}}, \citenamefont {{Holschneider}}, \citenamefont
  {{Huder}}, \citenamefont {{Hulot}}, \citenamefont {{Jager}}, \citenamefont
  {{Kloss}}, \citenamefont {{Korte}}, \citenamefont {{Kuang}}, \citenamefont
  {{Kuvshinov}}, \citenamefont {{Langlais}}, \citenamefont {{L{\'e}ger}},
  \citenamefont {{Lesur}}, \citenamefont {{Livermore}}, \citenamefont
  {{Lowes}}, \citenamefont {{Macmillan}}, \citenamefont {{Magnes}},
  \citenamefont {{Mandea}}, \citenamefont {{Marsal}}, \citenamefont {{Matzka}},
  \citenamefont {{Metman}}, \citenamefont {{Minami}}, \citenamefont
  {{Morschhauser}}, \citenamefont {{Mound}}, \citenamefont {{Nair}},
  \citenamefont {{Nakano}}, \citenamefont {{Olsen}}, \citenamefont
  {{Pav{\'o}n-Carrasco}}, \citenamefont {{Petrov}}, \citenamefont {{Ropp}},
  \citenamefont {{Rother}}, \citenamefont {{Sabaka}}, \citenamefont
  {{Sanchez}}, \citenamefont {{Saturnino}}, \citenamefont {{Schnepf}},
  \citenamefont {{Shen}}, \citenamefont {{Stolle}}, \citenamefont {{Tangborn}},
  \citenamefont {{T{\o}ffner-Clausen}}, \citenamefont {{Toh}}, \citenamefont
  {{Torta}}, \citenamefont {{Varner}}, \citenamefont {{Vervelidou}},
  \citenamefont {{Vigneron}}, \citenamefont {{Wardinski}}, \citenamefont
  {{Wicht}}, \citenamefont {{Woods}}, \citenamefont {{Yang}}, \citenamefont
  {{Zeren}},\ and\ \citenamefont {{Zhou}}}]{IGRF13_2021}%
  \BibitemOpen
  \bibfield  {author} {\bibinfo {author} {\bibfnamefont {P.}~\bibnamefont
  {{Alken}}}, \bibinfo {author} {\bibfnamefont {E.}~\bibnamefont
  {{Th{\'e}bault}}}, \bibinfo {author} {\bibfnamefont {C.~D.}\ \bibnamefont
  {{Beggan}}}, \bibinfo {author} {\bibfnamefont {H.}~\bibnamefont {{Amit}}},
  \bibinfo {author} {\bibfnamefont {J.}~\bibnamefont {{Aubert}}}, \bibinfo
  {author} {\bibfnamefont {J.}~\bibnamefont {{Baerenzung}}}, \bibinfo {author}
  {\bibfnamefont {T.~N.}\ \bibnamefont {{Bondar}}}, \bibinfo {author}
  {\bibfnamefont {W.~J.}\ \bibnamefont {{Brown}}}, \bibinfo {author}
  {\bibfnamefont {S.}~\bibnamefont {{Califf}}}, \bibinfo {author}
  {\bibfnamefont {A.}~\bibnamefont {{Chambodut}}}, \bibinfo {author}
  {\bibfnamefont {A.}~\bibnamefont {{Chulliat}}}, \bibinfo {author}
  {\bibfnamefont {G.~A.}\ \bibnamefont {{Cox}}}, \bibinfo {author}
  {\bibfnamefont {C.~C.}\ \bibnamefont {{Finlay}}}, \bibinfo {author}
  {\bibfnamefont {A.}~\bibnamefont {{Fournier}}}, \bibinfo {author}
  {\bibfnamefont {N.}~\bibnamefont {{Gillet}}}, \bibinfo {author}
  {\bibfnamefont {A.}~\bibnamefont {{Grayver}}}, \bibinfo {author}
  {\bibfnamefont {M.~D.}\ \bibnamefont {{Hammer}}}, \bibinfo {author}
  {\bibfnamefont {M.}~\bibnamefont {{Holschneider}}}, \bibinfo {author}
  {\bibfnamefont {L.}~\bibnamefont {{Huder}}}, \bibinfo {author} {\bibfnamefont
  {G.}~\bibnamefont {{Hulot}}}, \bibinfo {author} {\bibfnamefont
  {T.}~\bibnamefont {{Jager}}}, \bibinfo {author} {\bibfnamefont
  {C.}~\bibnamefont {{Kloss}}}, \bibinfo {author} {\bibfnamefont
  {M.}~\bibnamefont {{Korte}}}, \bibinfo {author} {\bibfnamefont
  {W.}~\bibnamefont {{Kuang}}}, \bibinfo {author} {\bibfnamefont
  {A.}~\bibnamefont {{Kuvshinov}}}, \bibinfo {author} {\bibfnamefont
  {B.}~\bibnamefont {{Langlais}}}, \bibinfo {author} {\bibfnamefont {J.~M.}\
  \bibnamefont {{L{\'e}ger}}}, \bibinfo {author} {\bibfnamefont
  {V.}~\bibnamefont {{Lesur}}}, \bibinfo {author} {\bibfnamefont {P.~W.}\
  \bibnamefont {{Livermore}}}, \bibinfo {author} {\bibfnamefont {F.~J.}\
  \bibnamefont {{Lowes}}}, \bibinfo {author} {\bibfnamefont {S.}~\bibnamefont
  {{Macmillan}}}, \bibinfo {author} {\bibfnamefont {W.}~\bibnamefont
  {{Magnes}}}, \bibinfo {author} {\bibfnamefont {M.}~\bibnamefont {{Mandea}}},
  \bibinfo {author} {\bibfnamefont {S.}~\bibnamefont {{Marsal}}}, \bibinfo
  {author} {\bibfnamefont {J.}~\bibnamefont {{Matzka}}}, \bibinfo {author}
  {\bibfnamefont {M.~C.}\ \bibnamefont {{Metman}}}, \bibinfo {author}
  {\bibfnamefont {T.}~\bibnamefont {{Minami}}}, \bibinfo {author}
  {\bibfnamefont {A.}~\bibnamefont {{Morschhauser}}}, \bibinfo {author}
  {\bibfnamefont {J.~E.}\ \bibnamefont {{Mound}}}, \bibinfo {author}
  {\bibfnamefont {M.}~\bibnamefont {{Nair}}}, \bibinfo {author} {\bibfnamefont
  {S.}~\bibnamefont {{Nakano}}}, \bibinfo {author} {\bibfnamefont
  {N.}~\bibnamefont {{Olsen}}}, \bibinfo {author} {\bibfnamefont {F.~J.}\
  \bibnamefont {{Pav{\'o}n-Carrasco}}}, \bibinfo {author} {\bibfnamefont
  {V.~G.}\ \bibnamefont {{Petrov}}}, \bibinfo {author} {\bibfnamefont
  {G.}~\bibnamefont {{Ropp}}}, \bibinfo {author} {\bibfnamefont
  {M.}~\bibnamefont {{Rother}}}, \bibinfo {author} {\bibfnamefont {T.~J.}\
  \bibnamefont {{Sabaka}}}, \bibinfo {author} {\bibfnamefont {S.}~\bibnamefont
  {{Sanchez}}}, \bibinfo {author} {\bibfnamefont {D.}~\bibnamefont
  {{Saturnino}}}, \bibinfo {author} {\bibfnamefont {N.~R.}\ \bibnamefont
  {{Schnepf}}}, \bibinfo {author} {\bibfnamefont {X.}~\bibnamefont {{Shen}}},
  \bibinfo {author} {\bibfnamefont {C.}~\bibnamefont {{Stolle}}}, \bibinfo
  {author} {\bibfnamefont {A.}~\bibnamefont {{Tangborn}}}, \bibinfo {author}
  {\bibfnamefont {L.}~\bibnamefont {{T{\o}ffner-Clausen}}}, \bibinfo {author}
  {\bibfnamefont {H.}~\bibnamefont {{Toh}}}, \bibinfo {author} {\bibfnamefont
  {J.~M.}\ \bibnamefont {{Torta}}}, \bibinfo {author} {\bibfnamefont
  {J.}~\bibnamefont {{Varner}}}, \bibinfo {author} {\bibfnamefont
  {F.}~\bibnamefont {{Vervelidou}}}, \bibinfo {author} {\bibfnamefont
  {P.}~\bibnamefont {{Vigneron}}}, \bibinfo {author} {\bibfnamefont
  {I.}~\bibnamefont {{Wardinski}}}, \bibinfo {author} {\bibfnamefont
  {J.}~\bibnamefont {{Wicht}}}, \bibinfo {author} {\bibfnamefont
  {A.}~\bibnamefont {{Woods}}}, \bibinfo {author} {\bibfnamefont
  {Y.}~\bibnamefont {{Yang}}}, \bibinfo {author} {\bibfnamefont
  {Z.}~\bibnamefont {{Zeren}}}, \ and\ \bibinfo {author} {\bibfnamefont
  {B.}~\bibnamefont {{Zhou}}},\ }\href {\doibase 10.1186/s40623-020-01288-x}
  {\bibfield  {journal} {\bibinfo  {journal} {Earth, Planets and Space}\
  }\textbf {\bibinfo {volume} {73}},\ \bibinfo {eid} {49} (\bibinfo {year}
  {2021})}\BibitemShut {NoStop}%
\bibitem [{\citenamefont {Wait}(2013)}]{wait2013electromagnetic}%
  \BibitemOpen
  \bibfield  {author} {\bibinfo {author} {\bibfnamefont {J.~R.}\ \bibnamefont
  {Wait}},\ }\href@noop {} {\emph {\bibinfo {title} {Electromagnetic Waves in
  Stratified Media: Revised Edition Including Supplemented Material}}},\
  \bibinfo {edition} {revised}\ ed.,\ edited by\ \bibinfo {editor}
  {\bibfnamefont {A.~L.}\ \bibnamefont {Cullen}}, \bibinfo {editor}
  {\bibfnamefont {V.~A.}\ \bibnamefont {Fock}}, \ and\ \bibinfo {editor}
  {\bibfnamefont {J.~R.}\ \bibnamefont {Wait}},\ International series of
  monographs on electromagnetic waves, vol.\ 3\ (\bibinfo  {publisher}
  {Pergamon (Elsevier)},\ \bibinfo {address} {Oxford; New York},\ \bibinfo
  {year} {2013})\ p.\ \bibinfo {pages} {620}\BibitemShut {NoStop}%
\bibitem [{\citenamefont {Bliokh}\ and\ \citenamefont
  {Nikolaenko}(1980)}]{Bliokh1980}%
  \BibitemOpen
  \bibfield  {author} {\bibinfo {author} {\bibfnamefont {P.~V.}\ \bibnamefont
  {Bliokh}}\ and\ \bibinfo {author} {\bibfnamefont {A.~P.}\ \bibnamefont
  {Nikolaenko}},\ }\href@noop {} {\emph {\bibinfo {title} {Schumann Resonances
  in the Earth-Ionosphere Cavity}}},\ edited by\ \bibinfo {editor}
  {\bibfnamefont {D.}~\bibnamefont {Llanwyn-Jones}},\ \bibinfo {series} {IEE
  Electromagnetic Waves Series}, Vol.~\bibinfo {volume} {8}\ (\bibinfo
  {publisher} {Peter Peregrinus Ltd.},\ \bibinfo {address} {Stevenage, UK},\
  \bibinfo {year} {1980})\ \bibinfo {note} {translated from Russian:
  \emph{Globalnye elektromagnitnye rezonansy v polosti
  Zemlia-ionosfera}}\BibitemShut {NoStop}%
\bibitem [{\citenamefont {{Galuk}}\ \emph {et~al.}(2019)\citenamefont
  {{Galuk}}, \citenamefont {{Kudintseva}}, \citenamefont {{Nickolaenko}},\ and\
  \citenamefont {{Hayakawa}}}]{Galuk_etal2019}%
  \BibitemOpen
  \bibfield  {author} {\bibinfo {author} {\bibfnamefont {Y.~P.}\ \bibnamefont
  {{Galuk}}}, \bibinfo {author} {\bibfnamefont {I.~G.}\ \bibnamefont
  {{Kudintseva}}}, \bibinfo {author} {\bibfnamefont {A.~P.}\ \bibnamefont
  {{Nickolaenko}}}, \ and\ \bibinfo {author} {\bibfnamefont {M.}~\bibnamefont
  {{Hayakawa}}},\ }\href {\doibase 10.1016/j.jastp.2019.105093} {\bibfield
  {journal} {\bibinfo  {journal} {Journal of Atmospheric and Solar-Terrestrial
  Physics}\ }\textbf {\bibinfo {volume} {194}},\ \bibinfo {eid} {105093}
  (\bibinfo {year} {2019})}\BibitemShut {NoStop}%
\bibitem [{\citenamefont {{Malta}}\ and\ \citenamefont
  {{Helay{\"e}l-Neto}}(2022)}]{Malta_Helayel-Neto2022}%
  \BibitemOpen
  \bibfield  {author} {\bibinfo {author} {\bibfnamefont {P.~C.}\ \bibnamefont
  {{Malta}}}\ and\ \bibinfo {author} {\bibfnamefont {J.~A.}\ \bibnamefont
  {{Helay{\"e}l-Neto}}},\ }\href {\doibase 10.1103/PhysRevD.106.116014}
  {\bibfield  {journal} {\bibinfo  {journal} {\prd}\ }\textbf {\bibinfo
  {volume} {106}},\ \bibinfo {eid} {116014} (\bibinfo {year} {2022})},\ \Eprint
  {http://arxiv.org/abs/2208.00527} {arXiv:2208.00527 [hep-ph]} \BibitemShut
  {NoStop}%
\bibitem [{\citenamefont {{Weber}}\ and\ \citenamefont {{de
  Boer}}(2010)}]{Weber_deBoer2010}%
  \BibitemOpen
  \bibfield  {author} {\bibinfo {author} {\bibfnamefont {M.}~\bibnamefont
  {{Weber}}}\ and\ \bibinfo {author} {\bibfnamefont {W.}~\bibnamefont {{de
  Boer}}},\ }\href {\doibase 10.1051/0004-6361/200913381} {\bibfield  {journal}
  {\bibinfo  {journal} {\aap}\ }\textbf {\bibinfo {volume} {509}},\ \bibinfo
  {eid} {A25} (\bibinfo {year} {2010})},\ \Eprint
  {http://arxiv.org/abs/0910.4272} {arXiv:0910.4272 [astro-ph.CO]} \BibitemShut
  {NoStop}%
\bibitem [{\citenamefont {{Sivertsson}}\ \emph {et~al.}(2018)\citenamefont
  {{Sivertsson}}, \citenamefont {{Silverwood}}, \citenamefont {{Read}},
  \citenamefont {{Bertone}},\ and\ \citenamefont
  {{Steger}}}]{Sivertsson_etal2018}%
  \BibitemOpen
  \bibfield  {author} {\bibinfo {author} {\bibfnamefont {S.}~\bibnamefont
  {{Sivertsson}}}, \bibinfo {author} {\bibfnamefont {H.}~\bibnamefont
  {{Silverwood}}}, \bibinfo {author} {\bibfnamefont {J.~I.}\ \bibnamefont
  {{Read}}}, \bibinfo {author} {\bibfnamefont {G.}~\bibnamefont {{Bertone}}}, \
  and\ \bibinfo {author} {\bibfnamefont {P.}~\bibnamefont {{Steger}}},\ }\href
  {\doibase 10.1093/mnras/sty977} {\bibfield  {journal} {\bibinfo  {journal}
  {\mnras}\ }\textbf {\bibinfo {volume} {478}},\ \bibinfo {pages} {1677}
  (\bibinfo {year} {2018})},\ \Eprint {http://arxiv.org/abs/1708.07836}
  {arXiv:1708.07836 [astro-ph.GA]} \BibitemShut {NoStop}%
\bibitem [{\citenamefont {{Sofue}}(2020)}]{Sofue2020}%
  \BibitemOpen
  \bibfield  {author} {\bibinfo {author} {\bibfnamefont {Y.}~\bibnamefont
  {{Sofue}}},\ }\href {\doibase 10.3390/galaxies8020037} {\bibfield  {journal}
  {\bibinfo  {journal} {Galaxies}\ }\textbf {\bibinfo {volume} {8}},\ \bibinfo
  {eid} {37} (\bibinfo {year} {2020})},\ \Eprint
  {http://arxiv.org/abs/2004.11688} {arXiv:2004.11688 [astro-ph.GA]}
  \BibitemShut {NoStop}%
\bibitem [{\citenamefont {{Holzworth}}\ \emph {et~al.}(1985)\citenamefont
  {{Holzworth}}, \citenamefont {{Kelley}}, \citenamefont {{Siefring}},
  \citenamefont {{Hale}},\ and\ \citenamefont
  {{Mitchell}}}]{Holzworth_etal1985}%
  \BibitemOpen
  \bibfield  {author} {\bibinfo {author} {\bibfnamefont {R.~H.}\ \bibnamefont
  {{Holzworth}}}, \bibinfo {author} {\bibfnamefont {M.~C.}\ \bibnamefont
  {{Kelley}}}, \bibinfo {author} {\bibfnamefont {C.~L.}\ \bibnamefont
  {{Siefring}}}, \bibinfo {author} {\bibfnamefont {L.~C.}\ \bibnamefont
  {{Hale}}}, \ and\ \bibinfo {author} {\bibfnamefont {J.~D.}\ \bibnamefont
  {{Mitchell}}},\ }\href {\doibase 10.1029/JA090iA10p09824} {\bibfield
  {journal} {\bibinfo  {journal} {\jgr}\ }\textbf {\bibinfo {volume} {90}},\
  \bibinfo {pages} {9824} (\bibinfo {year} {1985})}\BibitemShut {NoStop}%
\bibitem [{\citenamefont {{Takeda}}\ and\ \citenamefont
  {{Araki}}(1985)}]{Takeda_Araki1985}%
  \BibitemOpen
  \bibfield  {author} {\bibinfo {author} {\bibfnamefont {M.}~\bibnamefont
  {{Takeda}}}\ and\ \bibinfo {author} {\bibfnamefont {T.}~\bibnamefont
  {{Araki}}},\ }\href {\doibase 10.1016/0021-9169(85)90043-1} {\bibfield
  {journal} {\bibinfo  {journal} {Journal of Atmospheric and Terrestrial
  Physics}\ }\textbf {\bibinfo {volume} {47}},\ \bibinfo {pages} {601}
  (\bibinfo {year} {1985})}\BibitemShut {NoStop}%
\bibitem [{\citenamefont {{Richmond}}\ and\ \citenamefont
  {{Thayer}}(2000)}]{Richmond_Thayer2000}%
  \BibitemOpen
  \bibfield  {author} {\bibinfo {author} {\bibfnamefont {A.~D.}\ \bibnamefont
  {{Richmond}}}\ and\ \bibinfo {author} {\bibfnamefont {J.~P.}\ \bibnamefont
  {{Thayer}}},\ }\href {\doibase 10.1029/GM118p0131} {\bibfield  {journal}
  {\bibinfo  {journal} {Geophysical Monograph Series}\ }\textbf {\bibinfo
  {volume} {118}},\ \bibinfo {pages} {131} (\bibinfo {year}
  {2000})}\BibitemShut {NoStop}%
\bibitem [{\citenamefont {Cobb}\ and\ \citenamefont
  {Wells}(1970)}]{Cobb_Wells1970}%
  \BibitemOpen
  \bibfield  {author} {\bibinfo {author} {\bibfnamefont {W.~E.~C.}\
  \bibnamefont {Cobb}}\ and\ \bibinfo {author} {\bibfnamefont {H.~J.}\
  \bibnamefont {Wells}},\ }\href {\doibase
  10.1175/1520-0469(1970)027<0814:TECOOA>2.0.CO;2} {\bibfield  {journal}
  {\bibinfo  {journal} {Journal of the Atmospheric Sciences}\ }\textbf
  {\bibinfo {volume} {27}},\ \bibinfo {pages} {814} (\bibinfo {year}
  {1970})}\BibitemShut {NoStop}%
\bibitem [{\citenamefont {Kamra}\ and\ \citenamefont
  {Deshpande}(1995)}]{Kamra_Deshpande1995}%
  \BibitemOpen
  \bibfield  {author} {\bibinfo {author} {\bibfnamefont {A.~K.}\ \bibnamefont
  {Kamra}}\ and\ \bibinfo {author} {\bibfnamefont {C.~G.}\ \bibnamefont
  {Deshpande}},\ }\href {\doibase 10.1029/94JD02102} {\bibfield  {journal}
  {\bibinfo  {journal} {Journal of Geophysical Research: Atmospheres}\ }\textbf
  {\bibinfo {volume} {100}},\ \bibinfo {pages} {7139} (\bibinfo {year}
  {1995})}\BibitemShut {NoStop}%
\bibitem [{\citenamefont {Harrison}\ and\ \citenamefont
  {Carslaw}(2003)}]{Harrison_Carslaw2003}%
  \BibitemOpen
  \bibfield  {author} {\bibinfo {author} {\bibfnamefont {R.~G.}\ \bibnamefont
  {Harrison}}\ and\ \bibinfo {author} {\bibfnamefont {K.~S.}\ \bibnamefont
  {Carslaw}},\ }\href {\doibase 10.1029/2002RG000114} {\bibfield  {journal}
  {\bibinfo  {journal} {Reviews of Geophysics}\ }\textbf {\bibinfo {volume}
  {41}},\ \bibinfo {pages} {1012} (\bibinfo {year} {2003})}\BibitemShut
  {NoStop}%
\bibitem [{\citenamefont {{Toledo-Redondo}}\ \emph {et~al.}(2016)\citenamefont
  {{Toledo-Redondo}}, \citenamefont {{Salinas}}, \citenamefont {{Fornieles}},
  \citenamefont {{Port{\'\i}}},\ and\ \citenamefont
  {{Lichtenegger}}}]{Toledo-Redondo_2016}%
  \BibitemOpen
  \bibfield  {author} {\bibinfo {author} {\bibfnamefont {S.}~\bibnamefont
  {{Toledo-Redondo}}}, \bibinfo {author} {\bibfnamefont {A.}~\bibnamefont
  {{Salinas}}}, \bibinfo {author} {\bibfnamefont {J.}~\bibnamefont
  {{Fornieles}}}, \bibinfo {author} {\bibfnamefont {J.}~\bibnamefont
  {{Port{\'\i}}}}, \ and\ \bibinfo {author} {\bibfnamefont {H.~I.~M.}\
  \bibnamefont {{Lichtenegger}}},\ }\href {\doibase 10.1002/2015JA022083}
  {\bibfield  {journal} {\bibinfo  {journal} {Journal of Geophysical Research
  (Space Physics)}\ }\textbf {\bibinfo {volume} {121}},\ \bibinfo {pages}
  {5579} (\bibinfo {year} {2016})}\BibitemShut {NoStop}%
\bibitem [{\citenamefont {{Salinas}}\ \emph {et~al.}(2024)\citenamefont
  {{Salinas}}, \citenamefont {{Port{\'\i}}}, \citenamefont {{Navarro}},
  \citenamefont {{Toledo-Redondo}}, \citenamefont {{Albert}}, \citenamefont
  {{Castilla}},\ and\ \citenamefont {{Montagud-Camps}}}]{Salinas_etal2024}%
  \BibitemOpen
  \bibfield  {author} {\bibinfo {author} {\bibfnamefont {A.}~\bibnamefont
  {{Salinas}}}, \bibinfo {author} {\bibfnamefont {J.}~\bibnamefont
  {{Port{\'\i}}}}, \bibinfo {author} {\bibfnamefont {E.~A.}\ \bibnamefont
  {{Navarro}}}, \bibinfo {author} {\bibfnamefont {S.}~\bibnamefont
  {{Toledo-Redondo}}}, \bibinfo {author} {\bibfnamefont {I.}~\bibnamefont
  {{Albert}}}, \bibinfo {author} {\bibfnamefont {A.}~\bibnamefont
  {{Castilla}}}, \ and\ \bibinfo {author} {\bibfnamefont {V.}~\bibnamefont
  {{Montagud-Camps}}},\ }\href {\doibase 10.1016/j.cageo.2023.105499}
  {\bibfield  {journal} {\bibinfo  {journal} {Computers and Geosciences}\
  }\textbf {\bibinfo {volume} {183}},\ \bibinfo {eid} {105499} (\bibinfo {year}
  {2024})}\BibitemShut {NoStop}%
\bibitem [{\citenamefont {{Niknam}}\ and\ \citenamefont
  {{Simpson}}(2021)}]{Niknam_Simpson2021}%
  \BibitemOpen
  \bibfield  {author} {\bibinfo {author} {\bibfnamefont {K.}~\bibnamefont
  {{Niknam}}}\ and\ \bibinfo {author} {\bibfnamefont {J.}~\bibnamefont
  {{Simpson}}},\ }\href {\doibase 10.1109/JMMCT.2021.3136128} {\bibfield
  {journal} {\bibinfo  {journal} {IEEE Journal on Multiscale and Multiphysics
  Computational Techniques}\ }\textbf {\bibinfo {volume} {6}},\ \bibinfo
  {pages} {214} (\bibinfo {year} {2021})}\BibitemShut {NoStop}%
\bibitem [{\citenamefont {{Nickolaenko}}\ and\ \citenamefont
  {{Sentman}}(2007)}]{Nickolaenko_Sentman2007}%
  \BibitemOpen
  \bibfield  {author} {\bibinfo {author} {\bibfnamefont {A.~P.}\ \bibnamefont
  {{Nickolaenko}}}\ and\ \bibinfo {author} {\bibfnamefont {D.~D.}\ \bibnamefont
  {{Sentman}}},\ }\href {\doibase 10.1029/2006RS003473} {\bibfield  {journal}
  {\bibinfo  {journal} {Radio Science}\ }\textbf {\bibinfo {volume} {42}},\
  \bibinfo {eid} {RS2S13} (\bibinfo {year} {2007})}\BibitemShut {NoStop}%
\bibitem [{\citenamefont {{Navarro}}\ \emph {et~al.}(2008)\citenamefont
  {{Navarro}}, \citenamefont {{Soriano}}, \citenamefont {{Morente}},\ and\
  \citenamefont {{Port{\'\i}}}}]{Navarro_etal2008}%
  \BibitemOpen
  \bibfield  {author} {\bibinfo {author} {\bibfnamefont {E.~A.}\ \bibnamefont
  {{Navarro}}}, \bibinfo {author} {\bibfnamefont {A.}~\bibnamefont
  {{Soriano}}}, \bibinfo {author} {\bibfnamefont {J.~A.}\ \bibnamefont
  {{Morente}}}, \ and\ \bibinfo {author} {\bibfnamefont {J.~A.}\ \bibnamefont
  {{Port{\'\i}}}},\ }\href {\doibase 10.1029/2008JA013143} {\bibfield
  {journal} {\bibinfo  {journal} {Journal of Geophysical Research (Space
  Physics)}\ }\textbf {\bibinfo {volume} {113}},\ \bibinfo {eid} {A09301}
  (\bibinfo {year} {2008})}\BibitemShut {NoStop}%
\bibitem [{\citenamefont {{Silber}}\ \emph {et~al.}(2015)\citenamefont
  {{Silber}}, \citenamefont {{Price}}, \citenamefont {{Galanti}},\ and\
  \citenamefont {{Shuval}}}]{Silber_etal2015}%
  \BibitemOpen
  \bibfield  {author} {\bibinfo {author} {\bibfnamefont {I.}~\bibnamefont
  {{Silber}}}, \bibinfo {author} {\bibfnamefont {C.}~\bibnamefont {{Price}}},
  \bibinfo {author} {\bibfnamefont {E.}~\bibnamefont {{Galanti}}}, \ and\
  \bibinfo {author} {\bibfnamefont {A.}~\bibnamefont {{Shuval}}},\ }\href
  {\doibase 10.1002/2015JA021141} {\bibfield  {journal} {\bibinfo  {journal}
  {Journal of Geophysical Research (Space Physics)}\ }\textbf {\bibinfo
  {volume} {120}},\ \bibinfo {pages} {6036} (\bibinfo {year}
  {2015})}\BibitemShut {NoStop}%
\bibitem [{\citenamefont {{Nelson}}\ and\ \citenamefont
  {{Scholtz}}(2011)}]{Nelson_Scholtz2011}%
  \BibitemOpen
  \bibfield  {author} {\bibinfo {author} {\bibfnamefont {A.~E.}\ \bibnamefont
  {{Nelson}}}\ and\ \bibinfo {author} {\bibfnamefont {J.}~\bibnamefont
  {{Scholtz}}},\ }\href {\doibase 10.1103/PhysRevD.84.103501} {\bibfield
  {journal} {\bibinfo  {journal} {\prd}\ }\textbf {\bibinfo {volume} {84}},\
  \bibinfo {eid} {103501} (\bibinfo {year} {2011})},\ \Eprint
  {http://arxiv.org/abs/1105.2812} {arXiv:1105.2812 [hep-ph]} \BibitemShut
  {NoStop}%
\bibitem [{\citenamefont {{Fabbrichesi}}\ \emph {et~al.}(2020)\citenamefont
  {{Fabbrichesi}}, \citenamefont {{Gabrielli}},\ and\ \citenamefont
  {{Lanfranchi}}}]{Fabbrichesi_etal2020}%
  \BibitemOpen
  \bibfield  {author} {\bibinfo {author} {\bibfnamefont {M.}~\bibnamefont
  {{Fabbrichesi}}}, \bibinfo {author} {\bibfnamefont {E.}~\bibnamefont
  {{Gabrielli}}}, \ and\ \bibinfo {author} {\bibfnamefont {G.}~\bibnamefont
  {{Lanfranchi}}},\ }\href {\doibase 10.48550/arXiv.2005.01515} {\bibfield
  {journal} {\bibinfo  {journal} {arXiv e-prints}\ ,\ \bibinfo {eid}
  {arXiv:2005.01515}} (\bibinfo {year} {2020})},\ \Eprint
  {http://arxiv.org/abs/2005.01515} {arXiv:2005.01515 [hep-ph]} \BibitemShut
  {NoStop}%
\bibitem [{\citenamefont {{Arza}}\ \emph {et~al.}(2026)\citenamefont {{Arza}},
  \citenamefont {{Gong}}, \citenamefont {{Shu}}, \citenamefont {{Wu}},
  \citenamefont {{Yuan}},\ and\ \citenamefont
  {{Zhu}}}]{Arza_millichargedDM2026}%
  \BibitemOpen
  \bibfield  {author} {\bibinfo {author} {\bibfnamefont {A.}~\bibnamefont
  {{Arza}}}, \bibinfo {author} {\bibfnamefont {Y.}~\bibnamefont {{Gong}}},
  \bibinfo {author} {\bibfnamefont {J.}~\bibnamefont {{Shu}}}, \bibinfo
  {author} {\bibfnamefont {L.}~\bibnamefont {{Wu}}}, \bibinfo {author}
  {\bibfnamefont {Q.}~\bibnamefont {{Yuan}}}, \ and\ \bibinfo {author}
  {\bibfnamefont {B.}~\bibnamefont {{Zhu}}},\ }\href {\doibase
  10.1103/8xqd-dbrz} {\bibfield  {journal} {\bibinfo  {journal} {\prl}\
  }\textbf {\bibinfo {volume} {136}},\ \bibinfo {eid} {041001} (\bibinfo {year}
  {2026})},\ \Eprint {http://arxiv.org/abs/2501.14949} {arXiv:2501.14949
  [hep-ph]} \BibitemShut {NoStop}%
\bibitem [{\citenamefont {{Centers}}\ \emph {et~al.}(2019)\citenamefont
  {{Centers}}, \citenamefont {{Blanchard}}, \citenamefont {{Conrad}},
  \citenamefont {{Figueroa}}, \citenamefont {{Garcon}}, \citenamefont
  {{Gramolin}}, \citenamefont {{Kimball}}, \citenamefont {{Lawson}},
  \citenamefont {{Pelssers}}, \citenamefont {{Smiga}}, \citenamefont
  {{Sushkov}}, \citenamefont {{Wickenbrock}}, \citenamefont {{Budker}},\ and\
  \citenamefont {{Derevianko}}}]{Centers_etal2019}%
  \BibitemOpen
  \bibfield  {author} {\bibinfo {author} {\bibfnamefont {G.~P.}\ \bibnamefont
  {{Centers}}}, \bibinfo {author} {\bibfnamefont {J.~W.}\ \bibnamefont
  {{Blanchard}}}, \bibinfo {author} {\bibfnamefont {J.}~\bibnamefont
  {{Conrad}}}, \bibinfo {author} {\bibfnamefont {N.~L.}\ \bibnamefont
  {{Figueroa}}}, \bibinfo {author} {\bibfnamefont {A.}~\bibnamefont
  {{Garcon}}}, \bibinfo {author} {\bibfnamefont {A.~V.}\ \bibnamefont
  {{Gramolin}}}, \bibinfo {author} {\bibfnamefont {D.~F.~J.}\ \bibnamefont
  {{Kimball}}}, \bibinfo {author} {\bibfnamefont {M.}~\bibnamefont {{Lawson}}},
  \bibinfo {author} {\bibfnamefont {B.}~\bibnamefont {{Pelssers}}}, \bibinfo
  {author} {\bibfnamefont {J.~A.}\ \bibnamefont {{Smiga}}}, \bibinfo {author}
  {\bibfnamefont {A.~O.}\ \bibnamefont {{Sushkov}}}, \bibinfo {author}
  {\bibfnamefont {A.}~\bibnamefont {{Wickenbrock}}}, \bibinfo {author}
  {\bibfnamefont {D.}~\bibnamefont {{Budker}}}, \ and\ \bibinfo {author}
  {\bibfnamefont {A.}~\bibnamefont {{Derevianko}}},\ }\href {\doibase
  10.48550/arXiv.1905.13650} {\bibfield  {journal} {\bibinfo  {journal} {arXiv
  e-prints}\ ,\ \bibinfo {eid} {arXiv:1905.13650}} (\bibinfo {year} {2019})},\
  \Eprint {http://arxiv.org/abs/1905.13650} {arXiv:1905.13650 [astro-ph.CO]}
  \BibitemShut {NoStop}%
\bibitem [{\citenamefont {{Lisanti}}\ \emph {et~al.}(2021)\citenamefont
  {{Lisanti}}, \citenamefont {{Moschella}},\ and\ \citenamefont
  {{Terrano}}}]{Lisanti_etal2021}%
  \BibitemOpen
  \bibfield  {author} {\bibinfo {author} {\bibfnamefont {M.}~\bibnamefont
  {{Lisanti}}}, \bibinfo {author} {\bibfnamefont {M.}~\bibnamefont
  {{Moschella}}}, \ and\ \bibinfo {author} {\bibfnamefont {W.}~\bibnamefont
  {{Terrano}}},\ }\href {\doibase 10.1103/PhysRevD.104.055037} {\bibfield
  {journal} {\bibinfo  {journal} {\prd}\ }\textbf {\bibinfo {volume} {104}},\
  \bibinfo {eid} {055037} (\bibinfo {year} {2021})},\ \Eprint
  {http://arxiv.org/abs/2107.10260} {arXiv:2107.10260 [astro-ph.CO]}
  \BibitemShut {NoStop}%
\bibitem [{\citenamefont {Nakatsuka}\ \emph {et~al.}(2023)\citenamefont
  {Nakatsuka}, \citenamefont {Morisaki}, \citenamefont {Fujita}, \citenamefont
  {Kume}, \citenamefont {Michimura}, \citenamefont {Nagano},\ and\
  \citenamefont {Obata}}]{Nakatsuka:2022gaf}%
  \BibitemOpen
  \bibfield  {author} {\bibinfo {author} {\bibfnamefont {H.}~\bibnamefont
  {Nakatsuka}}, \bibinfo {author} {\bibfnamefont {S.}~\bibnamefont {Morisaki}},
  \bibinfo {author} {\bibfnamefont {T.}~\bibnamefont {Fujita}}, \bibinfo
  {author} {\bibfnamefont {J.}~\bibnamefont {Kume}}, \bibinfo {author}
  {\bibfnamefont {Y.}~\bibnamefont {Michimura}}, \bibinfo {author}
  {\bibfnamefont {K.}~\bibnamefont {Nagano}}, \ and\ \bibinfo {author}
  {\bibfnamefont {I.}~\bibnamefont {Obata}},\ }\href {\doibase
  10.1103/PhysRevD.108.092010} {\bibfield  {journal} {\bibinfo  {journal}
  {Phys. Rev. D}\ }\textbf {\bibinfo {volume} {108}},\ \bibinfo {pages}
  {092010} (\bibinfo {year} {2023})},\ \Eprint
  {http://arxiv.org/abs/2205.02960} {arXiv:2205.02960 [astro-ph.CO]}
  \BibitemShut {NoStop}%
\bibitem [{\citenamefont {{Press}}\ \emph {et~al.}(2002)\citenamefont
  {{Press}}, \citenamefont {{Teukolsky}}, \citenamefont {{Vetterling}},\ and\
  \citenamefont {{Flannery}}}]{Numerical_Recipes2002}%
  \BibitemOpen
  \bibfield  {author} {\bibinfo {author} {\bibfnamefont {W.~H.}\ \bibnamefont
  {{Press}}}, \bibinfo {author} {\bibfnamefont {S.~A.}\ \bibnamefont
  {{Teukolsky}}}, \bibinfo {author} {\bibfnamefont {W.~T.}\ \bibnamefont
  {{Vetterling}}}, \ and\ \bibinfo {author} {\bibfnamefont {B.~P.}\
  \bibnamefont {{Flannery}}},\ }\href@noop {} {\emph {\bibinfo {title}
  {{Numerical recipes in C++ : the art of scientific computing}}}}\ (\bibinfo
  {year} {2002})\BibitemShut {NoStop}%
\bibitem [{\citenamefont {{Balser}}\ and\ \citenamefont
  {{Wagner}}(1960)}]{Balser_Wagner1960}%
  \BibitemOpen
  \bibfield  {author} {\bibinfo {author} {\bibfnamefont {M.}~\bibnamefont
  {{Balser}}}\ and\ \bibinfo {author} {\bibfnamefont {C.~A.}\ \bibnamefont
  {{Wagner}}},\ }\href {\doibase 10.1038/188638a0} {\bibfield  {journal}
  {\bibinfo  {journal} {\nat}\ }\textbf {\bibinfo {volume} {188}},\ \bibinfo
  {pages} {638} (\bibinfo {year} {1960})}\BibitemShut {NoStop}%
\bibitem [{\citenamefont {{Ondr{\'a}{\v{s}}kov{\'a}}}\ \emph
  {et~al.}(2011)\citenamefont {{Ondr{\'a}{\v{s}}kov{\'a}}}, \citenamefont
  {{{\v{S}}ev{\v{c}}{\'\i}k}},\ and\ \citenamefont
  {{Kosteck{\'y}}}}]{Ondraskova_etal2011}%
  \BibitemOpen
  \bibfield  {author} {\bibinfo {author} {\bibfnamefont {A.}~\bibnamefont
  {{Ondr{\'a}{\v{s}}kov{\'a}}}}, \bibinfo {author} {\bibfnamefont
  {S.}~\bibnamefont {{{\v{S}}ev{\v{c}}{\'\i}k}}}, \ and\ \bibinfo {author}
  {\bibfnamefont {P.}~\bibnamefont {{Kosteck{\'y}}}},\ }\href {\doibase
  10.1016/j.jastp.2010.11.013} {\bibfield  {journal} {\bibinfo  {journal}
  {Journal of Atmospheric and Solar-Terrestrial Physics}\ }\textbf {\bibinfo
  {volume} {73}},\ \bibinfo {pages} {534} (\bibinfo {year} {2011})}\BibitemShut
  {NoStop}%
\bibitem [{\citenamefont {{S{\'a}Tori}}\ and\ \citenamefont
  {{Zieger}}(1996)}]{Satori_Zieger1996}%
  \BibitemOpen
  \bibfield  {author} {\bibinfo {author} {\bibfnamefont {G.}~\bibnamefont
  {{S{\'a}Tori}}}\ and\ \bibinfo {author} {\bibfnamefont {B.}~\bibnamefont
  {{Zieger}}},\ }\href {\doibase 10.1029/96JD00549} {\bibfield  {journal}
  {\bibinfo  {journal} {\jgr}\ }\textbf {\bibinfo {volume} {101}},\ \bibinfo
  {pages} {29,663} (\bibinfo {year} {1996})}\BibitemShut {NoStop}%
\bibitem [{\citenamefont {{Tatsis}}\ \emph {et~al.}(2020)\citenamefont
  {{Tatsis}}, \citenamefont {{Christofilakis}}, \citenamefont {{Chronopoulos}},
  \citenamefont {{Baldoumas}}, \citenamefont {{Sakkas}}, \citenamefont
  {{Paschalidou}}, \citenamefont {{Kassomenos}}, \citenamefont {{Petrou}},
  \citenamefont {{Kostarakis}}, \citenamefont {{Repapis}},\ and\ \citenamefont
  {{Tritakis}}}]{Tatsis_etal2020}%
  \BibitemOpen
  \bibfield  {author} {\bibinfo {author} {\bibfnamefont {G.}~\bibnamefont
  {{Tatsis}}}, \bibinfo {author} {\bibfnamefont {V.}~\bibnamefont
  {{Christofilakis}}}, \bibinfo {author} {\bibfnamefont {S.~K.}\ \bibnamefont
  {{Chronopoulos}}}, \bibinfo {author} {\bibfnamefont {G.}~\bibnamefont
  {{Baldoumas}}}, \bibinfo {author} {\bibfnamefont {A.}~\bibnamefont
  {{Sakkas}}}, \bibinfo {author} {\bibfnamefont {A.~K.}\ \bibnamefont
  {{Paschalidou}}}, \bibinfo {author} {\bibfnamefont {P.}~\bibnamefont
  {{Kassomenos}}}, \bibinfo {author} {\bibfnamefont {I.}~\bibnamefont
  {{Petrou}}}, \bibinfo {author} {\bibfnamefont {P.}~\bibnamefont
  {{Kostarakis}}}, \bibinfo {author} {\bibfnamefont {C.}~\bibnamefont
  {{Repapis}}}, \ and\ \bibinfo {author} {\bibfnamefont {V.}~\bibnamefont
  {{Tritakis}}},\ }\href {\doibase 10.1016/j.scitotenv.2020.136926} {\bibfield
  {journal} {\bibinfo  {journal} {Science of the Total Environment}\ }\textbf
  {\bibinfo {volume} {715}},\ \bibinfo {pages} {136926} (\bibinfo {year}
  {2020})}\BibitemShut {NoStop}%
\bibitem [{\citenamefont {{Mushtak}}\ and\ \citenamefont
  {{Williams}}(2002)}]{Mushtak_Williams2002}%
  \BibitemOpen
  \bibfield  {author} {\bibinfo {author} {\bibfnamefont {V.~C.}\ \bibnamefont
  {{Mushtak}}}\ and\ \bibinfo {author} {\bibfnamefont {E.~R.}\ \bibnamefont
  {{Williams}}},\ }\href {\doibase 10.1016/S1364-6826(02)00222-5} {\bibfield
  {journal} {\bibinfo  {journal} {Journal of Atmospheric and Solar-Terrestrial
  Physics}\ }\textbf {\bibinfo {volume} {64}},\ \bibinfo {pages} {1989}
  (\bibinfo {year} {2002})}\BibitemShut {NoStop}%
\bibitem [{\citenamefont {{Williams}}\ \emph {et~al.}(2006)\citenamefont
  {{Williams}}, \citenamefont {{Mushtak}},\ and\ \citenamefont
  {{Nickolaenko}}}]{Williams_Mushtak_Nickolaenko2006}%
  \BibitemOpen
  \bibfield  {author} {\bibinfo {author} {\bibfnamefont {E.~R.}\ \bibnamefont
  {{Williams}}}, \bibinfo {author} {\bibfnamefont {V.~C.}\ \bibnamefont
  {{Mushtak}}}, \ and\ \bibinfo {author} {\bibfnamefont {A.~P.}\ \bibnamefont
  {{Nickolaenko}}},\ }\href {\doibase 10.1029/2005JD006944} {\bibfield
  {journal} {\bibinfo  {journal} {Journal of Geophysical Research
  (Atmospheres)}\ }\textbf {\bibinfo {volume} {111}},\ \bibinfo {eid} {D16107}
  (\bibinfo {year} {2006})}\BibitemShut {NoStop}%
\end{thebibliography}
